%% file: hyper-chip.tex
\documentclass[a4paper,12pt]{article}
\usepackage{tikz}
\usetikzlibrary{positioning}
\usepackage[backend=biber,style=chem-angew,
			sorting=none,
			maxbibnames=100,giveninits=true,isbn=false,
			url=false,doi=true,alldates=year]{biblatex}
\usepackage[margin=1.65cm,tmargin=2cm,bmargin=2cm]{geometry}
\usepackage{amsmath}
\usepackage{bm}
\usepackage{mathptmx}
\usepackage{graphicx}
\usepackage{booktabs}
\usepackage{setspace}
\usepackage[pdfusetitle,linktocpage=true,colorlinks=true,citecolor=blue]{hyperref}
\usepackage{subfiles}
\usepackage{paralist}
\usepackage{braket}
\usepackage{multirow}
\usepackage{tabularx}
\usepackage{array}

\newenvironment{compactlist}{ % List with minimal white space to fit in small areas, e.g. table cell
    \begin{minipage}[t]{\linewidth}\begin{list}{$\bullet$}{\leftmargin=0.5em \rightmargin=0em \topsep=0em \parskip=0em \parsep=0em \listparindent=0em \partopsep=0em \itemsep=0pt \itemindent=0em \labelwidth=\leftmargin\labelsep+0.25em}
}{
    \end{list}\end{minipage}
}

\input{symbols}

\newcommand{\chemical}[1]{{\ensuremath{\mathrm{#1}}}}

\graphicspath{{./}{./Figures/}{../Figures/}}

\AtBeginDocument{%
}

\title{Synergies between Hyperpolarized NMR and Microfluidics: A Review}
\date{Edited by David Neuhaus and Geoffrey Bodenhausen}

\author{James Eills\thanks{Institute for Physics, Johannes Gutenberg University, D-55090 Mainz, Germany} \thanks{GSI Helmholtzzentrum für Schwerionenforschung GmbH, Helmholtz-Institut Mainz, 55128 Mainz, Germany} \thanks{Correspondence: eills@uni-mainz.de},
William Hale\thanks{Department of Chemistry, University of Florida, USA, 32611}, and
Marcel Utz\thanks{School of Chemistry, University of Southampton, UK SO17 1BJ} \thanks{Correspondence: Marcel.Utz@soton.ac.uk}}

\begin{document}

\maketitle

\begin{abstract}
Hyperpolarized nuclear magnetic resonance and lab-on-a-chip microfluidics are two 
dynamic, but until recently quite distinct, fields of research. Recent developments in both areas increased their synergistic overlap. By microfluidic integration, many complex experimental steps can be brought together onto a single platform. Microfluidic devices are therefore increasingly finding applications in medical diagnostics, forensic analysis, and biomedical research. In particular, they provide novel and powerful ways to culture cells, cell aggregates, and even functional models of entire organs. Nuclear magnetic resonance is a non-invasive, high-resolution spectroscopic technique which allows real-time process monitoring with chemical specificity. It is ideally
suited for observing metabolic and other biological and chemical processes in 
microfluidic systems. However, its intrinsically low sensitivity has limited its application. Recent advances in nuclear hyperpolarization techniques may change this: under special circumstances, it is possible to enhance NMR signals by up to 5 orders of magnitude, which dramatically extends the utility of 
NMR in the context of microfluidic systems. At the same time, hyperpolarization 
requires complex chemical and/or physical manipulations, which in turn may benefit
from microfluidic implementation. In fact,  many hyperpolarization methodologies rely on
processes that are more efficient at the micro-scale,
such as molecular diffusion, penetration electromagnetic radiation into the sample, 
or restricted molecular mobility on a surface. 
In this review we examine the confluence between the fields of hyperpolarization-enhanced NMR and microfluidics, and assess how these areas of research have mutually benefited one another, and will continue to do so.
\end{abstract}

\tableofcontents

\section{Introduction}
Nuclear magnetic resonance (NMR) is one of the most flexible experimental techniques
to study the structure and dynamics of matter. It provides a non-invasive 
means to observe systems in real-time with high chemical specificity. As a result, NMR spectroscopy has found
countless applications in chemistry, materials science, and the life sciences, and magnetic resonance
imaging (MRI) is widely used in medical diagnostics. This success is all the more impressive considering
that the sensitivity of  
magnetic resonance is quite limited compared to most other spectroscopic techniques. In part, this is due to the extreme
weakness of nuclear paramagnetism, which leads to low thermal equilibrium spin polarization on the order of $10^{-5}$. This means that even at the highest practical magnetic fields only one in $10^5$ spins contributes to the NMR signal. The problem of sensitivity becomes particularly acute when the amount of sample available is limited, as is generally the case in microfluidics.

Over the past two decades, the field
of microfluidics has made great strides. Lab-on-a-chip devices exploit microfabrication 
techniques such as soft lithography to implement highly complex fluidic circuits, which may include valves and
pumps, as well as passive fluidic elements such as diodes, resistors, and
capacitors. Since increased complexity comes at little additional cost with lithographically
fabricated structures, microfluidic devices can be used to parallelize assays 
for high throughput. Arguably the most
important aspect is the ability to engineer precisely-controlled conditions for
the growth and manipulation of small biological systems such as individual cells,
cell spheroids, or small organisms, in order to study their response to external
stimuli such as drugs, toxins, cytokines, or changes in nutrient provision.
Typical microfluidic systems operate with volumes ranging from picoliters up 
to tens of microliters. Solution-state NMR spectroscopy with
sample volumes around 1~$\mu$l or below is challenging due to the limited
sensitivity already mentioned. It is therefore unsurprising that NMR is not commonly used as
an observation/read out technique for microfluidics. So far, invasive optical readout methods based
on fluorescent probes, immunoassays, and electrochemical detection are far more
common. This is unfortunate since NMR is a non-invasive method, making it ideal 
for quantifying metabolic processes in living systems. Recent advances in microfluidic NMR technology have been
largely motivated by this possibility. However, even the most advanced micro-NMR
probes can only provide concentration limits of detection
\footnote{The concentration (cLOD) and number (nLOD) limits of detection are defined \cite{badilitaMicroscaleNuclearMagnetic2012}
as the concentration or number of spins per square root of measurement bandwidth that must resonate in order to produce a signal that is a factor of 3 above the noise. This is 
equivalent to the convention that scientific conclusions should be supported by results
that differ from their controls
by 3 standard deviations, corresponding to a statistical significance of better than 99.7\%.
In the context of pulsed liquid state NMR, the relevant bandwidth is usually given
by the inverse of the total measurement time, which in turn is the product
of the number of transients averaged and the relaxation delay between transients.}
around 1~mM$\sqrt{\text{s}}$
at 1~$\mu$l for \textsuperscript{1}H at magnetic fields around 14~T. Many compounds of interest in biological
systems are present at significantly lower concentrations, or only transiently visible.

\begin{figure}
	\begin{center}
			\includegraphics[width=8.4cm]{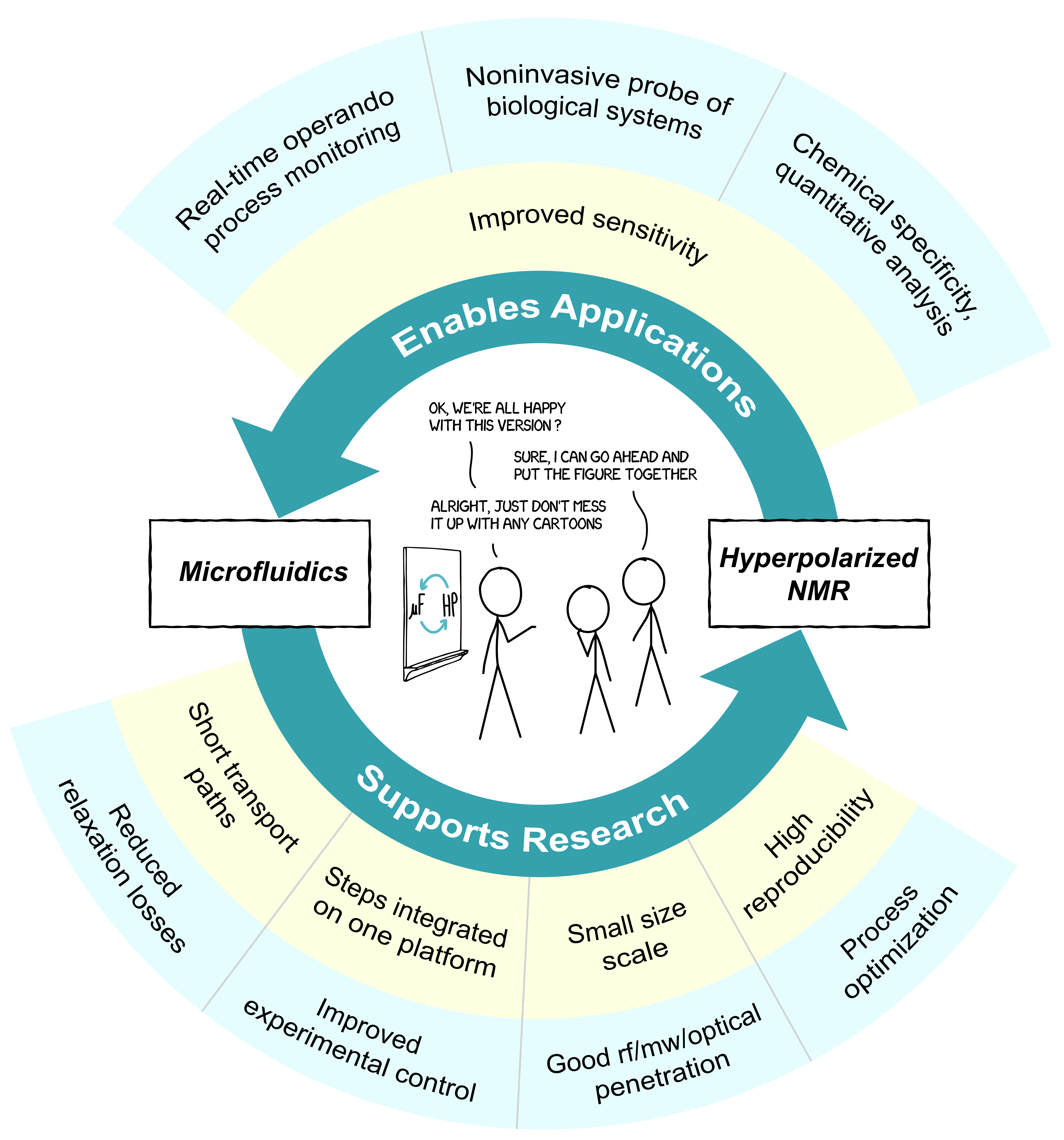}
	\end{center}
	\caption{The complementary nature of hyperpolarized NMR and microfluidics results in a positive feedback loop: (1) exerting microfluidic control over samples can boost the efficacy of hyperpolarization techniques, and; (2) the enhanced sensitivity from hyperpolarization means NMR spectroscopy, an immensely powerful method, is a more viable detection strategy in microfluidics.}
	\label{fig:concept}
\end{figure}

Microfluidic NMR could therefore benefit greatly from hyperpolarization techniques,
which offer several orders of magnitude increase in nuclear spin polarization, and hence sensitivity. 
This is illustrated in Fig.~\ref{fig:concept}. 
There are a number of hyperpolarization techniques which exploit different physical and chemical routes to produce molecules in states of much greater
nuclear spin alignment than could be achieved at thermal equilibrium, even in
a large magnetic field. It can therefore offer massively increased sensitivity. However, there are two important limitations:
\begin{enumerate}
    \item \emph{Relaxation} - By definition, hyperpolarization is a non-equilibrium phenomenon. This means that as soon
as a state of high nuclear spin alignment is created, it starts to relax
towards thermal equilibrium. The relevant longitudinal relaxation times are typically
on the order of seconds to tens of seconds, reaching several minutes in certain favorable cases.
    \item \emph{Lack of generality} - Although there now exist an increasing number of distinct hyperpolarization techniques, 
    they each only work under specific conditions, and tend to be applicable only to certain types of compounds. Often 
    additional steps of physical and/or chemical manipulation are required to transfer the polarization
from the hyperpolarized agent to an analyte of interest.
\end{enumerate}

Microfluidic technology can help on both counts. As will become
clear in the course of this review, many of the experimental steps required
for various hyperpolarization techniques benefit from integration into microfluidic
systems. In addition to this, if the microfluidic assay can be integrated into the same system, the distance and
therefore the time between production of the hyperpolarized agent and its use can be minimized.

At the time of writing, microfluidic lab-on-a-chip technology and hyperpolarized NMR are both
highly active fields of scientific investigation, each with the potential 
to benefit the other. This is exemplified by several ground-breaking
innovations emerging at their interface. For example,
the current record in mass sensitivity in NMR has been realized in a microfluidic
implementation of photo-chemically induced dynamic nuclear polarization (photo-CIDNP)~\cite{gomezIlluminationNanoliterNMRSpectroscopy2018}, closely
followed by parahydrogen-induced polarization (PHIP) on a microfluidic chip~\cite{eillsHighResolutionNuclearMagnetic2019}. 

\subsection{Overview}
In this paper we review the developments at the interface between the 
fields of microfluidics and hyperpolarized NMR. While the boundaries of what constitutes ``true'' microfluidic
technology are somewhat difficult to draw, we have
used a relatively narrow definition, restricted to systems that
not only use small sample volumes, but also exploit at least a small degree of
fluidic functionality. For example, experiments that use nanoliter quantities of sample
held statically in a quartz capillary would not qualify. On the other hand,
we do include some examples of work done at a somewhat larger volume scale (up to 100~$\mu$l or so),
where the concepts involved are clearly relevant for microfluidic implementation, and/or
the experiment could readily be scaled down.
We take ``hyperpolarized'' to mean any nuclear spin system strongly perturbed from its thermal equilibrium state that can yield enhanced NMR signals (a more quantitative analysis of this topic is given in Ref.~\cite{levittHyperpolarizationPhysicalBoundary2021}).

The review is organized as follows. We start by briefly introducing the fields of hyperpolarized magnetic resonance and microfluidics.
We then discuss the application of hyperpolarization to enhance signals for microfluidic detection, and 
the use of microfluidics to enhance existing hyperpolarization methodologies. These sections are generally organized according to 
hyperpolarization methodology, since the implementation of each method differs greatly, which leads to fairly distinct 
experimental realizations. In some cases, where there is considerable overlap, with several hyperpolarization methods 
sharing a similar experimental approach, we have deviated from this, and discuss the approach in a standalone section. For
example, both parahydrogen-induced polarization 
and optical pumping of noble gases can make use of microfluidics 
to enhance gas-liquid contact.

Sections~\ref{OpticalPumping} and~\ref{NVsection} describe optical pumping of electrons in alkali metal vapors and nitrogen-vacancy centers 
in diamond as sources of nuclear spin hyperpolarization. Both of these methods have also been employed as 
detection modalities for NMR experiments, in which the response of the electron spin(s) to an external magnetic 
field replaces the more conventional inductive detection using a coil. Although these strategies have been 
employed for NMR-detection at a microfluidic-scale~\cite{kitchingChipscaleAtomicDevices2018,shengMicrofabricatedOpticallypumpedMagnetic2017,kehayiasSolutionNuclearMagnetic2017,schwartzBlueprintNanoscaleNMR2019} we do not cover these topics in this review, since 
nuclear spin hyperpolarization is not necessarily involved.

\subsection{Hyperpolarization}
\label{sec:hyperpolarization}
The polarization ($P$) of an ensemble of spin-$1/2$ nuclei can be defined as:
\begin{equation}
    P=\frac{n_\alpha-n_\beta}{n_\alpha+n_\beta}
\end{equation}
where $n_\alpha$ and $n_\beta$ are the number of spins in the $\alpha$ and $\beta$ states, which are Zeeman states of the nuclear spins that correspond to orientation with or against the external magnetic field. The polarization of nuclear spins at ambient temperature is low even in high magnetic fields, due to the weak interaction of nuclear spins with magnetic fields. Many distinct hyperpolarization techniques exist to boost nuclear spin polarization in atoms and molecules, and hence the size of the NMR signals that they produce. These techniques usually (but not always) rely on an external source of polarization that can be transferred to the nuclear spins. Electrons, for example, are intrinsically many times more polarized than nuclear spins (by a factor of $\sim 10^3$ to $10^5$ depending on the nuclear isotope) in the same magnetic field. Their
polarization can be increased further in many cases by optical pumping with polarized light. By exploiting electron-nuclear spin couplings, this high degree of spin order can be transferred to the nuclear spins to enhance the NMR signals. Other techniques such as parahydrogen-induced polarization rely on the coupling between rotational and nuclear spin degrees of freedom that arise from exchange symmetry requirements of the 
wavefunction. Chemically-induced dynamic nuclear polarization uses differences in chemical reactivity of molecules depending on the nuclear spin state to produce hyperpolarized signals. In this review we only cover hyperpolarization techniques that have been used in combination with microfluidics. These are summarized in Table~\ref{tab:hyperpolarization}. 

\begin{table}[]
\tiny
    \begin{center}
    	\label{tab:HP}
		\singlespacing
    	\rotatebox{90}{
    \begin{tabular}{p{3cm}p{5cm}p{4cm}p{4.2cm}p{4.2cm}p{1cm}} \toprule
         \textbf{Method} & \textbf{Basic Principle} & \textbf{Rate of Production and Typical Polarization} & \textbf{Advantages} & \textbf{Limitations} & \textbf{Refs} \\ 
    \midrule 
      Overhauser dynamic nuclear polarization (Overhauser DNP) & 
      A solution containing free radicals is irradiated with microwaves to saturate the unpaired electron resonance. The Overhauser effect induces polarization transfer from the electron to nuclei in the solvent molecules via the fluctuating dipolar couplings.  & 
      Seconds, $<$0.5\% & 
      \begin{compactlist}
       \item[+] Low equipment demands 
      \item[+] Room-temperature operation
       \item[+] Simple technique with few experimental steps
       \end{compactlist} & 
       \begin{compactlist} 
       \item[--] Solutions are contaminated with free radicals
       \item[--] Short $T_1$ times due to free radicals
       \item[--] Limited achievable polarization levels
       \item[--] Efficiency decreases at high magnetic field
       \end{compactlist} & 
        \cite{raveraBasicFactsPerspectives2016,vanbentumPerspectivesDNPenhancedNMR2016} \\
    \midrule 
      Dissolution dynamic nuclear polarization (D-DNP) & 
      The sample is frozen as a glass with free radicals at cryogenic temperatures ($\sim$1.2~K) at high magnetic field. The unpaired electrons in the radicals are strongly polarized under these conditions, and microwave irradiation drives electron-nuclear polarization transfer. After polarization, the sample is rapidly dissolved in a warm solvent. & 
      Tens of minutes, usually $\sim$40\% polarization & 
      \begin{compactlist}
       \item[+] General method that can polarize many molecular targets
       \item[+] High polarization levels, approaching unity in some cases
       \end{compactlist} & 
       \begin{compactlist} 
       \item[--] Expensive, with significant equipment demands 
       \item[--] Repetition of experiments is slow
       \item[--] Single-shot technique
       \item[--] Radicals from the polarization step induce rapid relaxation
       \end{compactlist} & 
        \cite{bornetOptimizingDissolutionDynamic2016,janninApplicationMethodologyDissolution2019} \\
    \midrule 
      Spin-exchange optical pumping & 
      A hot vapor of alkali-metal atoms is optically pumped to polarize the unpaired electron spins. A noble gas is passed through the vapor and collisions between the noble gas atoms and alkali-metal atoms allows polarization transfer to the noble gas nucleus. & 
      Tens of seconds to tens of minutes depending on desired volumes and polarization levels, usually $\sim$40\% & 
      \begin{compactlist}
       \item[+] High polarization levels 
       \item[+] Long hyperpolarization lifetimes of the polarized gases
       \item[+] Hyperpolarized product not contaminated with polarization source
       \end{compactlist} & 
       \begin{compactlist} 
       \item[--] Limited to noble gas atoms
       \end{compactlist} & 
        \cite{meersmannHyperpolarizedXenon129Magnetic2015,barskiyNMRHyperpolarizationTechniques2017} \\
            \midrule 
      Parahydrogen-induced polarization (PHIP) & 
      Hydrogen gas is enriched in the para nuclear spin isomer by cooling to cryogenic temperatures over a catalyst. The \para-\Htwo gas is chemically reacted reversibly or irreversibly with other molecules to deliver the polarization.  & 
      Seconds, usually $\sim$1\dots10\% but higher in some cases & 
      \begin{compactlist}
       \item[+] Inexpensive and minimal equipment demands
       \item[+] Fast turnover rate and high rates of production
       \item[+] Para-enriched hydrogen gas can be stored for days
       \end{compactlist} & 
       \begin{compactlist} 
       \item[--] Chemical nature of method means specific reactions need to be designed
       \item[--] Hyperpolarized solutions are contaminated with remnants of the chemical reaction
       \end{compactlist} & 
        \cite{duckettApplicationParahydrogenInduced2012,hovenerParahydrogenBasedHyperpolarizationBiomedicine2018} \\
                    \midrule 
      Photo-chemically induced dynamic nuclear polarization (photo-CIDNP) & 
        A photosensitizer molecule in solution is irradiated with light to produce an unstable free radical. This excited molecule forms a radical pair with a target molecule, and the probability for the excited molecule to relax depends on the nuclear spin state of the target molecule. This spin selectivity leads to nuclear spin polarization.  & 
      Seconds, $<$1\% & 
      \begin{compactlist}
       \item[+] Inexpensive and minimal equipment demands
       \item[+] One-pot hyperpolarization, no sample alteration or transport required
       \end{compactlist} & 
       \begin{compactlist} 
       \item[--] Highly specific to chosen molecular system and sensitive to small changes
       \item[--] Samples can photo-degrade over time
       \end{compactlist} & 
        \cite{okunoPhotochemicallyInducedDynamic2017,goezChapterPhotoCIDNPSpectroscopy2009} \\
                            \midrule 
      Nitrogen vacancy (NV) centers in diamond & 
        A defect site in diamond in which two neighboring \chemical{^{13}C} atoms have been replaced with a single nitrogen atom is irradiated with light to generate electron polarization. This can then be transferred to coupled nuclear spins. This can be used to hyperpolarize diamond nanopowders in solution.  & 
      Seconds to tens of seconds, $<$1\% & 
      \begin{compactlist}
       \item[+] Method produces biologically-compatible hyperpolarized solutions
       \item[+] Low-cost and minimal equipment demands
       \end{compactlist} & 
       \begin{compactlist} 
       \item[--] Currently limited to polarizing nuclear spins within the diamond lattice (e.g., \chemical{^{14}N}, \chemical{^{15}N}, \chemical{^{13}C})
       \item[--] The diamonds are chemically inert which limits applications
       \item[--] Hyperpolarization is localized to the diamonds
       \end{compactlist} & 
        \cite{chenOpticalHyperpolarization132015,ajoyRoomTemperatureOptical2020} \\
        \bottomrule
    \end{tabular}
    }%rotatebox
    \caption{Overview of the hyperpolarization techniques addressed in this review.\label{tab:hyperpolarization}}
    \end{center}
\end{table}

A characteristic feature of hyperpolarization experiments is that once an atom or molecule is hyperpolarized, the non-equilibrium state relaxes back to thermal equilibrium with some characteristic relaxation time, often the longitudinal relaxation time $T_1$. This situation can be particularly problematic when remnants of the hyperpolarization process (such as molecules containing unpaired electrons) are in the solution and induce rapid nuclear spin relaxation. Protons are the most strongly magnetic of the stable nuclei, and the stronger coupling they experience with surrounding magnetic fields means relaxation times can be shorter than for less magnetic heteronuclei. Typical proton relaxation times for small molecules in solution range from seconds to a few tens of seconds, whereas heteronuclei such as \chemical{^{13}C} and \chemical{^{15}N} can have relaxation times
ranging from several tens of seconds to minutes. For this reason it is common in many hyperpolarization experiments to hyperpolarize protons, and then transfer the polarization to a heteronucleus where it is longer-lived. Bringing the hyperpolarization experiment into a microfluidic environment helps minimizing relaxation losses: if the transport paths and size-scale of the experiment are small, there is less time for relaxation to occur between hyperpolarization and detection.

A related problem is the destruction of polarization by the process of signal readout. Hyperpolarized
NMR experiments therefore often rely on small flip angles, which leave most of the polarization untouched,
at the expense of sensitivity. By contrast, microfluidic systems facilitate operation under continuous flow
conditions, where the hyperpolarized sample is continually replenished. In such systems, large
pulse angles can be used for excitation, leading to enhanced sensitivity.

\subsection{Microfluidics}
Microfluidic devices are miniaturized laboratories, often fabricated using lithographic
techniques similar to those used for semiconductor devices. They allow experimentation
with smaller samples, saving cost and enhancing throughput. A major motivation driving their
development is the integration of complex sample preparation procedures with analytics in 
a convenient and robust package, for example for genetic analysis of forensic samples, or
for point-of-care genomic diagnostics. Since its inception, progress in the field of
microfluidics has rested on three pillars:
\begin{enumerate}
    \item Introduction of novel materials and fabrication methods,
    \item Developing new ways of controlling the movement of fluids on the chip,
    \item Novel approaches to separation, detection, and analysis.
\end{enumerate}

An attractive aspect of microfluidic systems is their ability to integrate
analytical techniques. Indeed, some of the first microfluidic
systems were focused on miniaturizing chromatographic separation. This
work led to gas~\cite{terryGasChromatographicAir1979} and liquid chromatography~\cite{manzDesignOpentubularColumn1990} systems microfabricated on
silicon devices. The concept of integrated sample preparation with separation and
analysis on a single device has become known as a `miniaturized total analysis system' ($\mu$TAS)~\cite{manzMiniaturizedTotalChemical1990}. This has now matured into 
a lively field of research, which is beyond the scope of this review; only a few
examples are cited in the following (for an in-depth review, see for example~\cite{kovarikMicroTotalAnalysis2013}). 
Many other separation techniques have been integrated
into such $\mu$TAS devices, including capillary electrophoresis~\cite{harrisonCapillaryElectrophoresisSample1992,seilerPlanarGlassChips1993}, isotachophoresis~\cite{smejkalMicrofluidicIsotachophoresisReview2013}, and gel electrophoresis. These separation techniques
can be combined with immunoassays for detection of faint signals~\cite{heMicrofluidicPolyacrylamideGel2009}. An important research goal has been
the integration of polymerase chain reaction (PCR) for analysis of nucleic acids~\cite{khandurinaIntegratedSystemRapid2000,lagallyMonolithicIntegratedMicrofluidic2000,zhangPCRMicrofluidicDevices2006, ottesenMicrofluidicDigitalPCR2006,kaliskySinglecellGenomics2011}. In this case, microfluidic
implementation offers a functional advantage beyond that of compactness: the rapid
thermal cycling required for PCR can be done much faster at smaller scale.
This has led to impressive performance, with systems capable of sample-in/answer-out
cycle times of less than 20 minutes~\cite{easleyFullyIntegratedMicrofluidic2006,rouxIntegratedSampleinansweroutMicrofluidic2014}.

Once sample preparation and separation/analysis steps are integrated onto a single device, it is a logical next step to integrate the sample itself. This idea
has led to microfluidic culture systems for cells~\cite{yiMicrofluidicsTechnologyManipulation2006,bieleckaBioengineeredThreeDimensionalCell2017,d.castiauxReview3DCell2019}, cell aggregates, tissues~\cite{vanmidwoudMicrofluidicDevicesVitro2011},  
small organisms~\cite{qinMazeExplorationLearning2007,chronisMicrofluidicsVivoImaging2007},
and functional models of organs on a chip~\cite{sunOrganonaChipCancerImmune2019, zhangAdvancesOrganonachipEngineering2018}.
Microfabrication offers the ability to structure growth environments to support 
targeted interactions of specific cell types~\cite{bhatiaMicrofluidicOrgansonchips2014},
as well as the differentiation of stem cells into functional tissues~\cite{graczHighthroughputPlatformStem2015,wangMicroengineeredCollagenScaffold2017,bergstromStemCellDerived2015,ulmerHumanPluripotentStem2019}. Microfluidic culture systems have been used for NMR-based metabolomic analysis by extraction of culture medium aliquots from the chip, e.g., for bioartificial livers~\cite{shintuMetabolomicsonaChipPredictiveSystems2012} and
Hepatocyte cell lines~\cite{ouattaraMetabolomicsonachipMetabolicFlux2012} or plant cells
\cite{maischTimeresolvedNMRMetabolomics2016}. More recently, systems for direct NMR observation of the metabolism inside the culture device using micro NMR detectors have
been described~\cite{kalfeLookingLivingCell2015,patraTimeresolvedNoninvasiveMetabolomic2021}.

Microfluidic systems can be manufactured from a range of different materials, including
glass, hard (glassy) polymers such as polycarbonate or poly(methyl methacrylate), as
well as rubbery polymers, in particular, polydimethylsiloxane (PDMS). 
PDMS-based soft lithography~\cite{mcdonaldFabricationMicrofluidicSystems2000,whitesidesFlexibleMethodsMicrofluidics2001,ngComponentsIntegratedPoly2002}  offers an attractive route to highly integrated
microfluidic devices with active, pneumatically driven valves~\cite{melinMicrofluidicLargeScaleIntegration2007}.
PDMS is an inexpensive yet highly versatile biocompatible 
elastomer. It is optically transparent and gas permeable; biological systems can be readily cultured in PDMS chips under a controlled atmosphere. 
In addition, PDMS membranes can be integrated with hard materials such as
glass and glassy polymers in order to provide active as well as passive
functional elements such as fluidic diodes~\cite{leslieFrequencyspecificFlowControl2009}.

Microfluidic systems can benefit enormously from multi-phase flow, in which either
a second liquid phase or a solid phase is suspended in the primary working fluid.
Droplet microfluidics allows encapsulation of single cells into individual droplets,
which then become microscopic bioreactors~\cite{s.kaminskiDropletMicrofluidicsMicrobiology2016, kaminskiControlledDropletMicrofluidic2017,abalde-celaDropletMicrofluidicsHighly2018,dingRecentAdvancesDroplet2020}. Solid particles or quasi-solids such as cells can be manipulated independently from 
the fluid flow by acoustic fields~\cite{nilssonReviewCellParticle2009a,lenshofContinuousSeparationCells2010} and by inertial forces~\cite{carloInertialMicrofluidics2009,martelInertialFocusingMicrofluidics2014}. This has enabled efficient
separation of blood cells~\cite{antfolkContinuousFlowMicrofluidic2017}, as well as identification
of rare circulating tumor cells in blood~\cite{stottIsolationCirculatingTumor2010}.
% After this, $\mu$TAS devices began to find new applications in cell culture~\cite{huangGenerationManipulationHydrogel2017} and droplet microfluidics~\cite{}. 
% Today, microfluidics is a wide field of research that has many applications across the sciences from 3D cell
% culture and organ-on-a-chip~\cite{} to microfluidic droplets
%  and single cell
% analysis~\cite{fanFluorescentAnalysisBioactive2018,gaoRecentAdvancesSingle2019}.

% In NMR, the advantages of microfluidics continue to be employed. As described in \ref{Small_Scale},
% there is also inherent benefit in shrinking the size of the detector. The low volumes involved
% in experiments means that smaller amounts of costly labelled materials can be used for
% a protein experiment or metabolomic tracking. NMR also offers advantages in return.
% It's noninvasive, nondestructive nature makes is ideal for studying biological samples as well as
% complimenting other analysis techniques. NMR also provides quantitative, chemical resolution and
% can be used to monitor processes in real-time. However NMR still suffers from low
% sensitivity when compared to other, more common, methods in the microfluidics community. 

All microfluidic systems must solve the problem of extracting information
from the biological and/or chemical processes on the chip. To this effect,
analytical techniques are needed that can either be integrated into the chip (e.g., electrochemical detection), or 
accommodate the chip in their sampling area (e.g., fluorescence or Raman microscopy). In some
cases, it may be possible to remove aliquots from the chip for off-line analysis (e.g., by mass spectrometry), but due to the small size of the samples this is often not an option.
In any case, analytical
techniques must provide sufficient \emph{specificity} to distinguish signals that one wishes to detect from irrelevant background signals, as well
as \emph{sensitivity} to detect and quantify the species of interest above the noise. 
Most commonly, chemical specificity is achieved either through specific fluorescent labels, selective
oxidation electrodes, or by integrating chromatographic separation (most commonly, electrophoresis) into
the chip combined with a non-specific readout. 
Table~\ref{tab:detectors} summarizes typical microfluidic detection technologies.

A common feature of widely-used readout techniques in microfluidics is that they are invasive or destructive, requiring either permanent modification of the sample by introducing fluorescent
labels, chemical separation, or both. Few techniques are available that provide quantitative information while
experiments continue without at least partial loss of the system under investigation.
While taking aliquots from a culture is usually straightforward at the macroscale, it can
be problematic in microfluidic systems, where sample volumes are severely limited. As a non-invasive technique,
NMR spectroscopy provides a potential solution to this problem. NMR spectra of complex
mixtures can be deconvolved to quantify hundreds of analyte species simultaneously. 
However, sensitivity is a problem, particularly in the context of microfluidics.
While  conventional (non-hyperpolarized) NMR can detect analytes down to 10~$\mu$M or so, commonly used microfluidic detection methods offer
much higher sensitivities, albeit at the expense of generality and, in some cases, sample
destruction.
For example, Senel et al. reported a concentration limit of detection (cLOD) of 0.1~nM using electrochemical methods
to detect neurotransmitters in mice~\cite{senelMicrofluidicElectrochemicalSensor2020a}. Hou et al. used a cantilever
array to detect antibiotics and achieved a cLOD of 0.2~nM~\cite{houAptamerBasedCantileverArray2013}. Bowden et al. employed
a fiber-optic micro array to detect DNA at a cLOD of 10~aM~\cite{bowdenDevelopmentMicrofluidicPlatform2005}.

As a consequence, NMR spectroscopy has only rarely been used
in the context of microfluidics. 
Recent advances in the integration of efficient micro-NMR probes with microfluidic systems have 
started to change this situation. However, sensitivity is still a strongly limiting factor,
as discussed above. Emerging microfluidic hyperpolarization techniques are beginning to overcome
this limitation.

\begin{table}
\begin{center}
    \label{tab:detectors}
		\tiny
	%\rotatebox{90}{%
    \begin{tabular}{p{1.5cm}p{3.5cm}p{4cm}p{4cm}p{1cm}}\toprule
      \textbf{Method} & \textbf{Mechanism} & \textbf{Advantages} & \textbf{Limitations} & \textbf{Refs} \\ \midrule
      \multirow{2}{*}{\begin{minipage}[c]{\linewidth} \text{Electrochemical}\end{minipage}}
      & Detection is based on measuring changes in conductance, resistance and/or capacitance at the active surface of the electrodes.  
      &
      \begin{compactlist}
        \item[+] Real-time detection 
        \item[+] Low-cost microelectrode fabrication 
        \item[+] Miniaturization can increase signal to noise ratio 
      \end{compactlist} 
       & 
       \begin{compactlist}
        \item[--] Ionic concentrations before detection have to be controlled 
        \item[--] Detectors have a short shelf life as they can be sensitive to changes in temperature and pH
        \item[--] At nanoscale (50 nm or below) quantitative data may not be reliable
        \end{compactlist} 
       & 
       \cite{nesakumarMicrofluidicElectrochemicalDevices2019,deeguilazElectrochemicalDetectionViruses2020} \\ \midrule
      \multirow{2}{*}{\begin{minipage}[c]{\linewidth} \text{Mechanical}\end{minipage}} 
      & 
      Detection is based on the variation of the resonant frequency of a mechanical sensor or on the surface stress of a cantilever. 
      &
	  \begin{compactlist}
	    \item[+] Monolithic sensor integration 
	    \item[+] Label free detection \item[+] High sensitivity 
	    \item[+] High selectivity 
	  \end{compactlist} 
	  & 
	  \begin{compactlist}
	    \item[--] Viscous damping effects caused by the solvent in liquid samples can affect sensitivities
	    \item[--] Detection process can take on the order of 30 minutes as some rely on specific binding of molecules to a cantilever 
	    \item[--] Complex fabrication methods required
        \item[--] Complex microscopy setups can be required to observe effects 
      \end{compactlist} 
      & 
      \cite{basuMicroNanoFabricated2020,mathewReviewSurfaceStressBased2018,alundaReviewCantileverBasedSensors2020}\\ \midrule
      \multirow{2}{*}{\begin{minipage}[c]{\linewidth} Raman ~Spectroscopy\end{minipage}} 
      & 
      Detection is based on observing  scattered light which can be attributed to the various vibrational energy modes of a target molecule. 
      & 
      \begin{compactlist}
        \item[+] No sample preparation needed
        \item[+] Provides real-time detection 
        \item[+] Can provide chemical and structural information of a molecule 
        \item[+] Can identify molecules based on a Raman 'fingerprint' 
      \end{compactlist} 
      & 
      \begin{compactlist}
        \item[--] Instrumentation required is expensive 
        \item[--] Material must be transparent to the wavelength of light used 
        \item[--] The Raman effect is weak and requires sensitive and optimized equipment 
        \item[--] Sample heating through intense laser radiation can be destructive or interfere with the spectrum
      \end{compactlist} 
      & 
      \cite{j.jahnSurfaceenhancedRamanSpectroscopy2017,ochoa-vazquezMicrofluidicsSurfaceEnhancedRaman2019} \\ \midrule
      \multirow{2}{*}{\begin{minipage}[c]{\linewidth} UV-Visible ~Spectroscopy\end{minipage}} 
      & 
      Detection is based on the absorbance of the specific wavelength of light that excites electrons from the ground state to the first singlet excited state. 
      & 
      \begin{compactlist}
          \item[+] Minimal to no sample preparation required (for single molecule solutions) 
          \item[+] Simple and inexpensive instrumentation
          \item[+] Most organic molecules absorb in the UV-Vis region 
          \item[+] Quantitative measurement possible 
      \end{compactlist} 
      & 
      \begin{compactlist}
          \item[--] Mixtures of molecules can cause over lap in the spectrum
          \item[--] Spectra are not highly specific for different molecules
          \item[--] Material and solvent have to be chosen wisely based on the expected absorbance of the molecule of interest to mitigate spectral overlap
      \end{compactlist} & \cite{liSituSensorsFlow2021,b.myersInnovationsOpticalMicrofluidic2008} \\ \midrule
      \multirow{2}{*}{\begin{minipage}[c]{\linewidth} Fluorescence Spectroscopy\end{minipage}} & Detection is based on the wavelength, or wavelengths, of light emitted by a molecule after excitation by light irradiation. 
      & \begin{compactlist}
          \item[+] Fast detection
          \item[+] Can detect \emph{in vivo}
          \item[+] High sensitivity 
          \item[+] High specificity afforded by fluorescent tags
          \item[+] Miniaturization decreases signal to noise ratio
      \end{compactlist} 
      & 
      \begin{compactlist}
          \item[--] Conventional instrumentation is expensive
          \item[--] Specific fluorescent tags needed for each target molecule
          \item[--] Fluorescent tags may be toxic or mutagenic
          \item[--] Measurement can be noisy depending on the system
      \end{compactlist} 
      & 
      \cite{tianFluorescenceCorrelationSpectroscopy2011,zhaoApplicationsFiberopticBiochemical2020} \\ \midrule
	\multirow{2}{*}{\begin{minipage}[c]{\linewidth} \text{NMR}\end{minipage}} 
	& 
	Detection is based on using radio frequency pulses to excite nuclear spins and inductively detecting the nuclear spin precession. 
	&
	\begin{compactlist}\itemsep0pt
	    \item[+] Non-invasive, non-destructive detection 
	    \item[+] Quantitative, system level information 
	    \item[+] Can be used in tandem with existing microfluidic detection techniques
	\end{compactlist}  
	& 
	\begin{compactlist}\itemsep0pt 
	    \item[--] Limited sensitivity 
	    \item[--] Typical equipment setup is costly
		\item[--] Can usually only use non-magnetic materials in construction 
	\end{compactlist} 
	& 
	\cite{dupreMicroNanofabricationNMR2019,pjNuclearMagneticResonance2004} \\ \bottomrule
    \end{tabular}
    %} % rotatebox
\caption{A Summary of common detection methods in microfluidics listing some advantages and limitations in comparison to NMR. Adapted from Ref.~\cite{piresRecentDevelopmentsOptical2014}.}
    \label{tab:detectors}
\end{center}
\end{table}

\subsection{Nuclear Magnetic Resonance and its Detection at Small Scale}\label{SmallScale}

Nuclear magnetic resonance spectroscopy and imaging commonly rely on inductive detection
of nuclear spin precession. The sample is placed in a large 
static magnetic field, and is coupled, usually by
means of a passive circuit, to a transmission line, such that precession of the
nuclear spins produces an electromagnetic wave traveling away from
the sample towards a receiver, where it is amplified, demodulated, and digitized.
In most high-field NMR systems, the transmission line is coupled to the nuclear spins by
a coil surrounding the sample. The impedance of the coil is matched to the
transmission line by a passive circuit with a self-resonance frequency
close to the Larmor precession frequency of the nuclei, and sufficient
bandwidth to accommodate the entire expected NMR spectrum.
The sensitivity of this type of setup has been studied in detail by Hoult and
coworkers~\cite{houltSignaltonoiseRatioNuclear1976}. Hoult introduced the principle of reciprocity, which states that
the signal amplitude induced in a coil by a single precessing spin is proportional
to the magnetic field amplitude $\tilde B_1$ generated by an alternating current of unit magnitude in the same coil~\cite{houltSignaltonoiseRatioNuclear1976,houltPrincipleReciprocity2011}.
Since inductive detectors are invariably subject to thermal noise, 
the signal/noise ratio is determined by the ratio of
the rf magnetic field amplitude $\tilde B_1$ to the square root of the coil
resistance $R$.

\begin{figure}
	\begin{center}
			\includegraphics[width=12cm]{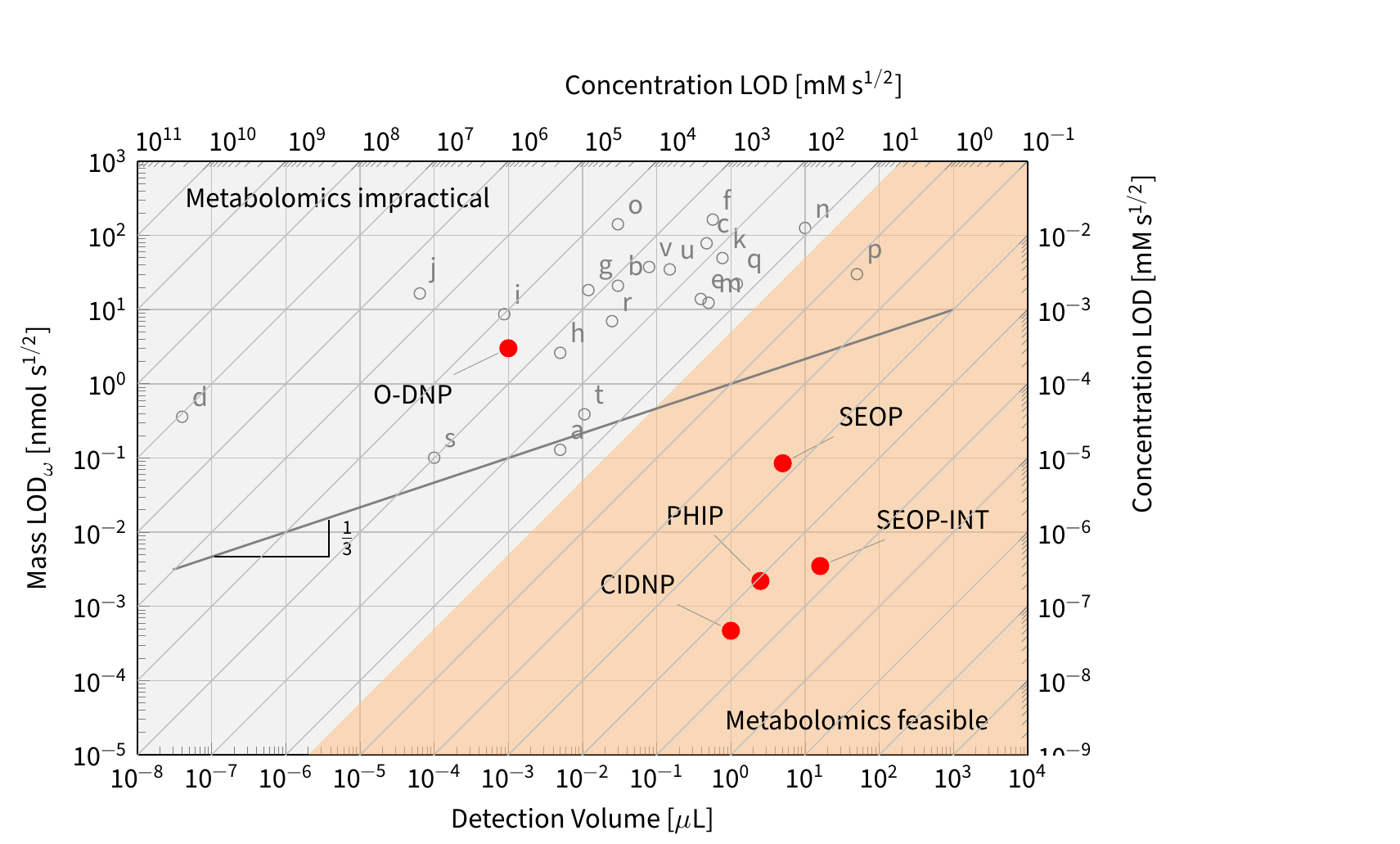}
	\end{center}
	\caption{Overview of reported sensitivity of micro-NMR detectors. Indicated
	are number limits of detection (nLOD), the number of spins required in the
	sample volume for a signal-to-noise ratio of 3. A general trend of
	$\mathrm{nLOD}\propto V^{1/3}$ can be observed, meaning that smaller
	detectors have better mass sensitivity. Open symbols represent conventional
	NMR with thermal equilibrium polarization (image reproduced from Ref.~\cite{badilitaMicroscaleNuclearMagnetic2012} with permission), filled symbols are examples of 
	microfluidic hyperpolarized NMR by different methods: PHIP: parahydrogen-induced hyperpolarization~\cite{eillsHighResolutionNuclearMagnetic2019}; 
	SEOP: spin-exchange optical pumping~\cite{causier3DprintedSystemOptimizing2015};
	SEOP-INT: spin-exchange optical pumping with internal detection~\cite{kennedyOptimizedMicrofabricatedPlatform2017};
	CIDNP: chemically-induced dynamic nuclear polarization~\cite{mompeanPushingNuclearMagnetic2018}; Overhauser DNP: Overhauser dynamic nuclear polarization~\cite{sahinsolmazSingleChipDynamicNuclear2020}.
	}
	\label{fig:sensitivity-ov}
\end{figure}

Sensitivity, or more precisely, the number limit of detection (nLOD) is determined by the minimum number of spins required to
produce a signal/noise ratio of at least 3~\cite{badilitaMicroscaleNuclearMagnetic2012} (see footnote above for a precise definition). At fixed concentration, the number
of spins contributing to the signal decreases linearly with the volume $V$. However,
inductive detection gets more efficient at smaller scale if the detector is scaled
down along with the sample, due to more intimate coupling of the spins to the detector.
A factor of $V^{-\alpha}$ is gained in this way, where $\alpha<1$ is the scaling exponent.
Estimates for $\alpha$ can be made based on theoretical arguments combining Ohm's and Ampère's law, the skin
depth at RF frequencies, and some assumptions on the detector geometry. Depending on
assumptions, one obtains values between $\alpha=\frac13$ and $\alpha=\frac12$. Experimental
data suggests values closer to $\alpha=\frac13$ (cf.~Fig.~\ref{fig:sensitivity-ov}),
leading to overall scaling
of the  signal/noise
ratio with $V^{\frac{2}{3}}$, and hence the concentration limit of detection as $\text{cLOD}\propto V^{-\frac{2}{3}}$.

% The mass sensitivity, also known as the number limit of detection (nLOD), is
% defined as the number of spins required to precess inside the detector to
% yield a signal/noise ratio of 3. Since the observed signal/noise ratio depends
% not only on the number of precessing spins, but also on the spectral width
% over which their resonances are distributed, the nLOD is measured in units of
% $\mathrm{mol/\sqrt{Hz}}=\mathrm{mol~\sqrt{\mathrm{s}}}$.
% The mass sensitivity of NMR circuits scales favourably as the detector size
% decreases.
The reduction of signal resulting from scaling down the sample volume can thus be
partially offset by scaling down the detector volume. 
This has been exploited in order to build highly sensitive NMR flow probes based on solenoid coils
with sample volumes down to $1$~pL, and mass limits of detection (otherwise known as the
number limit of detection, nLOD) around $100$~pmol$\sqrt{s}$
\cite{wu1HNMRSpectroscopyNanoliter1994,wuNanoliterVolumeSample1994,
olsonHighResolutionMicrocoil1HNMR1995,
laceyHighResolutionNMRSpectroscopy1999,laceyUnionCapillaryHighperformance2001,
seeberDesignTestingHigh2001},
as well as microfluidic NMR detection systems of various geometries~\cite{bartMicrofluidicHighResolutionNMR2009,badilitaMicroscaleNuclearMagnetic2012,zalesskiyMiniaturizationNMRSystems2014,finchOptimisedDetectorInsitu2016}. Among
other advantages, miniaturization offers the possibility to integrate multiple detectors in
the same NMR system~\cite{macnaughtanHighThroughputNuclearMagnetic2003a}.
Multiple detectors in close proximity require effective mutual radiofrequency decoupling.
Such miniaturized phased arrays have been reported~\cite{gobelPhasedArrayMicrocoils2015}, but a detailed description is beyond the scope
of this review.
% However, it is important to note that while the mass sensitivity of NMR detectors tends to
% improve as they become smaller, this is not enough to compensate for the smaller sample volume,
% if the samples are of fixed concentration. This is notably the case for many biological samples,
% in which the concentration of the analytes of interest is limited by biology. For this reason,
% the concentration sensitivity of conventional NMR detectors tends to \emph{decrease} at small scales.
% While it is routinely possible to detect and even quantify species at 50~$\mu$M
% in conventional NMR at the scale of 1~ml, the concentration limit of detection deteriorates to
% about 1~$mM\mathrm{\sqrt{s}}$ at the scale of 1~$\mu$l \cite{badilitaMicroscaleNuclearMagnetic2012}.
% \todo{I would very much enjoy some proportionalities here, to see how sensitivity depends on (1) sample volume, and (2) filling factor, at fixed concentration!}

Fig.~\ref{fig:sensitivity-ov}
summarizes the sensitivity of a large number of micro-NMR detectors (data points shown in light gray)
by correlating the detection volume to the achieved nLOD. While there is considerable scatter, the data points  broadly follow the theoretical scaling of nLOD with a gradient of $\frac13$, indicated by a solid line. The figure also contains an axis that indicates the 
concentration limit of detection, cLOD, with lines of constant cLOD running diagonally.
An arbitrary boundary is drawn at the level of $\text{cLOD}=5\;\mathrm{mM\sqrt{s}}$. This corresponds to a situation where an analyte present at 1~mM can be detected
within 10~minutes of measurement time with a signal/noise ratio of approximately 10:1. We consider this
to be the practical limit for metabolomic studies. As is evident from the figure, most conventional
micro-NMR systems lie outside this range. The red dots in Fig.~\ref{fig:sensitivity-ov} represent five 
different microfluidic NMR experiments using hyperpolarized samples which are discussed in greater detail in the 
following sections. Four of these lie deep inside the feasibility region;
the fifth, Overhauser dynamic nuclear polarization, lies outside this region because for technical reasons the experiment was carried out
in a low magnetic field. These studies aimed at a proof of principle, and did not
include any biological applications. However, it is clear from Fig.~\ref{fig:sensitivity-ov} that hyperpolarized 
NMR affords the possibility to reach far beyond the usual sensitivity limits and greatly expand 
the sphere of microfluidic experiments that can be tracked using NMR spectroscopy.

Microfluidic devices operating in an NMR spectrometer in order to provide
NMR observation of processes on a chip presents some additional engineering challenges.
Typical NMR spectroscopy cryomagnets are large cylindrical structures with bore
diameters around 50~mm, with a bore length up to 2~m. Conventional NMR
detectors are designed for cylindrical sample tubes (most commonly of 5~mm diameter).
While lab-on-a-chip devices themselves are 
typically small, they often require connections to ancillary hardware such as syringe
pumps, vacuum lines, and solenoid valve manifolds. These systems must remain outside
of the magnet, since they often contain magnetic components and are not designed to
operate in the presence of large magnetic fields. 
Microfluidic NMR therefore requires custom-designed micro-detectors that are 
designed to accommodate planar microfluidic devices while retaining a good filling
factor.

An additional constraint is that high-resolution NMR requires homogeneous
magnetic fields. Differences in magnetic susceptibility between the chip material
and the sample, as well as between different parts of the detector assembly can
lead to field distortions that broaden resonance lines and therefore lead to 
a loss in resolution and sensitivity. This can be managed by careful design of the microfluidic 
structures~\cite{ryanStructuralShimmingHighresolution2014}, as well as by doping of the fluid with suitable paramagnetic species~\cite{haleHighresolutionNuclearMagnetic2018}.
Obviously, chip materials that give strong NMR background signals must be
avoided. Unfortunately, this somewhat limits the use of PDMS,
since its elastomeric nature gives rise to strong 
background signals in both \textsuperscript{1}H and \textsuperscript{13}C NMR.
By contrast, glassy polymers such as poly(carbonate) and poly(methyl methacrylate)
work well, since their proton signals are 
extremely short-lived, and therefore easily separated from the desired
signals from species in solution. 
Glassy polymers are also amenable to efficient 
manufacturing and rapid-prototyping techniques such as laser cutting
and hot embossing. On the other hand, solvent compatibility of glassy polymers is quite poor, which
is a serious limitation in the context of hyperpolarization techniques. In these
cases, glass is a possible choice, but glass devices tend to be more
difficult to manufacture. 

\section{Dynamic Nuclear Polarization}
\label{sec:DNP}
\subsection{Introduction}
The gyromagnetic ratio of an electron is $\sim$660 times higher than that of a proton. Electrons 
in a sample can therefore be used to enhance the nuclear spin polarization if the electron polarization can be transferred to a 
coupled nuclear spin. Hyperpolarization experiments that exploit this source of polarization are collectively
known as dynamic nuclear polarization (DNP) experiments. There are a number of different DNP mechanisms, which
fall within two broad categories: solid-state DNP and solution-state DNP. In this review we are interested in 
the microfluidic implementation of hyperpolarization, which generally means that the hyperpolarized sample should 
be a fluid. To generate nuclear spin polarization directly in a solution, Overhauser DNP (Overhauser DNP) can be used. 
An alternative approach is to use solid-state DNP, followed by a dissolution step in which the hyperpolarized 
solid is rapidly dissolved to form a solution. This is known as dissolution DNP (D-DNP). Solid-state DNP relies on mechanisms such as the solid effect, 
the cross effect, and thermal mixing, which encompass different types of electron-nuclear interaction, and 
are discussed in greater detail in Refs.~\cite{guntherDynamicNuclearHyperpolarization2013} and~\cite{lillythankamonyDynamicNuclearPolarization2017}. 
Overhauser DNP and dissolution DNP experiments differ greatly in how the polarization is transferred to 
nuclear spins and the experimental approach, and so these approaches are examined separately in the following sections.

	\subsection{Overhauser Dynamic Nuclear Polarization}
	\label{overhauserDNP}
	Overhauser DNP 
	relies on cross-relaxation between the electron and nuclear spins~\cite{overhauserPolarizationNucleiMetals1953,carverPolarizationNuclearSpins1953,solomonRelaxationProcessesSystem1955}.
	Relaxation between a pair of energy levels requires a term in the Hamiltonian
	connecting the two levels that is modulated stochastically by a process with significant
	spectral density at the transition frequency. Dipolar couplings are averaged
	to zero in liquids due to rapid molecular tumbling and are therefore not
	directly reflected in solution-state NMR spectra. However, their stochastic modulation
	due to thermal motion still drives relaxation.
	The coherent mechanisms that drive DNP in solids rely on dipolar
	couplings between the electrons and nuclei, and
	are eliminated by the rapid molecular motion in the liquid state.
	However, the Overhauser effect, based on fluctuations of dipole-dipole 
	couplings inducing relaxation, remains active in the solution state, 
	and can be exploited to produce hyperpolarized liquids.

\begin{figure}
		\begin{center}
			\includegraphics[width=8.4cm]{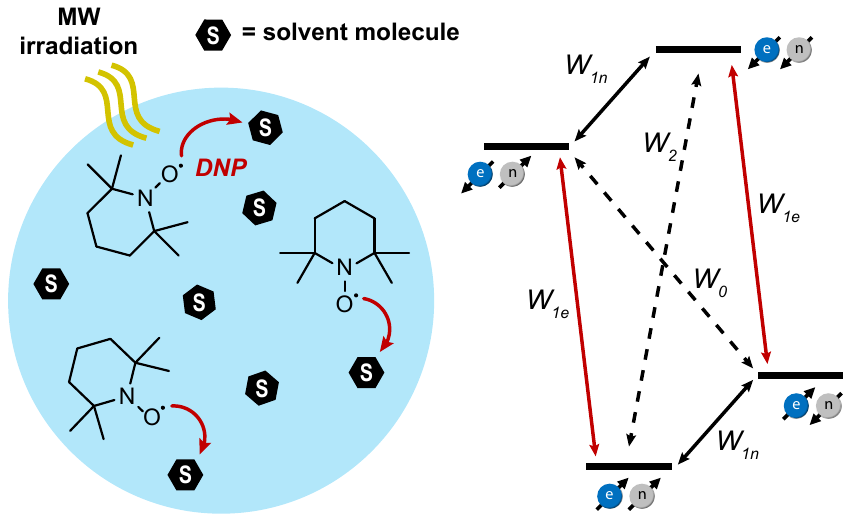}
		\end{center}
		\caption{Left: A solution containing free radicals (TEMPO is shown in the example) is placed in a magnetic field usually between 0.1 and 1~T and the electrons become polarized. The solution is irradiated with microwaves at the electron Larmor frequency to drive the Overhauser effect which causes electron polarization to be transferred to nuclear spins in the solvent molecules. Right: a simple energy level diagram for a two-spin system comprising an electron and a spin-1/2 nucleus, with zero-, single-, and double-quantum transition rates labeled $W_0$, $W_1$, and $W_2$, respectively. The transitions saturated by the microwaves are shown in red, and a difference in zero- and double-quantum relaxation rates can lead to an Overhauser effect.}
		\label{fig:ODNP}
	\end{figure}

	In such experiments, a solution of stable radicals (e.g., nitroxides such as (2,2,6,6-Tetramethylpiperidin-1-yl)oxyl, also known as TEMPO), typically at a concentration between 10 and 50~mM,
	is irradiated with microwaves at the electron Larmor frequency. 
	Sample volumes are limited to a few milliliters by the penetration depth of microwaves,
	which are strongly absorbed by many solvents.
	During irradiation, 
	polarization builds up on the solvent molecules, and an enhanced NMR signal is obtained after sufficient polarization time (typically some minutes).
	This process is illustrated in Fig.~\ref{fig:ODNP}.
	The enhancement ($E$) of the nuclear polarization due to Overhauser DNP can be
	expressed
	as~\cite{armstrongNewModelOverhauser2007}:
	\begin{equation}
		E = \frac{\langle I_z\rangle}{\langle I_0\rangle} = \left(1-\rho f s \frac{|\gamma_S|}{\gamma_I}\right),
	\end{equation}
	where $\langle I_z\rangle$ and $\langle I_0\rangle$ are the enhanced and thermal equilibrium
	nuclear polarizations, and $\gamma_S$ and $\gamma_I$ are the electron and
	nuclear magnetogyric ratios, respectively. $\rho$ is the coupling factor,
	given by the ratio of the cross-relaxation rate to the longitudinal nuclear relaxation
	rate caused by the electron spins. $f$ is a leakage factor, given by the ratio
	of the nuclear longitudinal relaxation rates of the solvent in the presence and
	absence of the stable radicals, and $s$ is the saturation factor, which quantifies
	the extent to which the electron transition is saturated by the microwave
	field. While $s$ and $f$ can be controlled by adjusting experimental conditions (concentration
	of the stable radical, microwave power, etc.), the coupling factor $\rho$ is
	an intrinsic property of the radical and the chosen solvent.
	Most commonly, $\rho$ is dominated by electron-nuclear dipolar couplings. In this
	case, the value of $\rho$ is positive, leading to negative enhancement factors.
	For magnetic fields exceeding about 1~T, the coupling factor for dipolar
	interactions drops rapidly.  This is due to the spectral density of dipolar fluctuations. These are driven by translational molecular diffusion, with correlation times of the order of $50\dots200$ps. Their spectral density therefore drops dramatically above (electron Larmor) frequencies of several GHz, corresponding to fields of about 1~T. Therefore, most implementations of Overhauser DNP
	have been at moderate magnetic fields such as 0.35~T. This has the
	additional advantage that microwave sources with sufficient power are readily
	available at the corresponding frequencies (X band, approximately 9~GHz).
	
	It is possible to combine the efficient and convenient Overhauser DNP at low
	magnetic fields ($<$0.5 T) with NMR detection at high fields ($>$2~T) by transporting
	the sample from a low-field DNP region to the high-field NMR magnet. To minimize
	losses from nuclear relaxation in solutions containing moderately high 
	concentrations of free radicals, the transport has to be rapid ($<$1~s), and should 
	take place in the presence of a guiding magnetic field to avoid faster relaxation at lower fields.
	Implementations of Overhauser DNP under flow have been demonstrated as early as 1989.
	Dorn et al.~\cite{dornFlowTransferBolus1989} have built a flow-through DNP system operating at 0.33~T connected
	to an NMR magnet operating at 4.7~T (corresponding to 200~MHz \textsuperscript{1}H Larmor
	frequency). Their system, operated in continuous flow, is driven by a high-performance
	liquid chromatography (HPLC) pump. Flow rates varied between 5 and 9~ml/min.
	Stable radicals were attached to a solid silica support maintained inside of the DNP
	magnet, such that the flowing liquid remained free from radicals. Modest
	enhancements of $E=-37$ were reported for benzene at first. Later improvements
	led to significantly higher values of $-330$ for \textsuperscript{1}H and $+2660$ for
	\textsuperscript{13}C
	~\cite{tsaiModelEstablishingUltimate1990}, the method
	was extended to immobilized nitroxide radicals~\cite{dornTransferOf1HAnd13C1991}, and an integrated LC-NMR system
	was demonstrated~\cite{stevenson13CDynamicNuclear1994}. The
	flow cell volumes in the electron paramagnetic resonance (EPR) and NMR part of the system were about 150~\muL. While
	these dimensions are not properly microfluidic, the work clearly demonstrates the
	potential of this approach for smaller scales.

\begin{figure}
		\begin{center}
			\includegraphics[width=8.4cm]{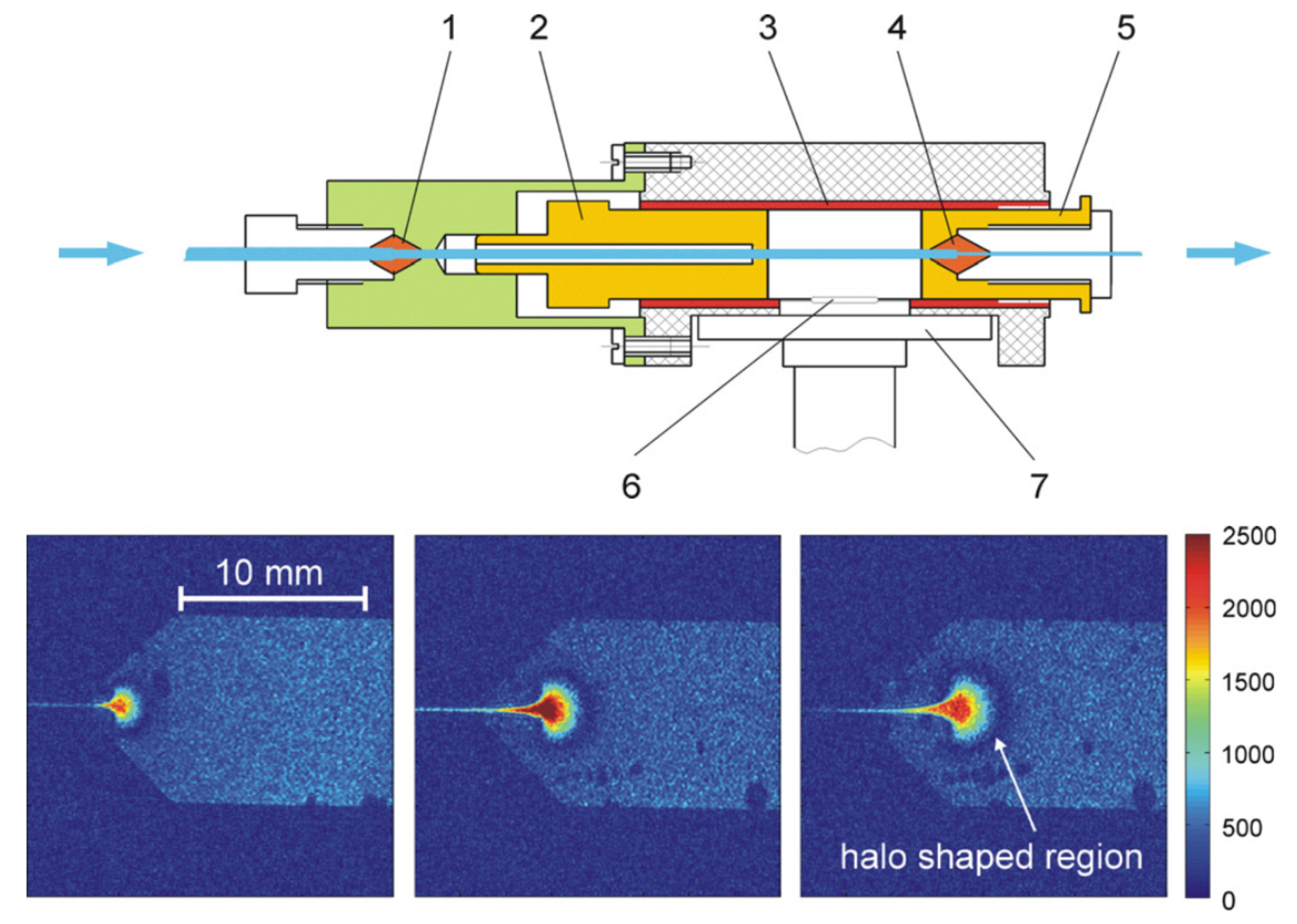}
		\end{center}
		\caption{Top: A Schematic drawing of the resonator design: (1) conical squeeze joint; (2) plunger; (3) hollow bore cylinder, I.D. 11~mm; (4) capillary necking; (5) plunger; (6) slit iris, 5.6~mm long, 0.20–0.30 mm~wide; (7) WR-19 waveguide flange. Bottom: Microfluidic Overhauser DNP. A solution of 24~mM  TEMPOL (4-hydroxy-2,2,6,6-tetramethylpiperidine-1-oxyl) in water
		is hyperpolarized while flowing
		through a microwave cavity (top) mounted inside of a 1.5~T imaging magnet.
	  The outflow is connected through a capillary to a planar microfluidic device
		where it is injected into the 90$^\circ$ corner on the left. Single-shot
		gradient-echo images have been taken at flow rates of 12, 20, and
		30~ml/h, respectively. Image adapted with permission from Ref.~\cite{krummenackerDNPMRIInbore2012}.}
		\label{fig:krummenacker2012}
	\end{figure}

	While the work of Dorn et al. focused on organic solvents such as benzene and
	\chemical{CDCl_3}, McCarney et al. proposed to use the same approach to produce
	hyperpolarized water for use as a contrast agent in magnetic resonance imaging
	~\cite{hanDynamicNuclearPolarizationEnhanced2009,mccarneyHyperpolarizedWaterAuthentic2007}.
	Using deuterated, \textsuperscript{15}N-labelled TEMPO immobilized on a sepharose
	matrix, water could be polarized at 0.35~T with enhancement factors up to --100 at
	flow rates up to 1.5~ml/min. The hyperpolarized water is injected into a phantom
	situated in a microimaging probe. Both the microwave cavity and the micro-imaging
	probe were kept in the 0.35~T $B_0$ field of a commercial X-band EPR spectrometer.
	The system was later adapted for use with a clinical 1.5~T MRI scanner, with
	the DNP chamber located in the fringe field at 0.35~T~\cite{lingwoodContinuousFlowOverhauser2010}. A completely integrated system
	that places the polarizer directly in the 1.5~T imaging magnet, in immediate
	proximity of the imaging target has been described by Krummenacker et al.~\cite{krummenackerDNPMRIInbore2012}. Due to the higher field, a higher microwave
	frequency of 45~GHz is required. This was achieved by coupling a microwave source
	outside of the magnet bore to the DNP cavity using a waveguide. Hyperpolarized
	water was thus injected into a microfluidic phantom cell of 0.4~mm thickness.
	Single-shot gradient images of the flowing solution are shown in Fig.~\ref{fig:krummenacker2012}. The hyperpolarized solution enters with negative
	polarization, and relaxes while flowing. A steady state in the spatial distribution of
	hyperpolarized molecules is thus established. The dark, halo-shaped region
	is the location where the polarization is zero, as the negatively-enhanced signals have 
	relaxed to the point where the signal exactly cancels out the positive thermal equilibrium polarization. 
	The system was later perfected by the same group, enabling higher flow rates, 
	in view of \emph{in vivo} applications~\cite{denysenkovContinuousflowDNPPolarizer2017}.

	The images shown in Fig.~\ref{fig:krummenacker2012} demonstrate
	that hyperpolarized MRI can be used to encode the
	flow velocity field experienced by the imaging liquid; the observed contrast
	is entirely due to the relative timing of convective/diffusive transport and
	longitudinal relaxation. This was also demonstrated in an almost simultaneous
	paper by Lingwood et al.~\cite{lingwoodOverhauserDynamicNuclear2012}, where
	conventional phase-encoded velocity imaging was compared with hyperpolarized imaging.

While the results discussed in the foregoing were obtained at modest magnetic fields
up to 1.5~T, modern NMR spectrometers operate at much higher magnetic fields, up to 24~T.
Implementation of Overhauser-enhanced NMR at such high fields is complicated on the
one hand by the need for high frequency microwave sources, waveguides, and
cavities, and on the other by a dramatic drop in the efficiency of the most
important cross-relaxation mechanism.
While Overhauser DNP based on dipolar electron-nucleus interactions is efficient
at low magnetic fields, the coupling factor for the dipolar interaction between stable radicals and
	solvent \textsuperscript{1}H nuclei tends towards zero for electron
	Larmor frequencies above 40~GHz (1.4~T) in liquids at room temperature.
	Therefore, the observed enhancement factors for \textsuperscript{1}H
	at typical NMR spectroscopy fields (between 5 and 24~T) are usually small. However,
	scalar electron-nuclear spin interactions can feature significant coupling factors
	for nuclei with a lower gyromagnetic ratio such as \textsuperscript{13}C, \textsuperscript{31}P, and
	\textsuperscript{15}N, and substantial
	Overhauser DNP enhancements at high magnetic fields (5~T and more)
	have been reported on this basis~\cite{griffinHighFieldDynamic2010,krummenackerLiquidStateDNP2012,levienNitroxideDerivativesDynamic2020,liuOnethousandfoldEnhancementHigh2017,loeningSolutionStateDynamicNuclear2002,neugebauerHighfieldLiquidState2014,orlandoDynamicNuclearPolarization2019,dubrocaQuasiopticalCorrugatedWaveguide2018,yoonHighFieldLiquidStateDynamic2018}, reaching up to 1000 in favorable cases~\cite{liuOnethousandfoldEnhancementHigh2017}.

For \textsuperscript{1}H, DNP enhancements at fields exceeding 9~T are much more
modest, with typical values ranging from -3 to about -30~\cite{denysenkovLiquidStateDNP2010,denysenkovLiquidStateDynamic2012,krummenackerLiquidStateDNP2012,neugebauerLiquidStateDNP2013}.
While these results are technically outside the
scope of this review since they were done on static samples, they nevertheless
demonstrate that high-field Overhauser DNP is possible at microfluidic size
scales.

Villanueva-Garibay et al.~have developed an Overhauser DNP system operating
at the intermediate field of 3.4~T (95~GHz electron Larmor frequency),
and reported DNP enhancements on protons of up to --100 in
samples of 10~mM TEMPO in water~\cite{villanueva-garibayPushingLimitLiquidstate2010a}.
The same system has been combined with a microfluidic supercritical fluid
chromatography system~\cite{taylerAnalysisMasslimitedMixtures2015a} that uses
a stripline resonator for efficient NMR detection~\cite{vanmeertenOverhauserDNPSupercritical2016}.
Supercritical \chemical{CO_2} is an attractive solvent for Overhauser DNP, since the microwaves are not absorbed, and the
rotational and translational correlation times are shorter than in water or
other liquid solvents. At high magnetic fields, this substantially increases
the efficiency of dipolar cross relaxation. While results in the truly supercritical
regime have not yet been reported, enhancements of \textsuperscript{1}H
NMR signals of up to --160 have
been achieved in a 1:1 mixture of toluene with \chemical{CO_2} containing
10~mM TEMPOL~\cite{vanmeertenOverhauserDNPSupercritical2016}. Similar enhancements
were achieved with aqueous samples at 15~MPa (150~bar).

The enhanced polarization offered by Overhauser DNP opens the prospect of compensating 
for the loss in sensitivity that results from using low magnetic fields. This is 
particularly attractive for compact NMR systems based on permanent magnets. Tabletop-sized
NMR systems~\cite{zalesskiyMiniaturizationNMRSystems2014} using permanent Halbach magnet arrays have recently become available
commercially, and offer spectral resolutions approaching 0.01~ppm at a fraction
of the size, investment, and running cost of a cryomagnet NMR system. However, permanent
magnets are limited to field strengths of less than 2~T, and the resulting sensitivity at thermal equilibrium polarization is insufficient for many microfluidic applications. Table-top Overhauser
DNP systems~\cite{armstrongPortableXbandSystem2008} offer a potential solution to this problem.
Recently, Kiss et al.~\cite{kissMicrofluidicOverhauserDNP2021} have presented a complete
NMR/EPR microprobe system that fits into the 11~mm gap of a commercial 0.5~T permanent magnet. 
Their system, complete with microfabricated shim coils, planar rf coils for NMR
pulses and detection, as well as a micro-stripline resonator and a 130~nL sample
volume, achieves a nLOD of 2.5~$\mathrm{\mu mol~s^{1/2}}$. 

\begin{figure}
    \centering
    \includegraphics[width=17.8cm]{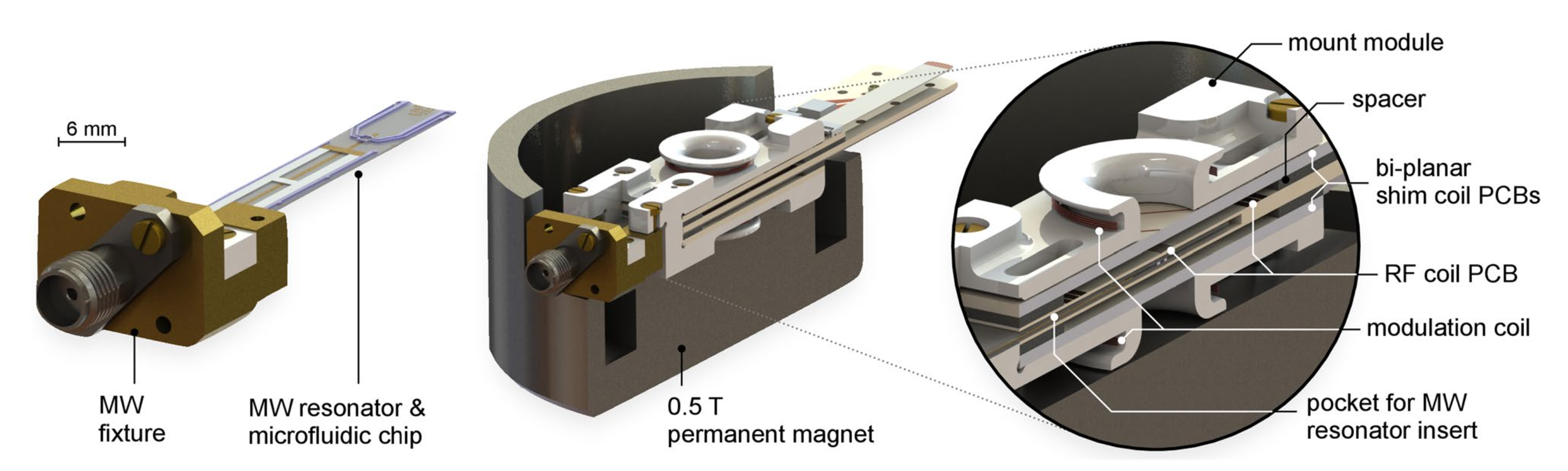}
    \caption{An Overhauser DNP probe that can fit inside a palm-sized permanent magnet. This setup was employed in Ref.~\cite{kissMicrofluidicOverhauserDNP2021} for the hyperpolarization and detection of \chemical{^1H} NMR signals. Image reproduced with permission from Ref.~\cite{kissMicrofluidicOverhauserDNP2021}.}
    \label{fig:kiss-2021-figure-1}
\end{figure}

Arguably the smallest-scale implementation of Overhauser DNP to date has made
use of a single-chip CMOS EPR spectrometer~\cite{sahinsolmazSingleChipDynamicNuclear2020}.
It has been shown that the circuitry required for EPR spectroscopy
can be integrated into a single, high-field compatible complementary metal-oxide
semiconductor (CMOS) chip~\cite{yalcinSinglechipDetectorElectron2008,andersKbandSinglechipElectron2012,kunstnerRapidScanElectron2021}.
Such integrated EPR spectrometers are extremely compact, with chip sizes of only a few millimeters, and can be operated with low power requirements. These devices integrate both the microwave electronics
and the detection coil/resonator on the same chip. A particularly ingenious
design uses an active feedback loop to continually tune a voltage-controlled microwave
oscillator; the frequency-dependent susceptibility due to the EPR response of
the sample is then reflected in the necessary control signal, from which
the EPR spectrum can be obtained~\cite{boeroRoomTemperatureStrong2013}.
Due to the inherently broadband nature of this approach, no field modulation is
required, dramatically reducing the dc power requirements for the experiment. Based
on this principle, Handwerker et al. have presented an integrated K-band (27~GHz) EPR
spectrometer that runs from a battery~\cite{handwerker2814GHzBatteryoperated2016}.

Such integrated EPR circuits could be of great interest in the context of
microfluidic hyperpolarization, as they could provide control and act as a
local microwave power source in miniaturized implementations of DNP. Interestingly,
it has been shown that such systems can operate at cryogenic temperatures~\cite{gualcoCryogenicSinglechipElectron2014}.
Recently, Solmaz et al. have reported an integrated CMOS single-chip EPR/NMR
system that operates at a nuclear proton Larmor frequency of 16~MHz (EPR frequency 10.7~GHz),
with a sample area of roughly 0.2~mm $\times$ 0.2~mm. The system has been
shown to produce Overhauser DNP enhancements up to about --50 in aqueous solutions of 10~mM TEMPOL~\cite{sahinsolmazSingleChipDynamicNuclear2020}.

In summary, Overhauser DNP is an attractive option for enhancing the sensitivity of microfluidic
NMR systems, particularly at low and moderate magnetic fields. It has the potential to directly operate on the working liquid
without the need for additional chemical reactions or solid-liquid phase transitions, which are
difficult to perform in a controlled manner at small scale.
Microfluidic implementations can alleviate problems arising from sample heating
and poor microwave penetration. 

\subsection{Dissolution-Dynamic Nuclear Polarization}
DNP in the solid state can yield much higher nuclear spin polarizations than Overhauser DNP. Overhauser DNP is performed on solution-state samples which cannot be cooled below the solution's freezing point. At 
room temperature, electron polarization, while higher than that of nuclei, is still small: and as an example, electron polarization at 5~T and 298~K is 1.1\%. In solid-state DNP the sample can be cooled to cryogenic temperatures at high field to yield near-unity electron polarization; at 5~T and 1~K the electron polarization is $\sim$100\%. The source is therefore much more strongly polarized, and in addition to this, the transfer of polarization from electrons to nuclei can be more efficient. Solid-state DNP employs microwave irradiation to coherently drive electron-nuclear polarization transfer, in contrast to Overhauser DNP which relies on incoherent cross-relaxation effects. The combination of these effects means that solid-state DNP can yield samples with nuclear spin polarization close to unity.

\begin{figure}
		\begin{center}
			\includegraphics[width=8.4cm]{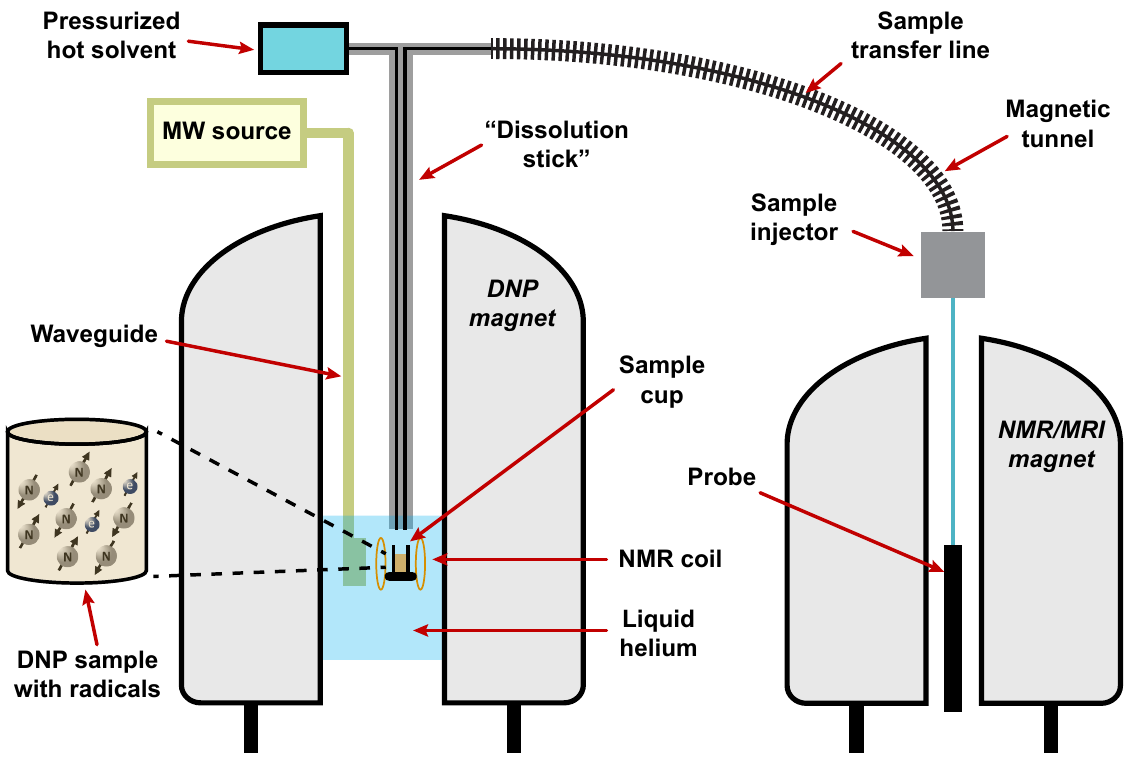}
		\end{center}
		\caption{An example of a standard dissolution DNP setup. The sample to be polarized is frozen together with free radicals as a glassy solid. The sample is cryogenically-cooled to $\sim$1.2~K in a magnetic field of $\sim$5~T to polarize the electron spins, and microwaves are applied to induce electron-nuclear polarization transfer. After sufficient nuclear spin polarization is established (which can be monitored \emph{in situ} with an NMR pick-up coil), a pressurized hot solvent is ejected onto the sample to dissolve the solid. Pressurized gas pushes the solution out from the DNP magnet and through a guiding magnetic field, at which point it can be collected for use.}
		\label{fig:D-DNP-apparatus}
	\end{figure}

In a D-DNP experiment a sample is frozen with free radicals (e.g., TEMPOL) dispersed within the amorphous solid matrix. The sample is cooled to cryogenic temperature ($\sim$1.2~K) in a high magnetic field (usually between 3 and 7~T) to generate electron polarization in the free radicals, which is then transferred to the nuclear spins in the sample via microwave irradiation. Once the nuclear spins are sufficiently polarized, a pressurized hot solvent (e.g., water at 150$^{\circ}$C and 7~bar) is injected to dissolve the sample and eject it from the cryostat, and the hyperpolarized solution containing the target molecule is collected. Unlike optical pumping, which is specific to noble gases, and PHIP, which relies on specific chemical reactions/interactions, D-DNP can polarize a broad range of molecules. An illustration of a dissolution DNP experiment is shown in Fig.~\ref{fig:D-DNP-apparatus}, and more comprehensive reviews of D-DNP are given in Refs.~\cite{bornetOptimizingDissolutionDynamic2016,janninApplicationMethodologyDissolution2019,leeDissolutionDynamicNuclear2016,kovtunovHyperpolarizedNMRSpectroscopy2018,pinonHyperpolarizationDissolutionDynamic2021}.

However, due to the necessary rapid temperature changes, D-DNP is difficult to scale, and is therefore not a natural friend of microfluidic NMR. Since the dissolution step requires the hot solvent to pass through a cryostat, large volumes are required to prevent the solvent freezing in the lines. Usually volumes on the order of 5 to 40~ml are used to dissolve $\sim$100 \muL~of hyperpolarized material. Dissolution DNP is a batch-mode technique, requiring approximately 20 to 60~minutes to fully polarize a sample due to the slow build-up rates under cryogenic conditions, before the sample is ejected from the polarizer for use. The process is then started again on a new sample. This means D-DNP cannot produce hyperpolarized samples in continuous-flow, and hence is unavailing as a method to produce microliter to nanoliter hyperpolarized sample volumes for observation in microfluidic NMR. The recent advent of `bullet-DNP'~\cite{kourilScalableDissolutiondynamicNuclear2019} might help to alleviate this shortcoming. In a bullet-DNP experiment, the hyperpolarized material is ejected from the polarizer as a solid and subsequently dissolved in a solvent; since the solvent does not pass through the cryostat, a lower volume can be used to dissolve the hyperpolarized solid, to produce samples at higher concentration.

A limitation of D-DNP is that the hyperpolarized fluid produced after the dissolution step usually contains the paramagnetic free radicals that were used for the polarization step. Even though they are in low concentration, the radical molecules induce rapid nuclear spin relaxation of the hyperpolarized molecules. This is particularly detrimental for \textsuperscript{1}H nuclei, so in most cases a lower-$\gamma$ X-nucleus is polarized since the lifetime is often longer. Radical-induced relaxation is especially efficient at low magnetic fields, so a magnetic guiding field is required during sample transport to mitigate relaxation losses~\cite{milaniMagneticTunnelShelter2015}. In addition to this, strategies for rapidly removing the radicals immediately following the dissolution step have been demonstrated, such as filtering polymer-supported radicals~\cite{vuichoudFilterableAgentsHyperpolarization2016}, phase extraction of the radicals from the aqueous phase into an organic solvent~\cite{harrisDissolutionDNPNMR2011}, and radical scavenging using ascorbic acid~\cite{mievilleScavengingFreeRadicals2010}. A particularly promising new method is to generate non-persistent UV-induced radicals for the polarization step, which are quenched upon sample warming during the dissolution step~\cite{eichhornHyperpolarizationPersistentRadicals2013}.

\begin{figure}
	\begin{center}
			\includegraphics[width=17.8cm]{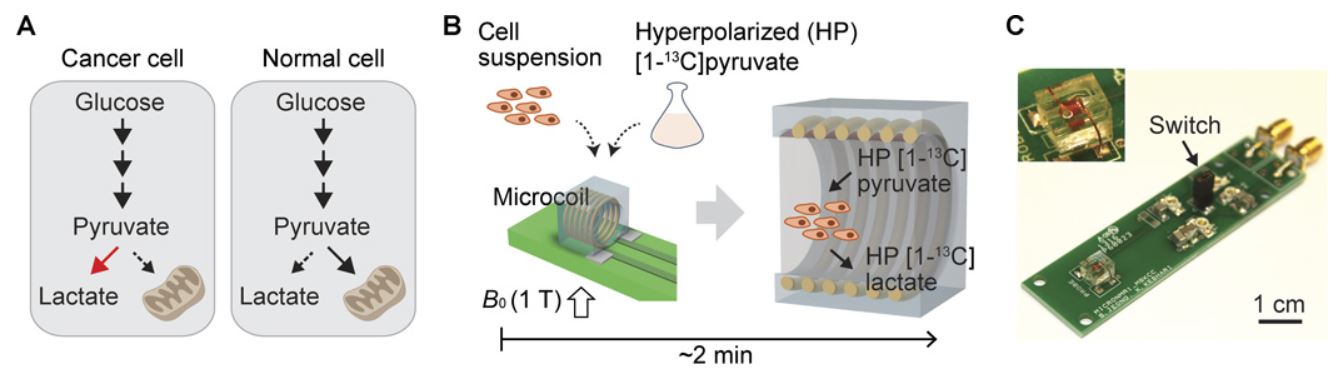}
	\end{center}
	\caption{The dissolution DNP microfluidics experimental setup of Jeong et al. (A) Metabolic differences (the Warburg effect) in pyruvate-to-lactate conversion between healthy and cancerous cells that can give image contrast. (B) Perfusion of hyperpolarized pyruvate with a cell suspension into the microfluidic chip. (C) The microfluidic chip used in experiments. Image adapted with permission from Ref.~\cite{jeongRealtimeQuantitativeAnalysis2017}.
	}
	\label{fig:jeong2017}
\end{figure}

Jeong et al. have shown that D-DNP can be combined with microfluidic sample control and detection using a hyperpolarized micromagnetic resonance spectrometer (see Fig.~\ref{fig:jeong2017})~\cite{jeongRealtimeQuantitativeAnalysis2017}. Their system contained a solenoid microcoil with a 2~\muL~internal volume, which could be frequency-tuned for either \proton~or \carbon~NMR in the 1.05~T external magnetic field. They hyperpolarized [1-\carbon]pyruvate using a commercial D-DNP polarizer, and mixed the hyperpolarized [1-\carbon]pyruvate solution with a cell suspension immediately prior to injection into a 100~\muL~reservoir. They used a syringe pump to push the sample into the detection region in a stop-flow manner, and the rate of pyruvate$\rightarrow$lactate metabolic flux was quantified. This work demonstrates clear advantages of a microfluidic approach over measurement in a conventional NMR tube:
\begin{enumerate}
    \item Usually small flip-angle pulses ($\leq$~30$^{\circ}$) are used to excite the nuclei, so as not to strongly perturb the hyperpolarized magnetization. If a 90$^{\circ}$ pulse were to be used, a larger signal could be acquired, but the hyperpolarized magnetization would be destroyed. By only exciting+detecting 2~\muL~of the sample at a time, the authors were able to use 60$^{\circ}$ flip-angle pulses for each acquisition without destroying the hyperpolarization of the entire sample.
    \item A significantly higher sensitivity (nLOD, not cLOD) was achieved: in macroscopic NMR samples, typically $\sim$10\textsuperscript{7} cells are required to generate a sufficient amount of lactate to be detected. By using their microfluidic system, the authors were able to detect metabolic flux using  $\sim$10\textsuperscript{4} cells.
\end{enumerate}

A closely-related analogue to a dissolution-DNP experiment was demonstrated by Sharma et al.
based on rapid melting of a DNP-polarized solid~\cite{sharmaRapidmeltDynamicNuclear2015}. In the rapid-melt DNP experiment, a 50 to 100~nL sample is contained within a capillary which is shuttled between a liquid nitrogen-cooled cryostat for the DNP process, a melting stage, and a stripline NMR coil for signal excitation and readout. This method circumvents the need for large volumes of solvent to produce solution-state samples of the polarized solid, and instead achieves this by melting the polarized solid in the melting stage immediately prior to signal readout. The sample is not chemically changed or destroyed in the experiment, so the same sample can be polarized and detected many times.  Several transients can therefore be added, increasing
sensitivity. It is also possible to acquire 2D \chemical{^1H-^1H} COSY (correlation spectroscopy) and \chemical{^1H-^{13}C} HSQC (heteronuclear single quantum correlation) spectra with this technique~\cite{vanmeertenRapidmeltDNPMultidimensional2020}. A notable advantage of this method is that it allows for the detection of \chemical{^1H} NMR signals since the time required for the sample to leave the cryostat, melt, and be measured is so short. As discussed previously the transport is generally challenging with D-DNP since protons often relax relatively quickly after the dissolution step due to the presence of paramagnetic free radicals. This is a remarkable achievement since in this method the radical concentration is not diminished in the melting step, so the \chemical{^1H} relaxation times are on the order of 1~s.

The advantages of D-DNP stem from the high polarization levels achievable and the generality. The drawback is in the expense and overhead to produce a single 5 to 40~ml bolus of hyperpolarized solution, which cannot be easily scaled down due to the necessity to balance heat loads and not have the solutions freeze in the cryogenically-cooled fluid path. Tens of minutes are required to polarize a single sample, and once ejected the hyperpolarization decays on a timescale of tens of seconds. The prospect of combining an array of microfluidic detectors so that many NMR experiments can be performed in tandem using one hyperpolarized sample remains an attractive future goal.

\section{Optical Pumping of $^{129}$Xe}
\label{OpticalPumping}
\subsection{Introduction}

\begin{figure}
    \begin{center}
    \includegraphics[width=8.4cm]{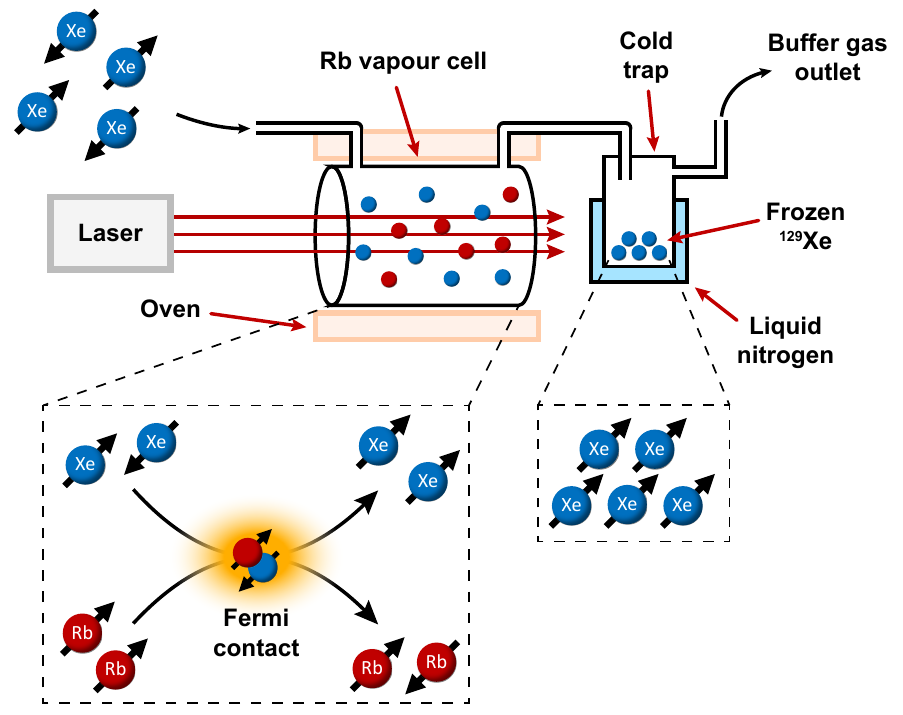}
    \end{center}
    \caption{An example of a spin-exchange optical pumping (SEOP) experimental setup. \Xe\ gas is flowed with a buffer gas into a heated rubidium vapor cell. A laser tuned to approximately 794.7~nm pumps the Rb $D_1$ transition with circularly-polarized light which leads to polarization of the rubidium valence electrons. Collisions between the rubidium and xenon atoms leads to the electron polarization being transferred to the xenon nuclei through the Fermi contact interaction. This mixture of hyperpolarized xenon and the buffer gas can then be passed through a cold trap that freezes out the \Xe\ as a pure hyperpolarized solid. From this point the \Xe\ can be warmed to induce sublimation and yield gaseous hyperpolarized \Xe.}
    \label{fig:Optical-Intro}
\end{figure}

Alkali metal atoms possess an unpaired valence electron which 
can be electronically excited by irradiation in the vapor phase with light \cite{kastlerOpticalMethodsAtomic1957}. Rubidium is the most 
commonly-used alkali metal for this purpose. It is heated to around 40 to 120$^{\circ}$C to produce an 
atomic vapor, and the $D_1$ transition of the valence electron
is optically excited with resonant circularly polarized light. This pumps the electrons exclusively from the $m_s=-1/2$ ground state to the $m_s=+1/2$ excited state (see Fig.~\ref{fig:Optical-Intro}). The electrons can then relax to the 
ground state with or without an accompanying spin flip, and so a steady-state is 
established. In this way, near-unity electron spin polarization is achievable with moderate 
irradiation power, usually between 20 and 200~W for a large-scale polarizer~\cite{nikolaouNearunityNuclearPolarization2013}, but as low as a few milliwatts in micro-scale polarizers. An applied magnetic field of typically a few millitesla provides a quantization axis along which the Rb electrons are polarized.
The hot alkali metal vapors are chemically highly reactive, but noble gas 
atoms (e.g., \chemical{\textsuperscript{3}He}, \chemical{\textsuperscript{83}Kr},
\chemical{\textsuperscript{129}Xe}) are sufficiently inert to be brought 
into contact with the vapor. Collisions between the species allow 
transfer of the electron polarization to the nuclei through the Fermi contact 
interaction. This can happen through two-body or three-body collisions, where 
the third body is a buffer gas
molecule (typically \chemical{N\textsubscript{2}}). The buffer gas provides
a pathway for nonradiative emission of energy as the excited electronic state of the
alkali metal atoms relaxes, which prevents the emission of unpolarized 
photons which would depolarize the atomic vapor. Hence, a noble gas can be hyperpolarized
by flowing it through a vapor cell while optically pumping the alkali metal atoms, a 
technique known as spin-exchange optical pumping (SEOP).

Typically a system would be run with a gas pressure on the order of a few atmospheres, with \Xe\ making up 2 to 90\% of the total pressure, the rest being vaporized rubidium and buffer gas.
A common approach to purify the \Xe\ is to cryogenically freeze it out of the 
gas mixture as a hyperpolarized solid. The solid \Xe\ can be stored in this state for hours~\cite{limesRobustSolid1292016}. 
After building up a suitable quantity, the \Xe\ is melted or sublimed to bring it back 
to a fluid state for applications.

Hyperpolarized \Xe\ has been used to study porous materials, and to extract information about the structure of the nano- and micro-scale pores~\cite{mouleAmplificationXenonNMR2003,seeleyRemotelyDetectedHighfield2004,knaggeAnalysisPorosityPorous2006}. In this type of work microfluidic control is not exerted over the hyperpolarized species, and the specific advantages of microfluidics are not utilized,
and hence this work is outside the scope of this review. A more comprehensive review of this topic is given in Ref.~\cite{weiland129XenonNMRReview2016}.

\subsection{Combined Optical Pumping and Detection on a Single Microfluidic Platform}
The use of vapor cells of optically pumped \Rb~to polarize nuclear spins of noble gas atoms is just one application; they are more commonly employed as highly sensitive magnetic field sensors~\cite{tierneyOpticallyPumpedMagnetometers2019}. In this context, they are referred to as optically pumped magnetometers (OPMs). Their operation requires low magnetic fields (typically less than 1~mT), and in this regime they have been employed for the detection of zero- to ultralow-field (ZULF) NMR signals~\cite{blanchardZeroUltralowFieldNMR2016,blanchardZeroUltralowfieldNuclear2020}. At such low fields the spin precession frequencies are small and inductive detection is inefficient.

\begin{figure}
	\begin{center}
			\includegraphics[width=8.4cm]{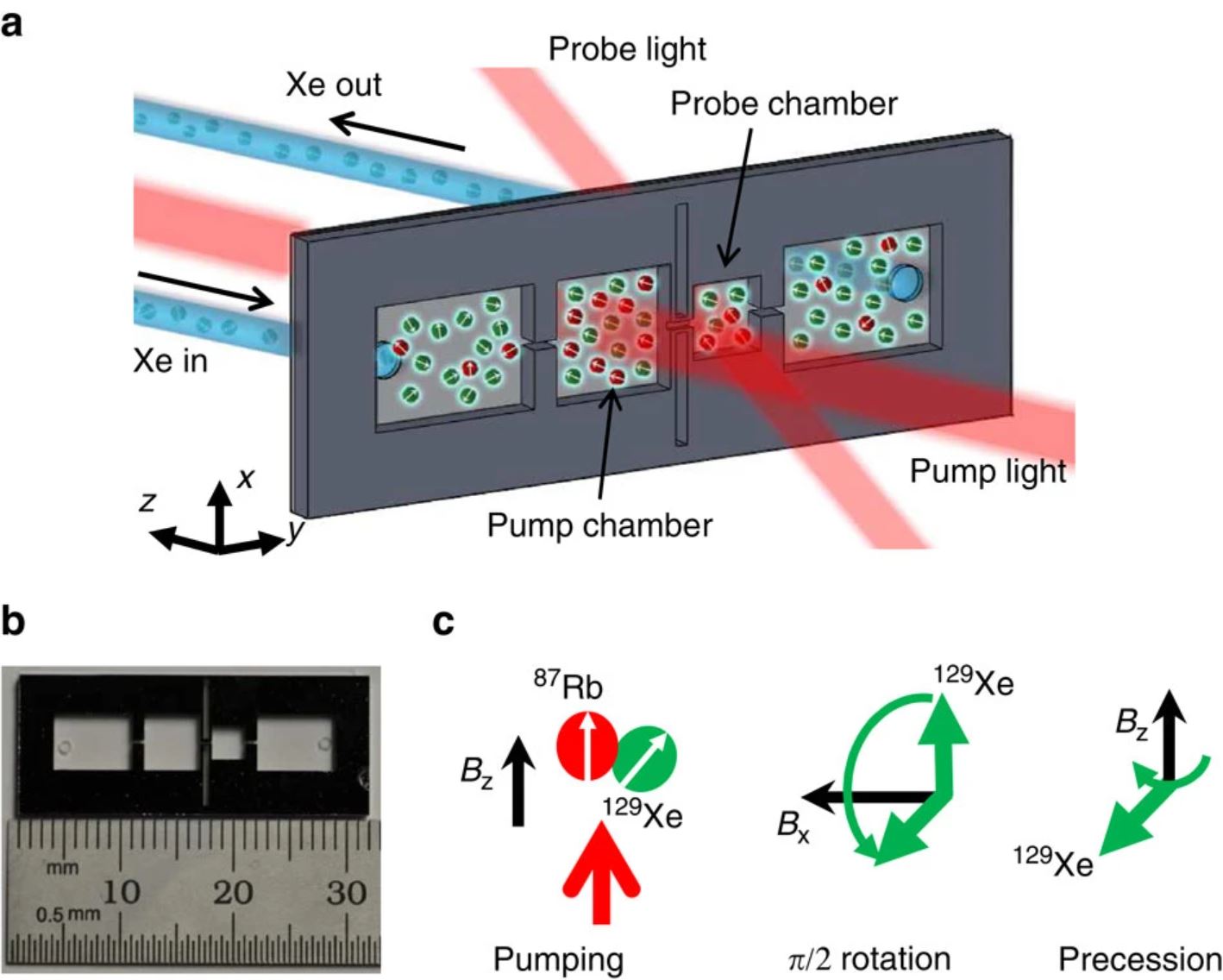}
	\end{center}
	\caption{The microfluidic chip used in the experiments by Jimenez-Martinez et al. which combines optical pumping and detection of \Xe\ on one chip. (a) A schematic of the chip. (b) Image demonstrating the size scale of the chip. (c) Pumping and probing sequence for \Xe. The chip was made by cutting chambers through a layer of silicon and sandwiching this between borosilicate glass slides. Image adapted with permission from Ref.~\cite{jimenez-martinezOpticalHyperpolarizationNMR2014}
	}
	\label{fig:jimenez-martinez2014}
\end{figure}

The combination of these two aspects of an optically pumped atomic vapor was recently shown on a single microfluidic platform~\cite{jimenez-martinezOpticalHyperpolarizationNMR2014} (see Fig.~\ref{fig:jimenez-martinez2014}). The authors used a microfluidic silicon chip 1~mm thick, sandwiched between borosilicate glass layers, with chambers cut out from the silicon wafer for gas flow, optical pumping of the rubidium, and optical detection. A schematic of the chip and the lasers used for optical excitation and readout is shown in Fig.~\ref{fig:jimenez-martinez2014}. The chip had a total volume of approximately 100~\muL, and the detection chamber itself had a volume of 9~\muL.

The chip was heated to between 120 and 150\degC~in order to create a Rb atom vapor, and a mixture of Xe and \Ntwo~gases (400 Torr \Ntwo~and 200 Torr Xe) was made to flow into the chip at rates from 1 to 30~\muL/s. The xenon was optically pumped in the pump chamber, and detected directly there or in the probe chamber. A 0.8~\muT~static magnetic field was applied and a short transverse magnetic field pulse was used to tip the nuclear spins and initiate spin precession, to generate observable signals for detection near $\sim$9~Hz.

Using this experimental setup, polarization levels of $\sim$0.7\% were determined for xenon in the detection chamber at the optimal flow rate of 5~\muL/s. The polarization levels were lower than expected, mostly due to rapid \Xe\ spin relaxation from wall collisions. While such collisions do not necessarily cause rapid relaxation, in the case at hand the particular process used for rubidium production led to barium chloride being formed, which deposited on the walls and induced \Xe\ relaxation.

\begin{figure}
	\begin{center}
			\includegraphics[width=8.4cm]{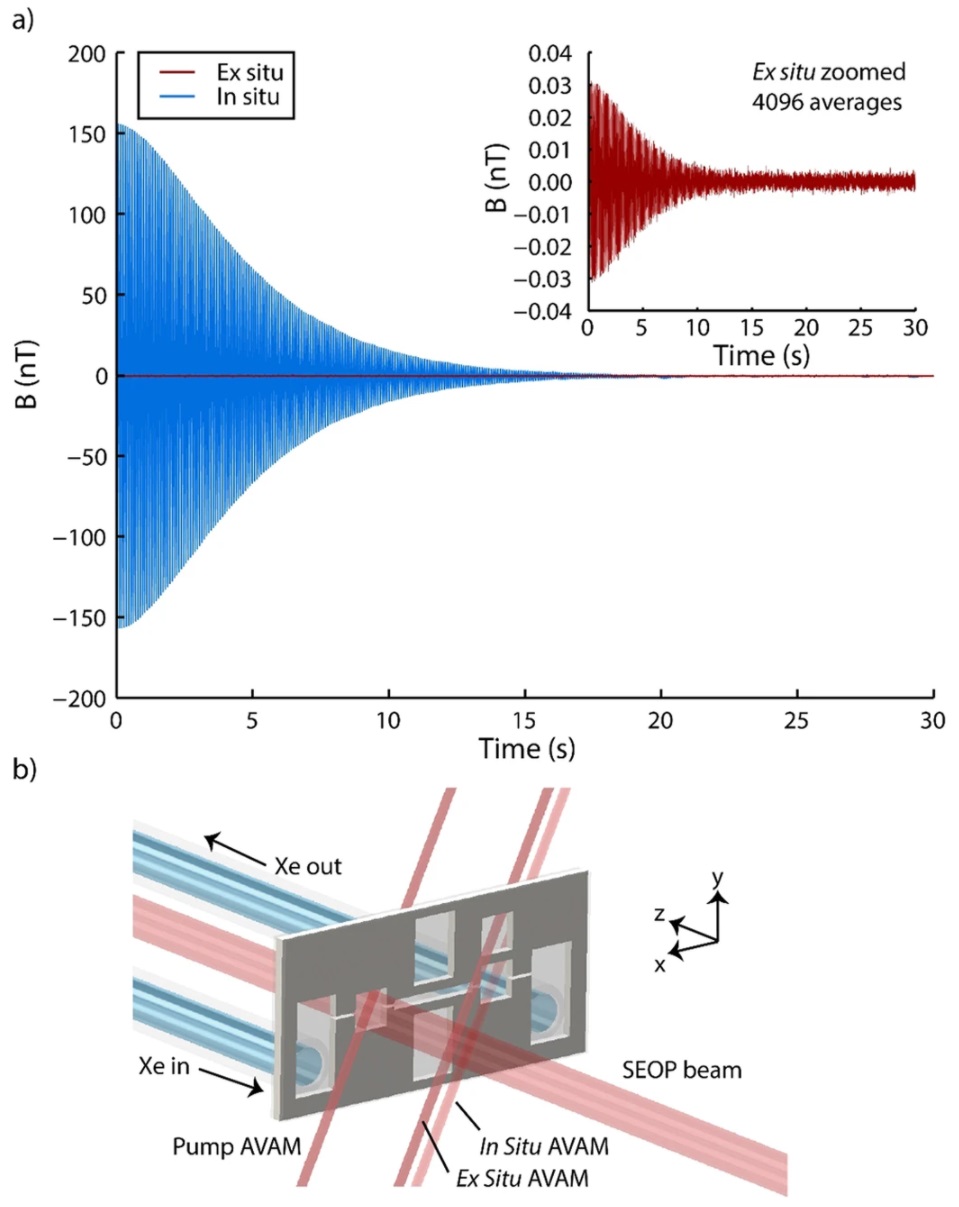}
	\end{center}
	\caption{Hyperpolarization of \Xe\ in a microfluidic chip from the experiments of Kennedy et al. a) A comparison between the \emph{in situ} and \emph{ex situ} NMR signals. b) The microfluidic chip used for experiments, with the \emph{in situ} and \emph{ex situ} alkali vapor atomic magnetometer (AVAM) chambers labelled. Image taken with permission from Ref.~\cite{kennedyOptimizedMicrofabricatedPlatform2017}.
	}
	\label{fig:kennedy2017}
\end{figure}

An improved version of this device was produced at a later date~\cite{kennedyOptimizedMicrofabricatedPlatform2017} which benefited from a different method to produce the rubidium (not resulting in the production of barium chloride), so that the walls of the chip were less contaminated. This led to longer spin relaxation times by a factor of $\sim$5, and hence to an improvement in \Xe\ polarization of up to 7\% under optimized conditions.
On this new device, a comparison was made between \emph{in situ} and \emph{ex situ} detection. \emph{In situ} refers to detection of the xenon in the presence of the Rb vapor, whereas \emph{ex situ} refers to detecting the xenon using an Rb vapor cell nearby but without the atoms being in direct contact. The \emph{in situ} and \emph{ex situ} detection chambers are shown in the microfluidic device in Fig.\ref{fig:kennedy2017}. By detecting \emph{ex situ} the Rb electrons respond to the dipole field produced by the \Xe\ nuclei. By detecting \emph{in situ} the NMR signals are higher by a factor of 5300: a factor of $\sim$10 is from the reduction in distance increasing the dipole-dipole interaction strength (the authors verified this with COMSOL magnetic field simulations), and a further factor of $\sim$500 is due to the Xe-Rb Fermi contact interaction~\cite{groverNobleGasNMRDetection1978}. This is illustrated in Fig.\ref{fig:kennedy2017}. A factor of 2 could be gained if the \emph{ex situ} OPM cell were placed near the chamber containing the polarized xenon. The authors note that the two detection methods are complimentary, as although \emph{in situ} detection is dramatically more sensitive, there are many practical applications in which one cannot use \emph{in situ} detection (i.e., when the xenon is to be dissolved in a solvent).

%\todo{Do we want to add LOD? 
%A representative LOD: 500 Torr Xe at natural abundance, 400 nT initial signal, 4 pT/sqrt[hz] noise floor in a 40 Hz bandwidth, 6s T1..}

% MU: On the basis of 40 Hz bandwidth, and 24.6% na for 129Xe, this corresponds
% to an nLOD of 3.53 pmol sqrt(s). Concentration limit of detection cLOD works out
% to 0.22 µM sqrt(s).
%

Despite the gain in polarization level in the improved device, the maximum polarization ($\sim$7\%) is still an order of magnitude lower than the \Xe\ polarization levels achievable in large-scale polarizers~\cite{meersmannHyperpolarizedXenon129Magnetic2015}. This is because of the relatively short xenon relaxation times due to the high surface-to-volume ratio in the microfluidic devices increasing wall collisions. This will be an important consideration for future work involving hyperpolarized gases in microfluidics, and solutions such as anti-relaxation wall coatings can be considered~\cite{straessleMicrofabricatedAlkaliVapor2014}.

\subsection{Microfluidic Enhancement of Gas-Liquid Contact}
\label{Xe-bubble-free}
The work with hyperpolarized xenon described so far has used xenon in the gas-phase, but there are many exciting 
applications of hyperpolarized \Xe\ as a biosensor in solution~\cite{schroderXenonNMRBiosensing2013}. This is motivated by relatively long solution-state relaxation times, the large chemical shift dispersion that means \Xe\ is highly sensitive to changes in its chemical environment, as well as being chemically inert and hence biologically compatible.
For applications as a biosensor,  \Xe\ needs to be brought into solution, but any process 
that requires bubbling or mechanical mixing of the \Xe\ gas with the solution introduces 
challenges to observing the samples with high-field NMR:
the presence of gas pockets in the solution disturbs the magnetic field homogeneity.
This results in spectral line broadening which has two important implications: the signal-to-noise 
ratio is reduced because the signal is spread over a wider frequency range, and resolution is reduced 
as signals from different species overlap. In the case of \Xe, the latter effect is often inconsequential 
because the chemical shift differences are typically large enough (hundreds or thousands of ppm) to prevent spectral overlap. This is certainly not the case in e.g., \textsuperscript{1}H NMR spectra where different chemical species 
can easily have sub-ppm chemical shift differences. Bubbling \Xe\ into 
the solution also brings up the question of experimental reproducibility, and introduces a 
necessary waiting period prior to detection for the solution to settle which prohibits 
the study of fast kinetic processes. The presence of bubbles in solution causes even greater problems 
in microfluidics, where they can disrupt the flow and block channels, and accumulate
in the detection chamber. To overcome these problems, bubble-free dissolution of gases into solutions through porous membranes has been implemented in a variety of ways~\cite{berthaultUseDissolvedHyperpolarized2020}.

In addition to the problems introduced by having gas pockets in the solution, the hyperpolarized 
\Xe\ relaxes during the experiment, so it is important to have efficient and rapid gas exchange. 
Microfluidic enhancement of the liquid-gas interface is a promising option. Liquid-gas interfaces are problematic in microfluidic systems and are best avoided, but gas-permeable solid materials and membranes can be used to mediate the gas transport into a liquid phase. The most famous example is PDMS (polydimethylsiloxane), an elastomeric gas-permeable solid polymer that can facilitate gas exchange while providing structural support into which fluid channels can be etched. However, there are many other gas-permeable solids that can be used for this purpose.

\begin{figure}
	\begin{center}
			\includegraphics[width=8.4cm]{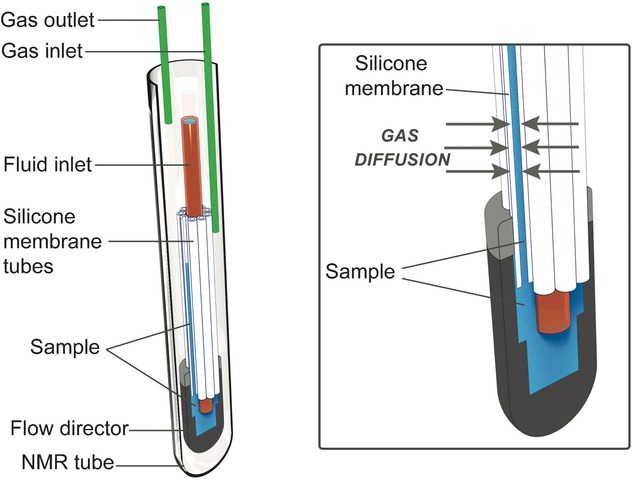}
	\end{center}
	\caption{The use of PDMS hollow-fiber membranes to dissolve hyperpolarized \Xe\ into viscous oriented media without the need for bubbling. Image reproduced with permission from Ref.~\cite{truxalNondisruptiveDissolutionHyperpolarized2016}.
	}
	\label{fig:truxal2016}
\end{figure}

In 2006 Baumer et al. demonstrated continuous delivery of hyperpolarized \Xe\ into aqueous solutions 
by flowing the gas through microfluidic polypropylene hollow-fiber membranes~\cite{baumerNMRSpectroscopyLaserPolarized2006}.
This type of membrane-based gas exchange was later expanded on by Cleveland et al. who produced 
a gas-exchange module to continuously infuse a flowing solution with 
hyperpolarized \Xe\ \cite{clevelandContinuouslyInfusingHyperpolarized2009}.
In 2009, Amor et al. studied a variety of commercially available polypropylene and
polymethylpentene hollow-fiber membrane materials for dissolving hyperpolarized \Xe\ into 
solutions, and acquired magnetic resonance images to map the spatial distribution of 
the \Xe\ \cite{amorNMRMRIBloodDissolved2011}.
In 2016, Truxal et al. showed that PDMS membranes can be used to dissolve 
hyperpolarized \Xe\ into viscous oriented liquid crystals without disrupting the 
media~\cite{truxalNondisruptiveDissolutionHyperpolarized2016}. They report that PDMS is advantageous 
over polypropylene because it is less prone to rupturing, and in their work they show an inverted 
setup in contrast with the previously-discussed works; the liquid is contained within the hollow-fiber 
membranes with the pressurized gas outside. This allows for higher gas pressures to be used, and 
lower media volumes (in this case $\sim$60~\muL). This is illustrated in Fig.~\ref{fig:truxal2016}.

\begin{figure}
	\begin{center}
			\includegraphics[width=8.4cm]{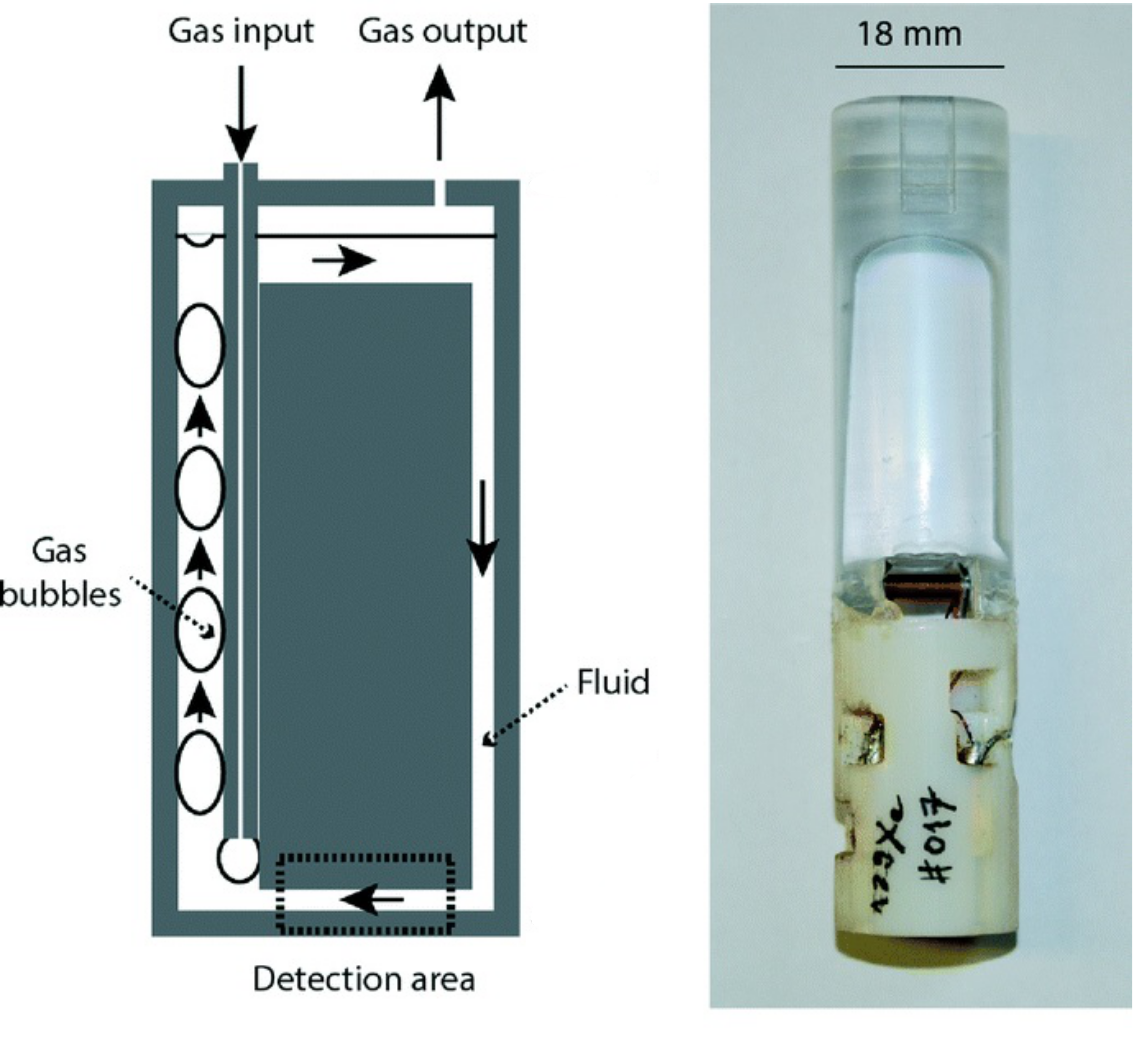}
	\end{center}
	\caption{The microfluidic device reported in Ref.~\cite{causier3DprintedSystemOptimizing2015} for dissolving hyperpolarized \Xe\ into solution and detecting the enhanced NMR signals. The gas bubbles act as a micropump, continually pumping fresh solution into the detection region. Reproduced from Ref.~\cite{causier3DprintedSystemOptimizing2015} with permission.
	}
	\label{fig:causier2015}
\end{figure}

A 3D-printed microfluidic system for both dissolving hyperpolarized \Xe\ into a solvent and detecting the NMR signals has been demonstrated by Causier et al.~\cite{causier3DprintedSystemOptimizing2015} (see Fig.~\ref{fig:causier2015}). The authors bubbled hyperpolarized \Xe\ into a solution of 0.5~mM cryptophane-2,2,2-hexacarboxylate in \chemical{D_2O} in the mixing chamber of the device. Through their buoyancy, the bubbles act as a micropump and push the sample into a 5~\muL~detection chamber, where the NMR signal was acquired using a solenoid microcoil. Two distinct NMR peaks were seen, corresponding to free \Xe\ in \chemical{D_2O}, and \Xe\ in the cryptophane cage. The authors were able to confirm that bubbles do not enter the volume encompassed by the microcoil. The advantage of this approach is that the xenon is dissolved close to the point of detection, meaning there is less time for spin relaxation to occur. This modular system is designed to be easily exchangeable, and clearly demonstrates the power of emerging 3D printing technologies for developing microfluidic devices for NMR. A downside is that the bubble-driven micropump cannot easily be scaled to different volumes, as its operation relies on a specific balance between surface tension, buoyancy and friction.

\section{Parahydrogen-Induced Polarization}
	\subsection{Introduction}
The \chemical{H_2} molecule contains two protons which support four 
nuclear spin states: a singlet state (parahydrogen, \textit{p}-\chemical{H_2}) with total 
spin $I=0$, and three triplet states (collectively called orthohydrogen, \textit{o}-\chemical{H_2}) 
with total spin $I=1$. The \chemical{H_2} molecule behaves 
as a quantum rotor, with quantized rotational states. The Pauli exclusion principle dictates that the overall wave function 
must remain antisymmetric with respect to exchange of the two nuclei. Therefore, rotational
states with even angular momentum quantum number, which are symmetrical with respect
to exchange, can only combine with the antisymmetric singlet spin state, while
odd rotational states combine with the three triplet states.
At room temperature, the even and odd rotational manifolds are evenly populated. At temperatures below 100~K, however,
Boltzmann statistics leads to significant excess population of the lowest rotational energy level,
which is even. Correspondingly, the para nuclear spin state is over-populated.
While this link between spin and rotational degrees of freedom exists also in other
diatomic molecules, hydrogen is special in the sense that the temperatures at which the overpopulation of the singlet state is reached is high enough to avoid freezing of the gas.

\begin{figure}
    \begin{center}
            \includegraphics[width=12cm]{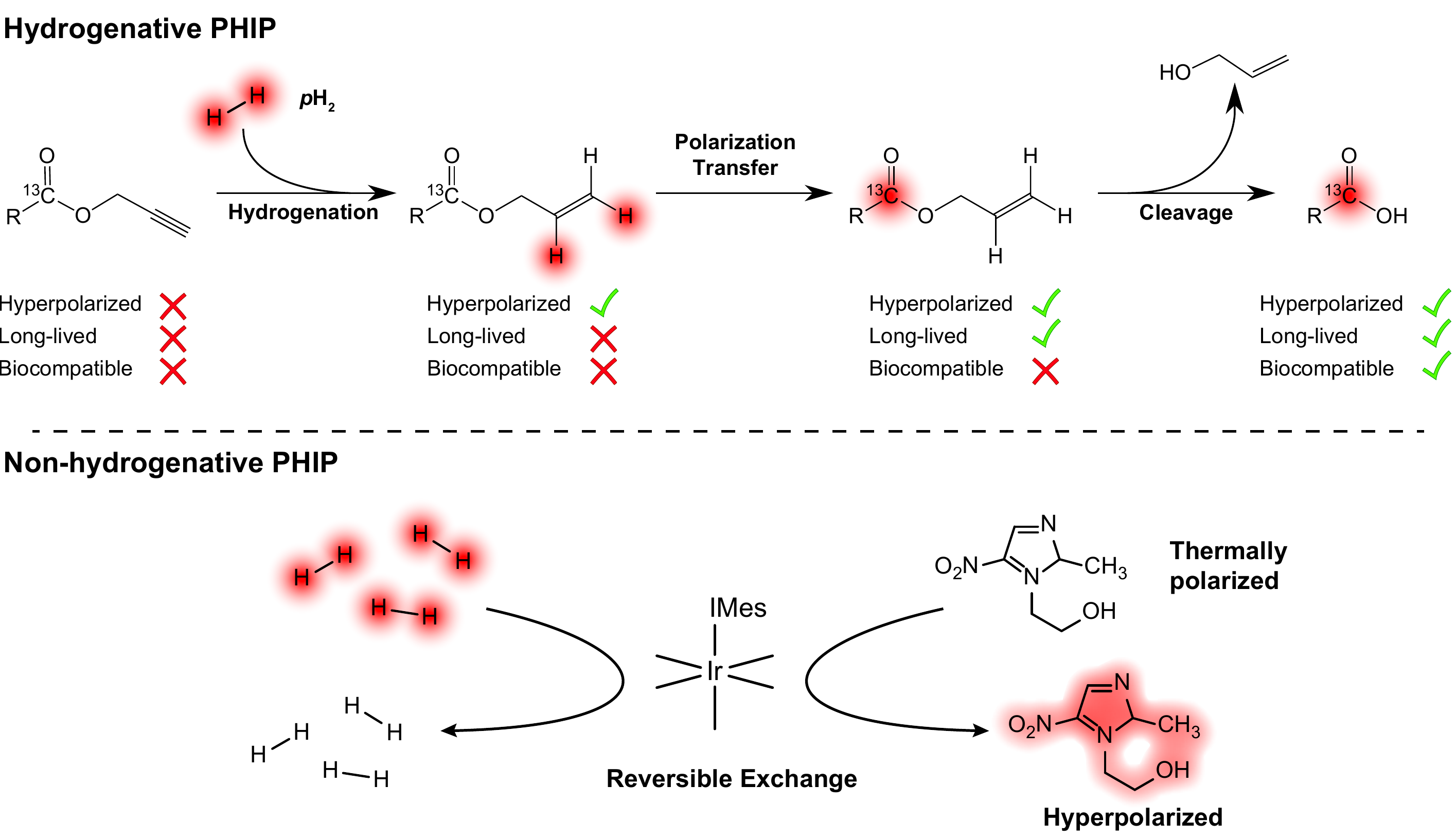}
    \end{center}
    \caption{Top: An example of hydrogenative PHIP being used in combination with the side-arm hydrogenation procedure to produce a hyperpolarized biomolecule. An unsaturated side arm is hydrogenated using \pHtwo\ to produce a proton-polarized molecule. The polarization is transferred via the $J$-coupling network to a nearby \carbon\ nucleus which relaxes more slowly. The side arm is then cleaved by ester hydrolysis to produce a biocompatible hyperpolarized species. Bottom: An example of non-hydrogenative PHIP being used to hyperpolarize metronidazole. \pHtwo\ and the substrate reversibly bind with an iridium catalyst, and the temporary $J$-coupling network allows polarization to flow from the \pHtwo\ to the metronidazole.}
    \label{fig:PHIP-Intro}
\end{figure}

Conversion between the even and odd rotational manifolds in molecular hydrogen is kinetically slow,
but can be accelerated by solid ferromagnetic catalysts such as iron oxide.
Below 25~K nearly 100\% para enrichment 
can be achieved in this way. 
Once the gas is separated from the catalyst, the parahydrogen can remain metastable for days at room temperature if stored in a 
suitable container~\cite{wagnerConversionRateParahydrogen2014}.
This non-equilibrium population distribution between the para and ortho states of \chemical{H_2}
is utilized by various techniques to hyperpolarize target molecules~\cite{bowersParahydrogenSynthesisAllow1987,eisenschmidParaHydrogenInduced1987}.
There are a number of reviews that describe parahydrogen-induced polarization (PHIP) 
in more detail~\cite{greenTheoryPracticeHyperpolarization2012,duckettImprovingNMRMRI2013,
hovenerParahydrogenBasedHyperpolarizationBiomedicine2018,reineriHowDesign13C2011,glogglerParahydrogenPerspectivesHyperpolarized2013}.

A notable advantage of PHIP as a hyperpolarization method is that the equipment demands are minimal, 
and the associated cost is low. Experiments 
typically involve bringing hydrogen gas at moderate pressure (between 1 and 20~bar) into 
a solution to undergo a chemical reaction/interaction with a target molecule. This is often done by bubbling the gas through the solution, or shaking an NMR tube 
which contains the solution under a pressurized parahydrogen atmosphere. Typical reaction temperatures range from room temperature to 150$^{\circ}$C. PHIP is not limited to solution-state hyperpolarization, and can also be used to generate hyperpolarized gases via heterogeneously-catalyzed gas-phase 
reactions~\cite{pokochuevaHeterogeneousCatalysisParahydrogenInduced}.

There are two categories of PHIP experiment: \textit{hydrogenative} PHIP in which 
a catalyst is used to chemically react parahydrogen with a precursor molecule to 
produce a hyperpolarized target, and \textit{non-hydrogenative} PHIP, in 
which parahydrogen and the target molecule, with the aid of a catalyst, form 
a transient complex which facilitates the polarization 
transfer from parahydrogen to the target. These two variants are shown in Fig.~\ref{fig:PHIP-Intro}, and in the following 
subsections we provide further introduction to hydrogenative PHIP and non-hydrogenative PHIP.

\subsubsection{Hydrogenative PHIP}
In hydrogenative PHIP experiments the hyperpolarization is initially on the parahydrogen-derived
hydrogen atoms~\cite{bowersParahydrogenSynthesisAllow1987}, but this limits the scope of signal 
observation, and the hyperpolarization lifetime is usually short. A common 
approach is to transfer the hyperpolarization to other nuclear spins in the molecule to widen 
the range of observable atoms, or as a means to store the hyperpolarization on a more slowly 
relaxing nucleus (such as \chemical{^{13}C} or \chemical{^{15}N}). The polarization transfer 
can be induced by rf pulses in high-field NMR experiments, or via low- to ultralow- magnetic 
field manipulations. See Ref.~\cite{stevanatoChapter11Converting2020} for a more in-depth discussion of this topic.

The most common examples of hydrogenative PHIP in the literature are the reduction of alkynes 
to alkenes and alkenes to alkanes~\cite{duckettApplicationParahydrogenInduced2012}. This is typically done using a transition metal 
catalyst that allows for \emph{pairwise} addition of parahydrogen to the product molecule, such 
that the parahydrogen protons are coupled throughout the process and remain in a correlated 
state. It should be noted that other hydrogenation reactions with parahydrogen have been
demonstrated; examples include hydroformylation reactions~\cite{perminOneHydrogenPolarizationHydroformylation2002,
godardNewPerspectivesHydroformylation2004}, and hydrogenation reactions in which the protons end up in geminal
positions on the product~\cite{harthunProofReversiblePairwise1996,dagysGeminalParahydrogeninducedPolarization2020}. 
When the product molecule is formed, if the parahydrogen protons are either chemically or magnetically 
inequivalent, the proton singlet state is broken and hyperpolarized signals can be observed~\cite{bowersParahydrogenSynthesisAllow1987}. Commonly the inequivalence is caused by 
a chemical shift difference between the two protons that is greater than their mutual $J$-coupling. However, 
if the protons are close to chemical or magnetic equivalence, the non-magnetic singlet state remains close to an eigenstate, 
and either electromagnetic pulses~\cite{eillsSingletOrderConversion2017} or further chemical reactions~\cite{zhangLongLived1HNuclear2014,eillsSingletContrastMagneticResonance2021} 
are needed to release the singlet order to generate observable hyperpolarized signals.

Once the hyperpolarized product molecule is formed, the protons typically relax on
a timescale of a few seconds to tens of seconds~\cite{emondtsPolarizationTransferEfficiency2017}. For this reason,
hydrogenative PHIP reactions are reliant on highly efficient catalysts in order for the
reaction to proceed on a similar timescale~\cite{itodaStructuralExplorationRhodium2019,
zhivonitkoCatalyticallyEnhancedNMR2014, hublerInvestigatingKineticsHomogeneous1999}.
These catalysts can be homogeneous or heterogeneous.

Homogeneous hydrogenative PHIP reactions are performed in a solvent, and use transition metal complexes as
catalysts (most commonly Rh-, Ru- or Pt-based) to which the hydrogen molecule and precursor
can bind~\cite{buljubasichParahydrogenInducedPolarization2013,tokmic13CNMRSignal2018}. This method offers
 high chemical reaction rates and good selectivity towards the desired product molecule.
The disadvantage is the difficulty in separating out the catalyst from the solution after the
reaction. For this reason, immobilized catalysts supported on solid surfaces are under
 investigation~\cite{skovpinParahydrogenInducedPolarizationHeterogeneous2011, kovtunovCatalysisNuclearMagnetic2016}.
Note that solution-state PHIP experiments require the dissolution of hydrogen gas~\cite{berthaultUseDissolvedHyperpolarized2020}, which can be the rate-limiting step 
and is one of the key areas in which microfluidics can provide
advantages~\cite{rothContinuous1H13C2010,lehmkuhlContinuousHyperpolarizationParahydrogen2018,
 eillsHighResolutionNuclearMagnetic2019,bordonaliParahydrogenBasedNMR2019} (see Section~\ref{Xe-bubble-free}).

Hydrogenative PHIP can also be catalyzed by active sites on solid-phase catalysts, to which
hydrogen molecules and reactant molecules can transiently bind and
react~\cite{kovtunovParahydrogenInducedPolarizationHeterogeneous2013}.
Heterogeneous catalysts can be readily separated
out of solution at the end of the chemical reaction. However, producing high levels of polarized product molecules is challenging, because the
hydrogen atoms can migrate across the catalyst surface and become uncorrelated, causing the
spin order to be lost before the reaction is complete.
Examples include rhodium/platinum supported on
titania which has been used to produce hyperpolarized gases (i.e., ethane, propane)~\cite{sharmaStronglyHyperpolarizedGas2012, kovtunovParahydrogeninducedPolarizationAlkyne2010},
and ligand-capped nanoparticles for producing hyperpolarized liquids~\cite{glogglerNanoparticleCatalystHeterogeneous2015}.

The range of target molecules which can be polarized by hydrogenative PHIP is severely
limited, since they must be
products of hydrogenation reactions. 
This restriction was somewhat alleviated by the
development of the side-arm hydrogenation procedure~\cite{reineriParaHydrogenInducedPolarization2015}. 
A target molecule is functionalized with an unsaturated side-arm moiety which can be hydrogenated 
with \pHtwo, followed by polarization transfer to the target part of the molecule. In a final 
step, the sidearm is chemically cleaved by hydrolysis to liberate the hyperpolarized target. 
Side-arm hydrogenation has greatly expanded the catalogue of molecules that can be hyperpolarized 
using parahydrogen.

\subsubsection{Non-hydrogenative PHIP}

Non-hydrogenative PHIP is commonly referred to as Signal Amplification 
By Reversible Exchange (SABRE)~\cite{adamsReversibleInteractionsParaHydrogen2009,barskiySABREChemicalKinetics2019}. This encompasses a range of chemical reactions in 
which parahydrogen and a target molecule reversibly bind to an iridium catalyst, with spin order
transfer between the two molecules able to occur during the transient period in
which they are bound to the catalyst and experience mutual spin-spin couplings
\cite{kovtunovHyperpolarizedNMRSpectroscopy2018, adamsReversibleInteractionsParaHydrogen2009}.
Unlike hydrogenative PHIP, SABRE does not require that the polarizable targets
be products of hydrogenation reactions; instead, the technique is limited to
polarizing molecules which can transiently bind to the iridium metal center.
To date this has mostly been demonstrated for molecules containing a nitrogen
atom which can directly bond to the iridium center through its lone pair of electrons. 
Polarization levels of a few percent are typically achieved for protons, and up to a few tens of 
percent for nitrogen-15, which has longer spin relaxation times, allowing for larger steady-state 
polarization levels to be established~\cite{shchepinHyperpolarizingConcentratedMetronidazole2019}. A practical 
advantage of non-hydrogenative PHIP over hydrogenative PHIP is that the reactions are reversible, so the material is not 
depleted during the experiment. A sample can therefore be recycled for many experiments. The recent 
development of SABRE-Relay, which uses a secondary exchange mechanism to polarize the molecule of interest, has somewhat broadened the range of molecules which can be polarized 
via SABRE~\cite{richardsonQuantificationHyperpolarisationEfficiency2018,roySABRERelayVersatileRoute2018,ialiUsingParahydrogenHyperpolarize2018}.

\subsection{Microfluidic Implementation of PHIP}

In solution-state PHIP/SABRE, the first step is to dissolve parahydrogen into the reaction mixture. This
is typically done by bubbling the gas through the solution or vigorous shaking~\cite{cavallariMetabolicStudiesTumor2019} of the sample
to maximize the liquid/gas surface area-to-volume ratio, and hence facilitate efficient gas exchange, but 
these methods present the same drawbacks discussed in Section~\ref{Xe-bubble-free}. As was 
the case with \Xe, improved strategies for dissolving parahydrogen into solutions without 
introducing gas bubbles are necessary.

One early example of bubble-free hydrogenative PHIP was the use of polypropylene hollow-fibre membranes (Celgard X50) as
described by Roth et al.~\cite{rothContinuous1H13C2010}. The hollow-fibre membranes (OD = 300 \mum, ID = 200 \mum) had a pore size of 0.03 \mum~and were embedded in an NMR tube
and pressurized with \chemical{pH_2} gas. The large gas-liquid surface area provided by the membranes
led to efficient gas exchange, and they showed hydrogenation of 2-hydroxyethyl acrylate to 2-hydroxyethyl propionate.
Roth et al.~\cite{rothContinuous1H13C2010} hyperpolarized the \chemical{^1H} and \chemical{^{13}C} nuclei, reporting a 2000-fold enhancement for protons and 5500-fold enhancement for
carbon, at 7~T. The authors demonstrated the stability of this experimental setup by acquiring a \chemical{^1H-^1H} COSY
2D NMR spectrum, which required stable hyperpolarized signals for 7~minutes.

\begin{figure}
	\begin{center}
			\includegraphics[width=7.4cm]{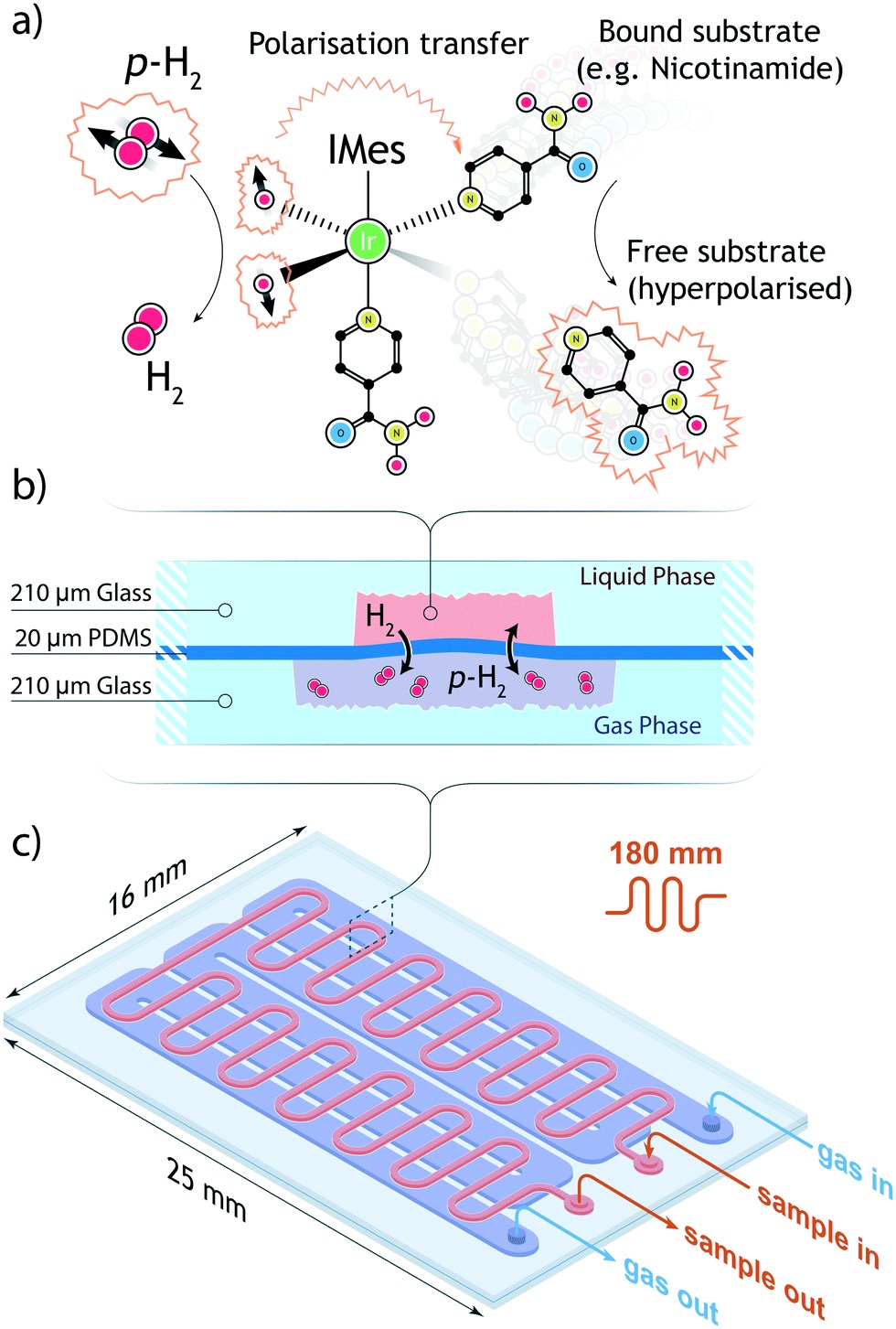}
	\end{center}
	\caption{SABRE implemented on a microfluidic chip. (a) The SABRE process. (b) Schematic of the gas–liquid contact channel. (c) The two fluid channels meander for enhanced contact area. On the liquid side, the total channel length is 180~mm and the enclosed volume is 4.8 \muL. On the gas side, the channel length is 120~mm and the enclosed volume is 20.2~\muL. Reproduced from Ref.~\cite{bordonaliParahydrogenBasedNMR2019} with permission.
	}
	\label{fig:Bordonali-fig}
\end{figure}

Lehmkuhl et al. used a commercial `MicroReactor' to perform SABRE experiments in the Earth's field to
hyperpolarize pyridine and nicotinamide~\cite{lehmkuhlContinuousHyperpolarizationParahydrogen2018}.
The reactor had serpentine liquid and gas channels atop one another, separated by a gas-permeable
poly(ethylene oxide)-poly(butylene terephthalate) copolymer membrane to facilitate efficient gas exchange.
The solutions were flowed into a 1~T benchtop NMR
spectrometer for signal detection at flow rates between 1 and 10~ml/min. A  maximum increase of 1330-fold in \chemical{^1H} signal was 
obtained from pyridine at 3 ml/min, compared to a thermal equilibrium signal acquired under the same conditions. The combination of membrane hydrogenation and flow chemistry paved the way for future microfluidic efforts.

Bordonali et al. presented the first PHIP-capable microfluidic device incorporated into 
an NMR probe head, featuring membrane assisted \emph{in situ} dissolution of \pHtwo\ into solution~\cite{bordonaliParahydrogenBasedNMR2019}. An alveolus was formed by two layers of glass (L~=~25~mm, W~=~16~mm) 210 \mum-thick, 
with gas and solution micro-channels etched onto each surface as shown in Fig.~\ref{fig:Bordonali-fig}. The layers of glass were separated by a 20 \mum-PDMS membrane to allow perfusion of \pHtwo\ from the 
gas channel to the solution channel. A syringe pump outside the magnet delivered the solution at flow rates between 20 and 300~\muL/min and \pHtwo\ gas was supplied at 0.2~MPa (2~bar). The detection in the probe head was performed in a 0.56~\muL~chamber using a micro-Helmholtz pair 
(R = 0.6~mm) in an 11.7~T magnetic field. To test the efficacy of their setup they performed SABRE reactions at high field. As SABRE is typically performed at low magnetic fields to generate polarization, the maximum substrate enhancements observed for pyridine, nicotinamide, and metronidazole under continuous flow were 4, 1.2 and 4.6 respectively. However, the dihydride signal produced by the \pHtwo~bound to the iridium complex has a distinct shift for each substrate that it is co-bound with. Using this dihydride signal as a chemosensor, the presence of the three substrates at a minimum concentration of 50~\muM~was detected in 128 scans, corresponding to a concentration limit of
detection of approximately $\text{cLOD}=0.5~\mathrm{mM\sqrt{s}}$.

This highlights one challenge of microfluidic NMR: most commercial high-field NMR magnets are designed
with a small homogeneous region to accommodate standard 5 or 10~mm NMR tubes, but many microfluidic
devices do not fit entirely within this volume. For high-resolution NMR the detection chamber should be
held in the homogeneous region, but if other parts of the microfluidic device extend outside this region
it can be difficult or impossible to apply rf pulses to manipulate the nuclear spins in these parts of the device.
In the case of SABRE, it would be helpful to apply rf irradiation to the spins in the channels prior
to detection, so that coherent polarization transfer could be implemented with techniques such as radiofrequency SABRE (RF-SABRE)~\cite{pravdivtsevRFSABREWayContinuous2015} or spin-lock induced crossing SABRE (SLIC-SABRE)~\cite{knechtEfficientConversionAntiphase2019}.

\begin{figure}
	\begin{center}
			\includegraphics[width=8.4cm]{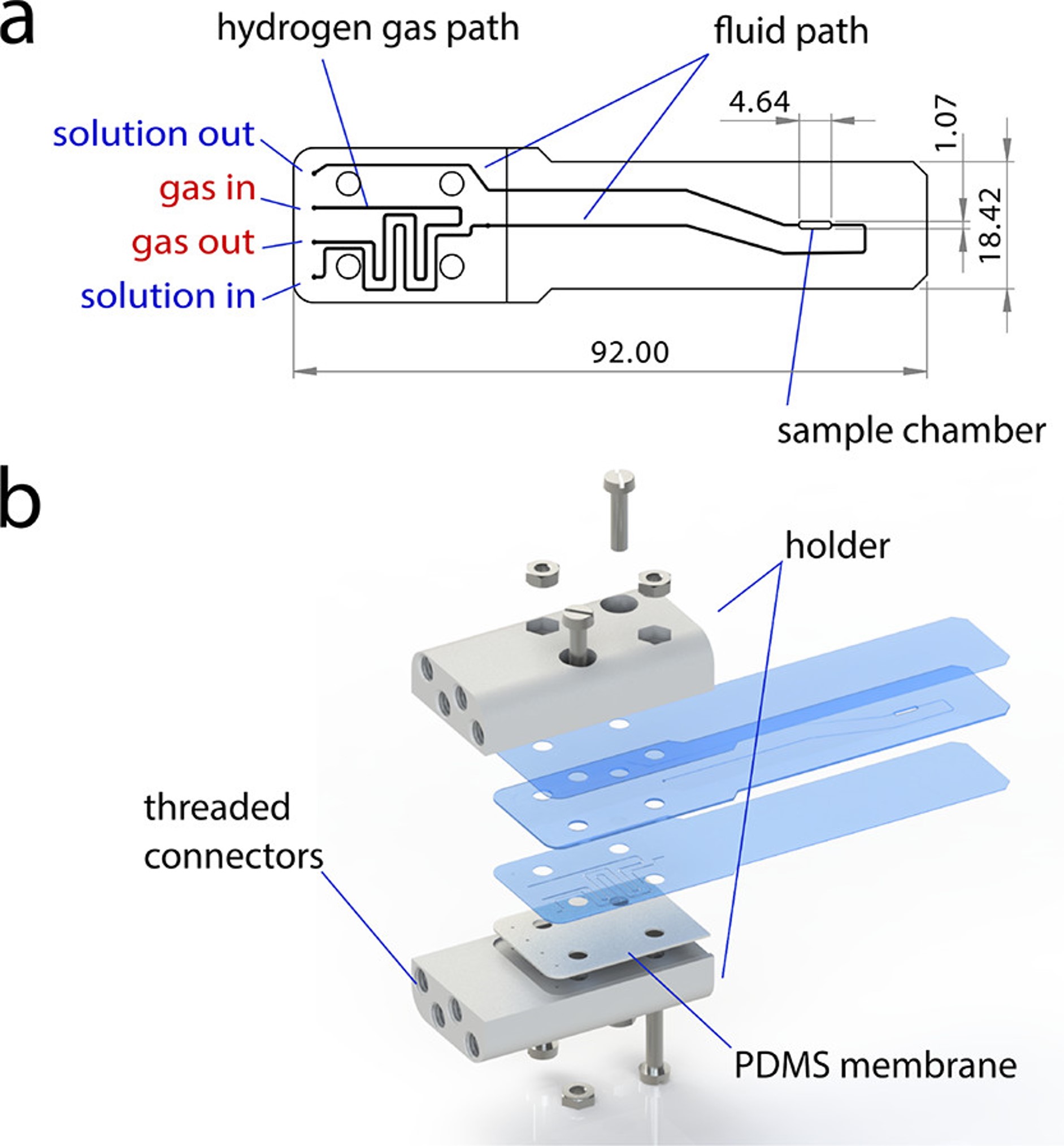}
	\end{center}
	\caption{(a) Lay-out and (b) design of the microfluidic device used for PHIP-on-a-chip experiments in Ref.~\cite{eillsHighResolutionNuclearMagnetic2019}. The threaded connectors at the head of the chip holders allowed capillaries to be connected to provide gas and solution flow in and out of the chip. All measurements are shown in millimeters, and the volume of the detection chamber was 2.5~\muL. Image taken with permission from Ref.~\cite{eillsHighResolutionNuclearMagnetic2019}.}
	\label{fig:Micro-on-Mem-1}
\end{figure}

Eills, Hale and co-workers made the first demonstration of hydrogenative PHIP on a microfluidic device, an experiment known as PHIP-on-a-chip~\cite{eillsHighResolutionNuclearMagnetic2019}.
The microfluidic device (L~=~92~mm, W~=~22~mm) consisted of layers of poly(methyl methacrylate) (PMMA) with serpentine gas- and
solution-flow channels laser-cut into the material that run side-by-side on the surface, and a PDMS
membrane over the channels to act as a fluid seal and a gas conduit. This design is advantageous
since high \chemical{pH_2} pressures can be employed; designs in which the gas and solution channels
cross on opposite sides of the PDMS membrane are limited in the \chemical{pH_2} pressures that can
be used since the PDMS deforms and blocks the channels~\cite{bordonaliParahydrogenBasedNMR2019}.
The layers were sandwiched between two 3D-printed holders that contained ports for capillary tubing to deliver
liquid and gas to the chip. The \pHtwo~pressure was fixed at 0.5~MPa (5~bar) and flow rates between 1 and 20~\muL/min
were used. The microfluidic device is shown in Fig.~\ref{fig:Micro-on-Mem-1}. In this work
the authors observed the hydrogenation of propargyl acetate to allyl acetate in
PASADENA experiments~\cite{bowersParahydrogenSynthesisAllow1987},
in an 11.7~T high-field NMR magnet. The resulting \chemical{^1H} signal enhancements were on the order of 10$^3$
which allowed for the detection of molecules at micromolar concentrations, with an $\text{nLOD} = 2.2 \pm 0.4$~pmol~$\sqrt{s}$. The stability of the hyperpolarized NMR
signals in this continuous-flow setup was sufficient to allow for the acquisition of 2D
NMR spectra at natural \chemical{^{13}C} abundance without introducing $T_1$ noise.

An important advantage afforded by the microfluidic implementation of PHIP is the minimization of the time between the generation of hyperpolarization, i.e., when \pHtwo\ is first
dissolved into the solution to initiate the chemical reaction, and when the hyperpolarized NMR signals are detected. This minimization is crucial in limiting losses due to relaxation, especially in the case of proton
experiments. Another key advantage is the enhanced reproducibility of the experiments when \pHtwo is dissolved into solution via a permeable membrane as opposed to bubbling of the gas through a solution. The better experimental control has allowed, for example, the detailed kinetic study of hydrogenation
reactions~\cite{ostrowskaSpatiallyResolvedKinetic}. An important consideration for microfluidic PHIP is the choice of chip material. While some PHIP experiments can be performed in aqueous solution, many use organic solvents such as methanol, acetone and chloroform, which can dissolve or warp some plastics. However, microfluidic devices can be readily fabricated from solvent-resistant materials
such as poly(imides) or glass.

Multi-step procedures for producing chemically pure solutions of biomolecules via PHIP at concentrations and polarizations competitive with D-DNP have been recently developed~\cite{knechtRapidHyperpolarizationPurification2021}. A good example is the side-arm
hydrogenation procedure~\cite{reineriParaHydrogenInducedPolarization2015} which has become an important cornerstone of PHIP, but this requires: (0) synthesis of the precursor molecule, (1) hydrogenation of a precursor molecule, (2) polarization transfer from the protons to a heteronucleus, and (3) cleavage of the side-arm and purification by efficient mixing and separation of the organic PHIP solution with an aqueous solution. We expect lab-on-a-chip platforms to soon be used as a convenient environment for these multi-step PHIP experiments, since they are ideally suited for the combination of a number of experimental steps close to the point of sample detection.

\subsection{Gas-Phase Microreactors}
Heterogeneously-catalyzed gas-phase PHIP has been implemented in microreactors with volumes 
of tens of microliters, to study the spatiotemporal 
progression of hydrogenation reactions. These microreactors 
provide a convenient means to monitor reactions at a small scale, which is advantageous over 
larger scale (tens of milliliters and above) reactors for the following reasons:
\begin{enumerate}
    \item Less material is required,
    \item The experimentalist has better control over the reactor temperature,
    \item It is both easier and safer to pressurize small-scale reactors,
    \item The material has a shorter residence time in the reactor, which is especially 
relevant for the case of hyperpolarized molecules.
\end{enumerate}
By studying hydrogenation reactions in microreactors it is possible to extract
information about how the reactivity correlates with the packing structure and density of the catalyst particles.
Hydrogenation reactions are widely used in the petrochemical industry as in the Fischer-Tropsch process, and in the fine chemicals industry to selectively reduce unsaturated molecules. Improving the selectivity of these reactions and increasing yield is highly desirable.
PHIP is particularly well-suited for these mechanistic studies, since the \chemical{H_2} addition should be carried out pairwise for hyperpolarized signals
to be generated on the product molecules. This can therefore act as a
marker for single-atom hydrogenation sites, which are a goal in the field of heterogeneous catalysis
as they can provide improved reaction selectivity and efficiency~\cite{buruevaSingleSiteHeterogeneousCatalysts2019,liIntroductionHeterogeneousSingleAtom2020}.

This principle was demonstrated by Bouchard et al. in their study of different types of hydrogenation reactors~\cite{bouchardNMRImagingCatalytic2008}.
Reactor 1 contained a silica gel support with Wilkinson's catalyst (\chemical{RhCl(PPh_3)_3}) attached to the
surface, and reactor 2 contained a loosely packed Wilkinson's catalyst powder. They studied the hydrogenation
of propylene to propane in these microreactors, and a \chemical{^1H} signal enhancement of 300 enabled the
authors to resolve velocity flow maps and map the spatial distribution of the active catalyst centers
in the catalyst beds.

Despite the mechanistic information that can be obtained by studying catalytic microreactors with PHIP,
the heterogeneous nature of the reaction media leads to spectral line broadening which limits spectral
resolution. Another limitation of microreactors in the relatively
large coils used for imaging, which leads to low mass sensitivity. A solution to both of these problems
is to use remote detection - that is to perform the reaction in one location, spatially encode the NMR signals,
and detect in a separate location.

\section{Remote Detection}
    \subsection{Introduction}
The microfluidic detection strategies discussed so far require the use of an inductive detector 
positioned near or around the sample. In many systems 
such as porous materials or living organisms this is not possible, or geometric 
constraints of the experiment could make inductive detection inefficient. In other systems 
containing many microfluidic features to probe, it would be inconvenient to implement and 
separately tune an array of LC circuits for signal acquisition. To circumvent this, remote 
detection strategies have been introduced. The magnetization of the nuclear spins in a sample can be 'encoded' by using rf pulses and magnetic field gradients to essentially map out the distribution of nuclei in space. The information-rich NMR signal is stored as longitudinal magnetization, and
carried to the detector by the flowing liquid. This principle is known as remote detection.
In this way, 
the detector can be made much smaller than the microfluidic device, offering higher sensitivity,
while still allowing the entire device to be mapped. One downside of this technique 
is that the spins are continually relaxing during physical transport to the detector, so the 
distance should be minimized. Remote detection has been used on thermally-polarized spin systems, 
and in macroscopic samples, but in this review we will focus specifically on the cases 
involving remote detection of hyperpolarized species to probe microfluidic devices.

\subsection{Remote Detection and Micro-Imaging with Parahydrogen}

In 2010 Telkki et al. showed that hyperpolarized gases can be used to image microfluidic structures~\cite{telkkiMicrofluidicGasFlowImaging2010}.
They produced hyperpolarized propane gas outside the NMR magnet by flowing a propene/\chemical{pH_2}
gas mixture over rhodium catalysts supported on solid silica in a tubular oven. The gas was flowed into
the NMR magnet through capillaries into a microfluidic chip with an internal volume of between 1 and 2~\muL, and imaging coils were used for spatial
encoding. The gas was then detected by a solenoid microcoil around the outlet capillary encasing a 53~nL
sample, which afforded a much higher mass sensitivity than could be obtained with the large coils surrounding
the microfluidic chip. The imaging coils were used to spatially encode the signals in two dimensions, and time-of-flight (TOF) information was acquired by using a train of detection pulses. By combining these schemes, they were able to generate flow profile and
velocity maps, which exposed manufacturing defects in some of the microfluidic devices. By combining the gains
in sensitivity from remote detection and PHIP, they could report $10^4$ to $10^5$ enhancements in signal compared to
thermal equilibrium experiments.

A further advance came from the same group in 2012, when they showed that the reactor could also be scaled down
to a microfluidic scale and brought into the encoding coils inside the NMR magnet, such that the microreactor
itself could be imaged~\cite{zhivonitkoCharacterizationMicrofluidicGas2012}. Three reactors with 0.25 to 2.5~\muL~internal volumes were studied,
which were simply capillary tubes of differing geometry. This allowed extraction of detailed mass transport information
about the gas flow, as well as the spatial dependence of the reaction progression. They were also able to
quantitatively observe gas adsorption in the porous catalyst material by measuring flow profiles to extract the gas velocity through the catalyst beds, with lower velocity indicating higher adsorption.

\begin{figure}
	\begin{center}
			\includegraphics[width=8.4cm]{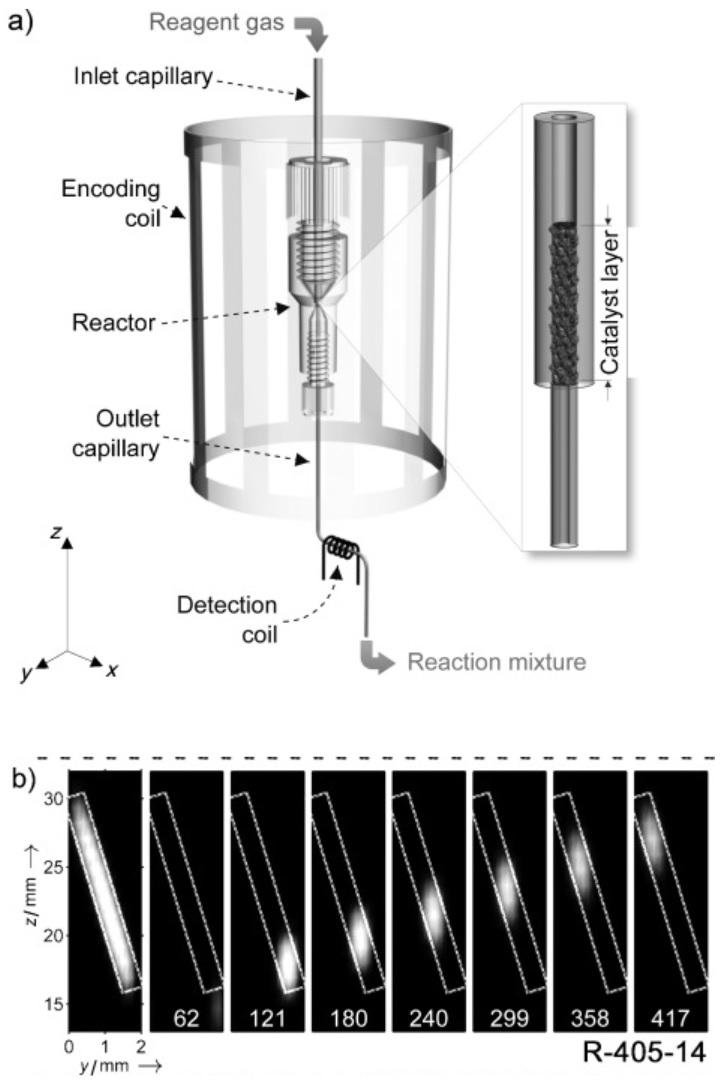}
	\end{center}
	\caption{a) An experimental setup for imaging a microreactor using remote detection. A parahydrogen/propene gas mixture flows through a catalyst bed, and reacts to form hyperpolarized propane. This process is imaged by the encoding coils, and the signal is detected remotely by the solenoid microcoil as the gas flows out of the reactor. b) Time-of-flight images of hyperpolarized propane gas in a capillary reactor with diameter 405~nm and length 14~mm. Image modified with permission from Ref.~\cite{zhivonitkoCharacterizationMicrofluidicGas2012}.}
	\label{fig:gas-phase-1}
\end{figure}

The same group showed in 2013 that a more complex microfluidic chip device could replace the simple capillary microreactors~\cite{zhivonitkoRemoteDetectionNMR2013}.
In these experiments the catalyst was a 100~nm layer of Pt sputtered onto the inner surface of the
microfluidic device. The authors were able to collect TOF images to measure mass transport of the reaction mixture (molecules of propene and propane) in the microfluidic chips, as well as image where the reaction was taking place. However, these experiments were all done with normal \chemical{H_2} gas. When parahydrogen was used, no PHIP-enhanced NMR signals were observed, which is likely due to
the hydrogen atoms losing their correlated spin state as they rapidly diffuse across the metal surface while
adsorbed. This demonstrates the additional mechanistic information that can be obtained by working
with parahydrogen.
    
	\subsection{Remote Detection with Hyperpolarized \chemical{^{129}Xe}}

Hyperpolarized \Xe\ is a particularly good candidate for remote detection experiments 
because in the gas phase the $T_1$ relaxation times are usually on the order of minutes, 
and can exceed 10~minutes in suitably prepared systems~\cite{zengWallRelaxationSpin1983}.

In 2005 McDonnell et al. demonstrated the use of remote detection to measure the flow rate of hyperpolarized \Xe\ gas in a microfluidic chip~\cite{mcdonnellNMRAnalysisMicrofluidic2005}. The experiment was performed in one 7~T magnet, using a probe with a large coil of 30mm diameter surrounding the microfluidic chip for signal encoding with rf pulses, and a solenoid microcoil of 219~nL volume to detect the signal after outflow from the chip. This resulted in a gain in sensitivity by a factor of 150 as compared to direct detection using the encoding coils, due to the smaller detection coil (see Section~\ref{SmallScale}).
The chemical shift experienced by \Xe\ inside the microfluidic chip was
encoded using a pair of $\pi/2$ pulses in the encoding coil, separated
by an evolution interval $t_1$. The chemical shift experienced by \Xe\ inside the microfluidic chip was encoded using a pair of $\pi/2$ pulses in the encoding coil, separated by a delay $t_1$ which was incremented between transients. Plotting the signal (magnetization) as a function of $t_1$ produces a free-induction decay which can be Fourier transformed to yield an NMR spectrum. This method allows for the detection of chemical/physical processes occurring in the encoding region while taking advantage of the higher sensitivity of the miniature detection coil. 
The spectral resolution was limited to tens of hertz due to the short sample residence time in the detection coils. While this could be in principle
be improved by reducing the flow rate, relaxation losses place a lower limit 
on the flow rate. 

Hilty et al. expanded on this work by acquiring images of the hyperpolarized \Xe\ flow profile using gradient pulses in multiple axes in the encoding area, and studied how the flow profile varied for different chip structures~\cite{hiltyMicrofluidicGasflowProfiling2005}.
In 2006 Han et al. reported a micro-solenoid rf probe for signal detection, and used this to map the flow profile of hyperpolarized \Xe\ in a macro-scale glass phantom. This work focused on the ability to perform the encoding and signal detection at different magnetic field strengths~\cite{hanAuxiliaryProbeDesign2006}.

\begin{figure}
	\begin{center}
			\includegraphics[width=8.4cm]{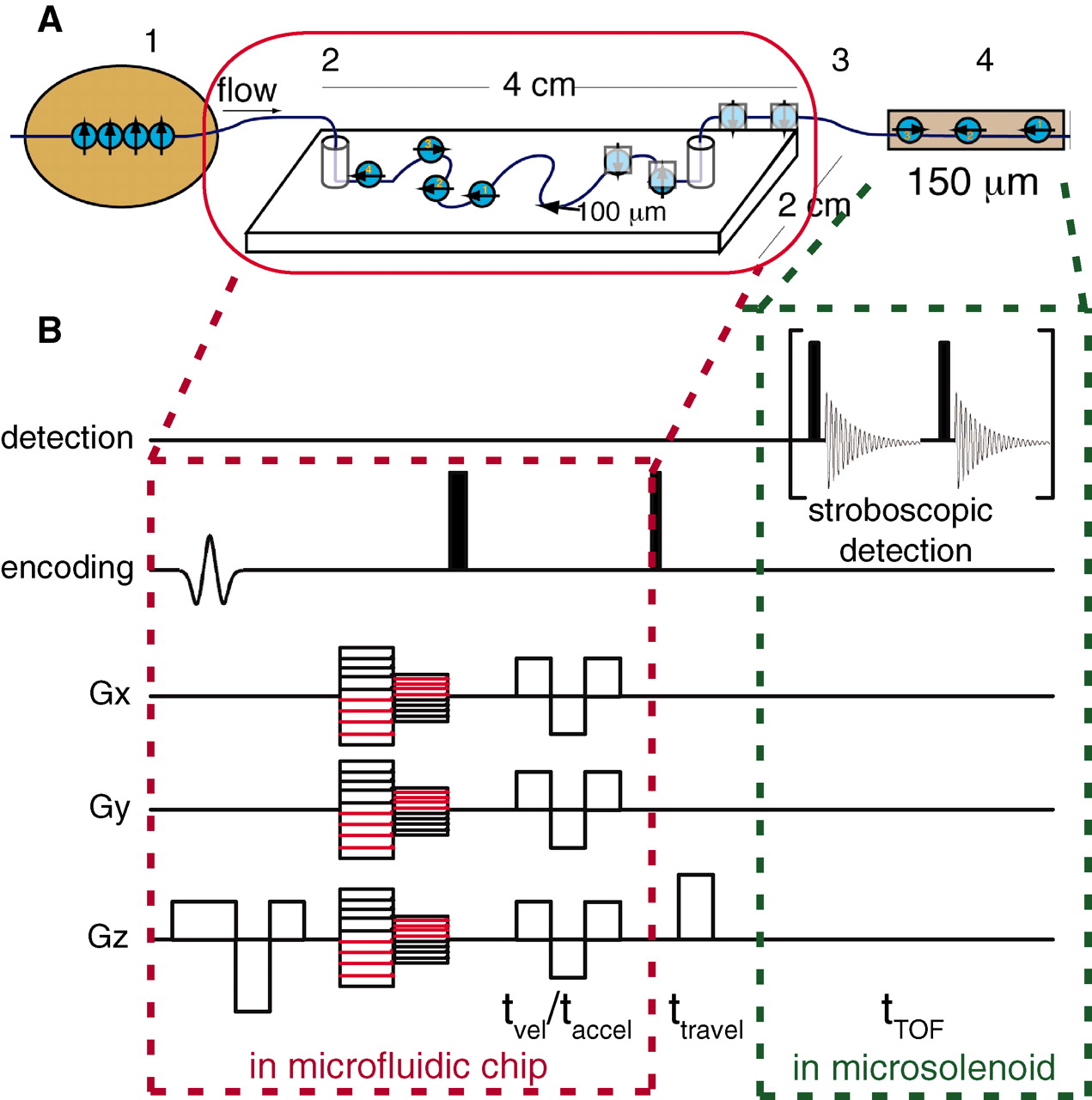}
	\end{center}
	\caption{An illustration of a remote detection MRI experiment adapted with permission from Ref.~\cite{bajajZoomingMicroscopicFlow2010}. (A) Hyperpolarized \Xe\ gas flowing through a microfluidic chip for signal encoding, and into a micro-solenoid coil for signal acquisition. (B) An example pulse sequence for a 3D remote detection experiment.
	}
	\label{fig:bajaj2010}
\end{figure}

In 2010, Bajaj et al. employed imaging gradients and a Fourier phase-encoding pulse sequence to spatially map a hyperpolarized \Xe\ sample in a microfluidic chip during continuous flow~\cite{bajajZoomingMicroscopicFlow2010}. Their experiment is illustrated in Fig.~\ref{fig:bajaj2010}. The signal was acquired from a $\sim$25~pL sample in a microsolenoid coil, and from this they were able to reconstruct images showing the gas flow profiles from the microfluidic chip. Since the sample was sparsely distributed within a large volume imaging probe during encoding, they implemented compressed sensing (nonuniform sampling) to maximize the information encoded in a given time.

\section{Photo-Chemically Induced Dynamic Nuclear Polarization}
\label{photoCIDNP-section}
\subsection{Introduction}
Chemically induced dynamic nuclear polarization (CIDNP) is a
hyperpolarization method that relies on a radical pair mechanism for spin sorting in chemical reactions
involving radicals~\cite{kapteinChemicallyInducedDynamic1969,clossMechanismExplainingNuclear1969}. Spin sorting refers to molecules in different nuclear spin states behaving differently in a chemical process: this might be that they react to form different product molecules, or that they experience different rates of spin relaxation. The name CIDNP should not be misread to suggest similarity of the process with DNP, which commonly refers to methods that transfer polarization from electron to nuclear spins. No such transfer takes place in CIDNP. Rather, "dynamic nuclear polarization" is used here in its original, broader meaning - nuclear polarization arising transiently from a non-equilibrium process.

\begin{figure}
	\begin{center}
			\includegraphics[width=8.4cm]{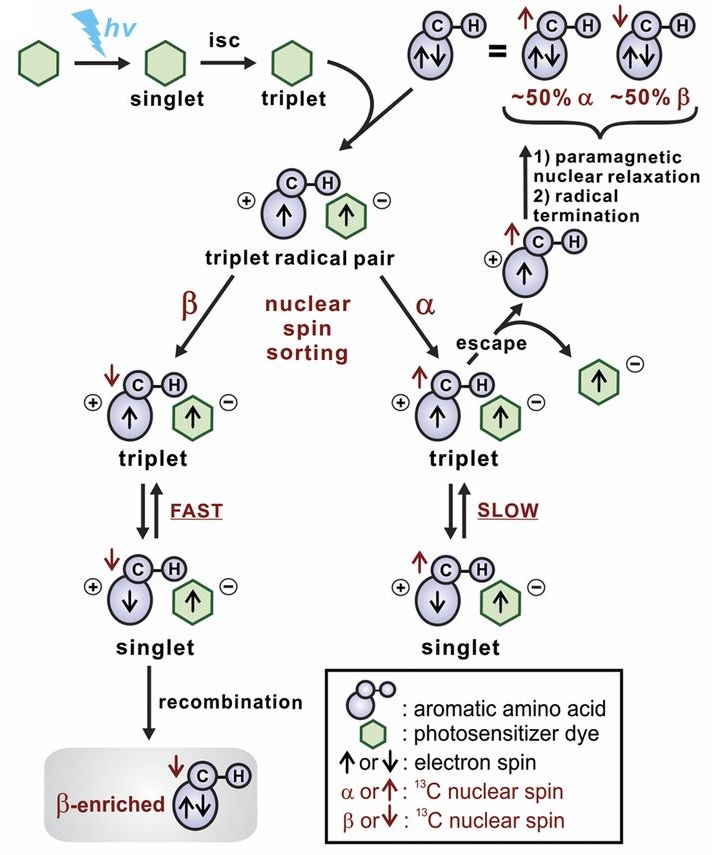}
	\end{center}
	\caption{An illustration of the photo-CIDNP effect used to hyperpolarize an aromatic amino acid, adapted with permission from Ref.~\cite{okunoLaserCryogenicProbeassisted2019}.
	}
	\label{fig:okuno2019}
\end{figure}

There is a particular variant of the experiment known as photo-CIDNP in which 
the radicals are generated by irradiating the sample with light, as shown in  Fig.~\ref{fig:okuno2019}. 
The first step is to electronically excite a photosensitizer molecule (a `dye')
with light to an excited electronic singlet state, which 
is converted to an excited triplet state through intersystem crossing. At this point 
another molecule (the `target') can donate an electron to the photosensitizer so they form a 
radical pair. In order for the target molecule to be released from the
radical pair in its original chemical form, the electron pair needs to recombine.
However, an electronic singlet state is required for recombination.
The likelihood of recombination is
influenced by the nuclear spin states of the target and dye molecules through their hyperfine
coupling to the electrons, and in Fig.~\ref{fig:okuno2019} the recombination process 
is shown as fast if a coupled nuclear spin on the target molecule is in the $\beta$ state, and slow if it 
is in the $\alpha$ state. Recombination of the electron pair and 
release of the target molecule is favored for certain nuclear spin states. If the 
recombination does not occur, the target molecule will remain a paramagnetic radical,
and the unpaired electron can cause rapid nuclear spin relaxation, or the
molecule can react further to form a different chemical species. The different
fates of the molecules depending on their nuclear spin states means that simply
by irradiating a suitable sample with light, nuclear spin polarization can be 
generated on the target (and dye) molecules.

\subsection{Microfluidic Enhancement of Photo-CIDNP}

\begin{figure}
	\begin{center}
			\includegraphics[width=8.4cm]{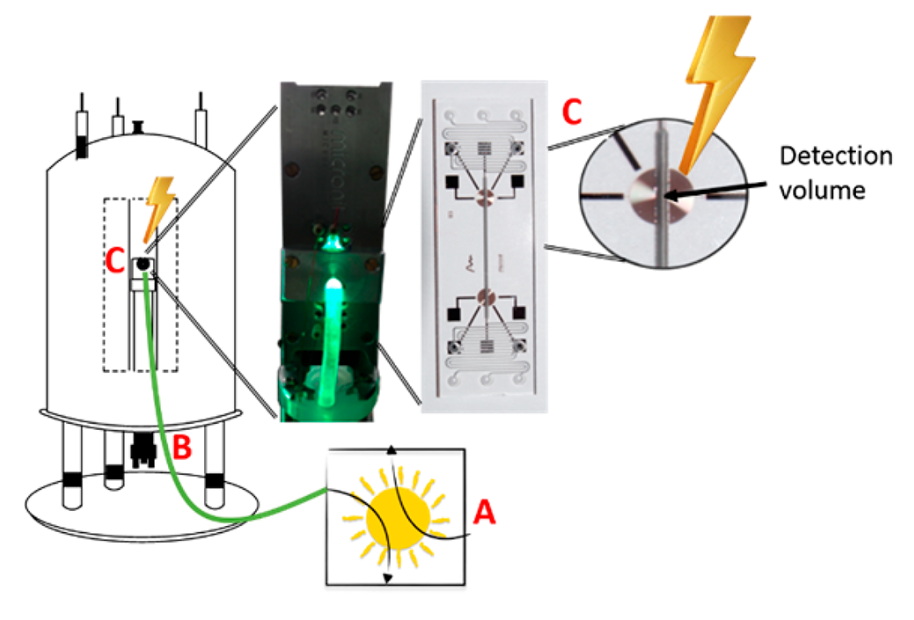}
	\end{center}
	\caption{A microfluidic setup used to study light-induced chemical reactions with NMR. An external light source (A) is guided by an optical fibre (B) into the microfluidic chip (C) to induce a chemical reaction that is detected by the microcoil. Image taken with permission from Ref.~\cite{gomezIlluminationNanoliterNMRSpectroscopy2018}}
	\label{fig:cidnp-1}
\end{figure}

The rate of photochemical reactions depends on photon absorption;
a high photon flux is desirable, but this is challenging for a large sample
because it is obviously opaque at the relevant photosensitizer
absorption wavelengths. Such reactions can be followed with NMR~\cite{hallOnlineMonitoringPhotocatalytic2017}, but when
macroscopic sample volumes are used (i.e., a sample in a 5~mm NMR tube), they
suffer from several problems:
\begin{enumerate}
\item The light intensity falls off exponentially with the distance from the source (usually an optical fibre) within the sample,
\item The presence of an optical fibre in the rf coils can induce magnetic-field distortions,
\item Over time the light can cause photodegradation of the photosensitizer molecules,
\item The light can induce undesirable local heating effects.
\end{enumerate}
Implementation of the photochemical reaction on a microfluidic device can in principle overcome all of these limitations:
\begin{enumerate}
\item Working at small volume scales increases the proportion of the sample that is irradiated, and allows for more homogeneous irradiation,
\item The light source can be positioned outside the detection microcoils without sacrificing photon flux,
\item The ability to provide fresh sample under continuous flow circumvents problems associated with sample degradation,
\item Light-induced heating is homogeneous through the sample and can be better controlled.
\end{enumerate}
An illustration of a microfluidic setup for studying light-induced reactions by NMR is shown in Fig.~\ref{fig:cidnp-1}, and a more thorough review on the topic of microfluidics in photochemistry can be found in Ref.~\cite{cambieApplicationsContinuousFlowPhotochemistry2016}.

The first demonstration of a CIDNP experiment performed on a
microfluidic chip came in 2018~\cite{mompeanPushingNuclearMagnetic2018}.
In one experiment, the authors flowed a solution of 0.8~$\mu$M flavin 
mononucleotide (the dye) and 0.8~$\mu$M 4-fluorophenol (the target) at 5~$\mu$l/min through a 
microfluidic chip, and by irradiating the sample with light they were 
able to generate and detect photo-CIDNP signals from the 4-fluorophenol 
in a 1~$\mu$l sample chamber. The signal enhancement was on the order of 
$10^3$ which allowed for a striking nLOD for an NMR
experiment of just under 1~pmol~Hz$^{-1/2}$.
With this experimental setup the authors were
able to explore more advantages of working on a microfluidic platform. In one set
of experiments, they worked with solutions of 1~mM flavin (the dye) and 5~mM N-acetyl-L-tyrosine 
(AcTyr, the target), and showed that the photo-CIDNP signals for a static sample in
the chip decay approximately linearly with time due to photodegradation of the
photosensitizer, the signal being reduced to half the maximum intensity after
$\sim$10~minutes. By repeating the experiment with the sample continually
flowing at a rate of 5~$\mu$l/min, the signal intensity was stationary at
the maximum value, as fresh reaction solution was continually supplied. 

The authors also performed experiments in which they flowed two solutions into the 
microfluidic chip and allowed them to mix prior to irradiation and detection. In one 
experiment they mixed an AcTyr/flavin solution with a second solution containing $\beta$-cyclodextrin. 
In the combined solution the $\beta$-cyclodextrin was in excess of the AcTyr, and 
they observed the photo-CIDNP-enhanced NMR peak shift, broaden, and become less intense. This indicated 
that the $\beta$-cyclodextrin formed a complex with AcTyr, reducing the ability of the 
AcTyr molecules to freely interact with the flavin, so photo-CIDNP was hindered. 
This was a clear demonstration of an advantage of microfluidic implementation: the ability to mix 
solutions immediately prior to detection enables probing chemistry on short timescales.

%\begin{figure}
%	\begin{center}
%			\includegraphics[width=9cm]{Fig1mompeanPushingNuclearMagnetic2018.JPG}
%	\end{center}
%	\caption{A visual representation of a microfluidic photo-CIDNP experiment, with excitation of the photosensitizer molecule with light to induce nuclear spin hyperpolarization, and detection of the resulting signals with a solenoid microcoil. Image adapted with permission from Ref.~\cite{mompeanPushingNuclearMagnetic2018}.\todo{Labels 1-4 not addressed in caption}
%	}
%	\label{fig:mopean2018}
%\end{figure}

Rapid mixing of solutions (at a volume scale of a few hundreds of
microliters) and subsequent
observation of the photo-CIDNP signals can be used to probe protein dynamics
on a timescale of tens to hundreds of milliseconds~\cite{horeStoppedFlowPhotoCIDNPObservation1997,mokRapidSampleMixingTechnique2003}.
This is an area in which microfluidics can provide better experimental
control; reactions can be probed by mixing solutions on a chip prior
to detection, and the reaction timescale varied by controlling the flow rates.

Unfortunately, photo-CIDNP is not a truly general hyperpolarization method, in comparison 
to e.g., D-DNP or even PHIP, since the mechanism relies on specific molecular targets that can 
participate in the radical-pair mechanism. This has limited photo-CIDNP to molecules with 
low ionization energies such as aromatics. Note that this includes all aromatic amino acids, as well 
as methionine and some amino acid derivatives~\cite{horePhotoCIDNPBiopolymers1993}. As a result, photo-CIDNP has mostly been used to study protein structure 
and protein-ligand interactions~\cite{kuhnPhotoCIDNPNMRSpectroscopy2013}, but not as a method  to produce hyperpolarized solutions of arbitrary molecules.

%Reactions have been studied with time resolution in the order of hundreds of milliseconds using photo-CIDNP in the work of wirmer et al.\cite{wirmerMillisecondTimeResolved2001}, but this was done for a light-initiated chemical reaction which doesn't require mixing of two fluids. Additionally, this was done using a static sample of 160~$\mu$l, and so is outside the scope of this review.

\section{Nitrogen-Vacancy Centers in Diamond}
\label{NVsection}
\subsection{Introduction}
\begin{figure}[!h]
	\begin{center}
			\includegraphics[width=8.4cm]{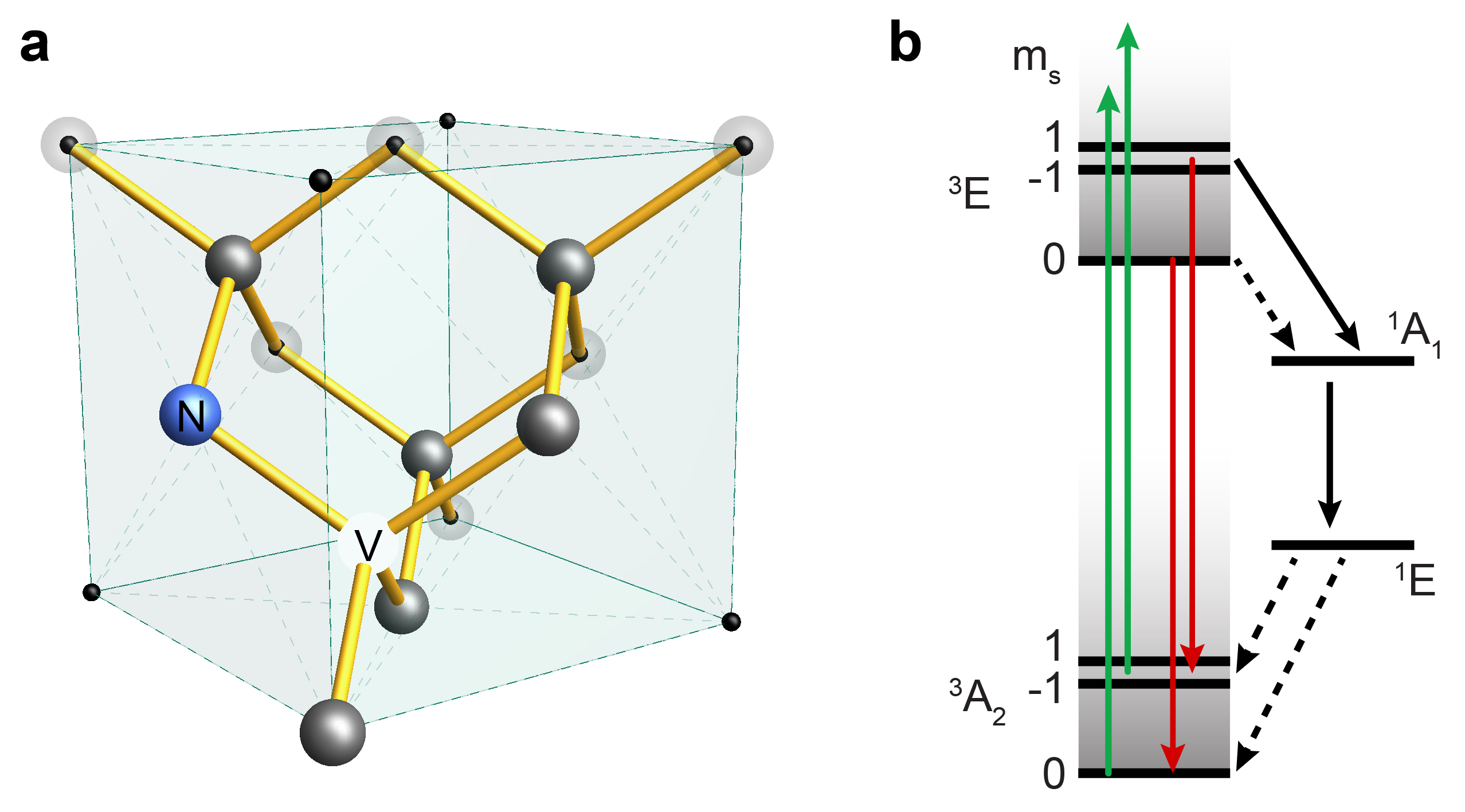}
	\end{center}
	\caption{(a) The NV center in diamond. (b) A simplified energy level diagram showing the optical excitation from the ground electronic $^3A_2$ state to the excited $^3E$ state (green arrows), and the possible relaxation pathways. Dashed lines represent transitions with a lower probability of occurring, and filled lines represent higher probability transitions. Optical pumping with 532~nm light leads to a steady-state overpopulation of the ground $m_s=0$ magnetic sublevel. Image adapted with permission from Ref.~\cite{chatzidrososMiniatureCavityEnhancedDiamond2017}.
	}
	\label{fig:chen2019}
\end{figure}

Diamonds are a crystalline form of carbon which can contain defects of various types within the lattice structure. In one of these, known as nitrogen-vacancy (NV) centers, two neighboring carbon atoms are replaced by a single nitrogen atom, leaving one of the sites vacant, as shown in Fig.~\ref{fig:chen2019}. In synthetic diamonds, NV centers can be incorporated deliberately either during the synthesis, or implanted afterwards~\cite{bassoNanodiamondsSynthesisApplication2020}. They can be uncharged (NV\textsuperscript{0}), or negatively charged (NV\textsuperscript{-}), the latter being of greater interest for NMR applications \cite{suterSinglespinMagneticResonance2017}. All references to `NV-centers' in this work actually refer specifically to NV\textsuperscript{--}-centers. In its ground electronic state, and NV center contains two unpaired electrons (total spin $F=1$) which can be in one of three triplet states ($m_s=0,\pm 1$). This system can be optically pumped to the excited triplet state with green light (532~nm). From this point, the system can either relax directly to its initial state, or relax via an excited singlet state, which favors relaxation to the ground $m_s=0$ magnetic sublevel. After approximately a microsecond of irradiation a steady state is established, with the $m_s=0$ sublevel overpopulated at a polarization level of up to 80\%~\cite{robledoSpinDynamicsOptical2011}. This process is known as `initializing' the NV system. In this way the electron spins can be hyperpolarized at room temperature via irradiation with a single laser, and without the need for cryogenic or high temperature equipment, or high magnetic fields.

The electron spin polarization can then be transferred to nuclear spins to which the electrons are coupled via the hyperfine interaction, such as \textsuperscript{13}C, \textsuperscript{14}N or \textsuperscript{15}N, either at natural abundance or in an isotopically-enriched sample~\cite{alvarezLocalBulk132015}. This is typically done by matching the external magnetic field to the ground-state level anticrossing (GSLAC) or excited-state level anticrossing (ESLAC) condition during the optical pumping period~\cite{jeongMinireviewNuclearSpin2016}, or by irradiating the diamond with microwaves after optical initialization, which can give efficient transfer at variable magnetic field~\cite{alvarezLocalBulk132015,ajoyOrientationindependentRoomTemperature2018}. Recent advances have allowed for the production of suspensions containing \chemical{^{13}C}-polarized diamond nanoparticles with polarization levels on the order of 1\%~\cite{ajoyRoomTemperatureOptical2020}. It has been shown that the \chemical{^{13}C} spins can have a $T_1$ of more than 60~s, and these nanoparticle suspensions have now been applied as a contrast agent for \emph{in vivo} imaging~\cite{waddingtonPhaseEncodedHyperpolarizedNanodiamond2019}. This work does not yet involve microfluidic control, but it is plausible that a microfluidic platform would be an ideal environment on which to form and observe these effects.

\subsection{Hyperpolarizing Liquids on the Surface of a Diamond}
A compelling step beyond hyperpolarizing the nuclear spins within the diamond is the possibility to hyperpolarize a liquid sample on the surface, a concept known as NV-driven dynamic nuclear polarization, or NV-DNP~\cite{abramsDynamicNuclearSpin2014,tetienneProspectsNuclearSpin2021}. This method has many advantages:
\begin{enumerate}
    \item Electron polarization can be generated at room temperature in a weak magnetic field without requiring cryogenic cooling in a large magnetic field,
    \item The hyperpolarization step does not require chemical modification of the target molecule,
    \item The method is general, and does not require a specific chemical structure of the target molecule,
    \item The solid diamond particles can be easily separated from the fluid sample after the polarization step.
\end{enumerate}
The drawback is the short length scale of the effect; the sample should be close to the polarized spins in the diamond to maximize the coupling, meaning this endeavour is likely to be suitable for microfluidic technology.

\begin{figure}
	\begin{center}
			\includegraphics[width=8.4cm]{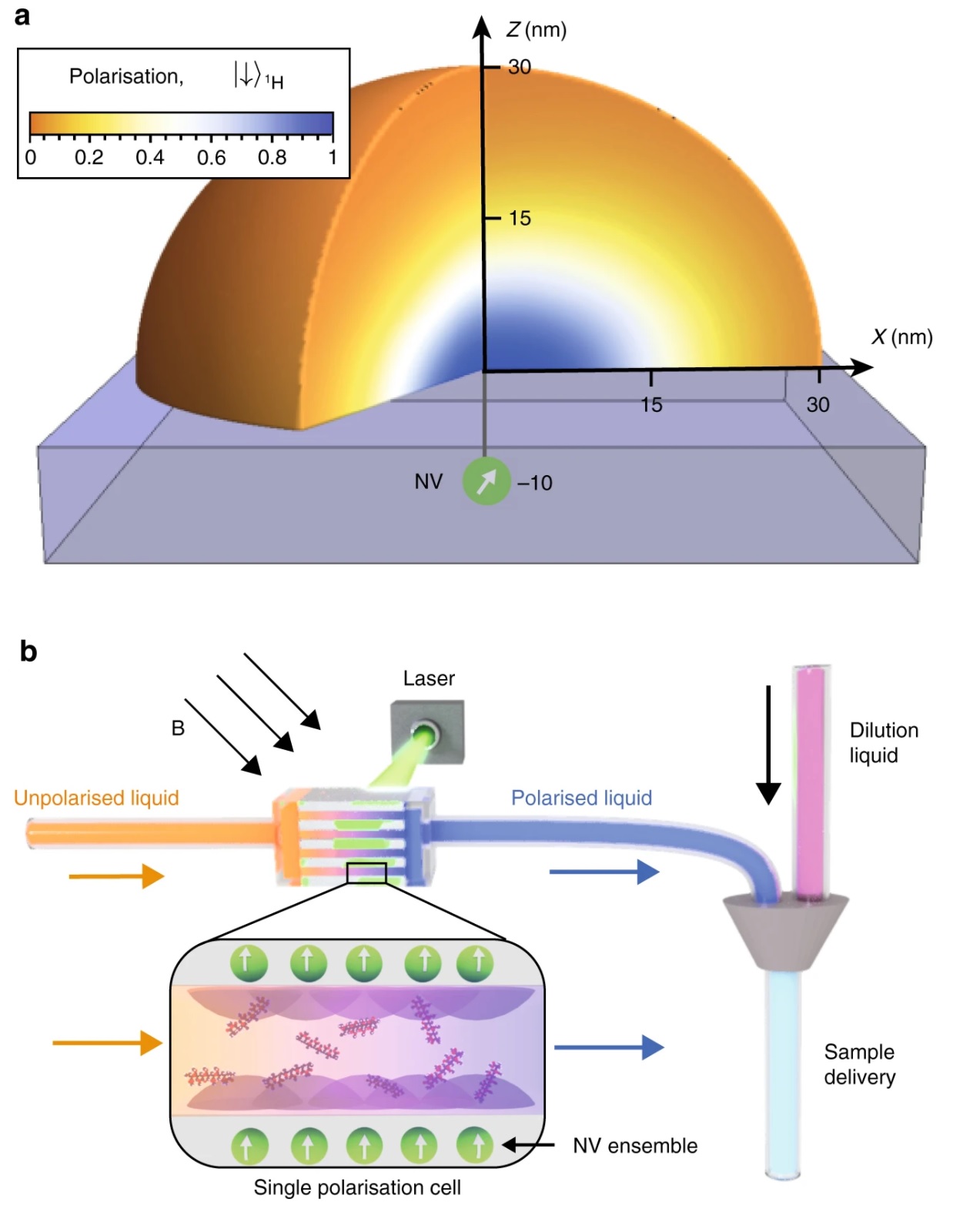}
	\end{center}
	\caption{Hyperpolarization of a liquid on the surface of a diamond. (a) Calculated \chemical{^1H} spin polarization generated via cross relaxation in a sample of PMMA on the surface of a diamond. (b) Illustration of a hypothetical experimental setup for generating hyperpolarized fluids using optically-pumped NV centers in diamond as the polarization source. Image adapted with permission from Ref.~\cite{broadwayQuantumProbeHyperpolarisation2018}.
	}
	\label{fig:broadwayQuantumProbeHyperpolarisation2018}
\end{figure}

Indirect observation of \chemical{^1H} hyperpolarization in a PMMA (poly(methyl methacrylate)) layer baked onto the surface of a diamond has been shown using cross-relaxation-induced polarization~\cite{broadwayQuantumProbeHyperpolarisation2018}. In this method no microwave or rf irradiation is required for polarization transfer. Instead, an external magnetic field is applied to match the nuclear (proton) Zeeman splitting to the energy difference between the $m_s=0$ and $m_s=-1$ electron spin states. This is possible because at zero field the electron $m_s=0$ state is $\sim$2.87~GHz lower in energy than the $m_s=\pm 1$ states (because of the zero-field splitting). By applying a magnetic field to lower the energy of the $m_s=-1$ state, the splitting between $m_s=0$ and $m_s=-1$ can be tuned to the nuclear (proton) Zeeman frequency. Fulfilling this matching condition allows efficient electron-nuclear cross relaxation. An impressive 50\% polarization level on $\sim 10^6$ \chemical{^1H} spins within a sample volume of approximately (26~nm)$^3$ has
been achieved in this way (see Fig.~\ref{fig:broadwayQuantumProbeHyperpolarisation2018}). 

In another work, microscope immersion oil was hyperpolarized on the surface of a diamond~\cite{shagievaMicrowaveAssistedCrossPolarizationNuclear2018} via microwave-assisted cross-polarization, which is achieved by driving electron Rabi oscillations with a microwave field to match the nuclear spin Larmor frequency. This is more generally known as a nuclear spin orientation via electron spin locking (NOVEL) experiment~\cite{henstraNuclearSpinOrientation1988}. The authors were able to experimentally verify the `no-slip' condition often invoked in
fluid dynamics by showing that diffusion of the molecules in close proximity to the diamond surface is hindered compared to the diffusion in the bulk. High viscosity fluids were used to ensure that the diffusion of molecules on the surface was slow, in order to maximize the time spent by a molecule in the vicinity of the NV centers to allow efficient polarization transfer.

To date, no \emph{direct} measurements have been made of an external fluid hyperpolarized via NV-DNP, i.e., using an external sensor. The \emph{indirect} detection described relies on observing a difference in photoluminescence decay rate of the NV when it is on- or off-resonance with the external proton spin bath.

Nanodiamonds can also be used to hyperpolarize liquids via Overhauser DNP~\cite{waddingtonNanodiamondenhancedMRISitu2017}. Paramagnetic defects on the surface of the diamond particles are the source of unpaired electrons that couple to the nuclear spins of the solvent, and saturating the electron transitions can produce a nuclear Overhauser effect that polarizes the solvent. Unlike other experiments with NV centers discussed above, this type of experiment does not utilize the near-unity electron polarization that can be generated by optical pumping. Nevertheless, \chemical{^1H} signal enhancements by a factor of 4 were achieved in bulk water samples containing a nanodiamond powder. As described in Section~\ref{overhauserDNP}, there are many benefits of scaling this method down to a microfluidic scale.

There is a large body of work in which NV centers in diamond are used to detect microfluidic solutions on the surface~\cite{ziemHighlySensitiveDetection2013a,smitsTwodimensionalNuclearMagnetic2019b}, but this work does not involve nuclear spin hyperpolarization, or even NMR, and is therefore outside the scope of this review.

\section{Summary and Outlook}

In summary, the foregoing review has shown that microfluidic lab-on-a-chip technology and hyperpolarized NMR, both
highly dynamic fields of research in their own right, have much to gain from each other. Microfluidic
implementation of chemical conversions and liquid sample manipulation can be advantageous for hyperpolarization
by enhancing electromagnetic radiation penetration in samples, and minimizing transport distances, allowing time savings that can be crucial if relaxation losses 
need to be minimized. On the other hand, microfluidic culture systems can benefit greatly from non-invasive
NMR quantification of metabolic processes, but this is limited by sensitivity. Hyperpolarization provides
a possible solution to this problem.
NMR microcoil technology has improved considerably over the last decade, and customized micro-NMR probes specifically designed for use with microfluidic chips are becoming available. 

Lab-on-a-chip research is predominantly concerned with the study of biological systems, but until recently dissolution dynamic nuclear polarization (D-DNP) was the only hyperpolarization technique capable of reliably producing biocompatible solutions of hyperpolarized molecules with sufficient polarization to allow for metabolic imaging. However, D-DNP yields solutions with volumes of tens of milliliters, and is therefore not easy to integrate effectively with microfluidic experiments. Ongoing research is beginning to overcome these limitations, such as the production of low-volume D-DNP solutions, and parallelization of many microfluidic detectors to take advantage of the large sample volumes. In addition to this, other hyperpolarization techniques are now available to aid in lab-on-a-chip metabolic imaging: parahydrogen-induced polarization can now produce biologically-compatible solutions of hyperpolarized molecules, and hyperpolarized diamond nanopowders can be generated in aqueous solution. These advances have greatly widened the scope for using hyperpolarization in microfluidic systems, and the corresponding improvement in sensitivity makes it much more feasible to use NMR spectroscopy as the detection method for lab-on-a-chip experiments.

The evolution of these two fields is mutually-beneficial, and we see a positive feedback loop with microfluidics supporting developments in hyperpolarized NMR, and hyperpolarized NMR enabling more applications of microfluidics to study chemical and biological systems. Improvements in microfabrication technologies widens the scope of hyperpolarization experiments that can be implemented on a microfluidic chip. 
To maximize future progress, it is therefore important for researchers active in both fields
to be aware of the possibilities and developments in the other. We hope that the present review
may make a modest contribution in this regard. The reward for successfully harnessing the synergies between
microfluidics and hyperpolarization could be very large indeed, and bring about non-invasive, detailed quantification
of life processes in microfluidic culture systems. Combined with the power of microfluidics to 
enhance experimental throughput, reliability, and to lower cost, this could have a profound impact
on the life sciences, for example in drug discovery, drug safety testing, and in personalised medicine.

\section{Acknowledgments}
J.E. would like to acknowledge funding from the European Union’s Horizon 2020 research and innovation programme under the Marie Skłodowska-Curie Grant Agreement No. 766402.
M.U. gratefully acknowledges support from the EU H2020 FETOPEN project "TISuMR" (Grant Agreement No. 737043). The authors thank Dr.~Giorgos Chatzidrosos and Dr.~Kirill Sheberstov for helpful comments on the manuscript, and J.E. would like to thank Prof.~Igor Koptyug and Prof.~Dmitry Budker for stimulating discussions about hyperpolarization-enhanced NMR and catalysis. W.H. thanks Prof.~Russ Bowers for helpful discussions and mentorship.

\printbibliography

\section{Glossary}
AVAM: \indent Alkali vapour atomic magnetometer\\
CIDNP: \indent Chemically-induced dynamic nuclear polarization\\
cLOD: \indent Concentration limit of detection\\
COSY: \indent Correlation Spectroscopy\\
D-DNP: \indent Dissolution dynamic nuclear polarization\\
dc: \indent Direct current
DNA: \indent Deoxyribonucleic acid\\
DNP: \indent Dynamic nuclear polarization\\
EPR: \indent Electron paramagnetic resonance\\
ESLAC: \indent Excited-state level anti-crossing\\
GSLAC: \indent Ground-state level anti-crossing\\
HPLC: \indent High-performance liquid chromatography\\
HSQC: \indent Heteronuclear single quantum correlation\\
I.D.: \indent Inner diameter\\
LC-NMR: \indent Liquid chromatography nuclear magnetic resonance\\
MRI: \indent Magnetic resonance imaging\\
$\mu$TAS: \indent Miniaturized total analysis system\\
nLOD: \indent Number limit of detection\\
NMR: \indent Nuclear magnetic resonance\\
NOVEL: \indent Nuclear spin orientation via electron spin locking\\
NV: \indent Nitrogen vacancy\\
NV-DNP: \indent Nitrogen vacancy dynamic nuclear polarization\\
OPM: \indent Optically-pumped magnetometer\\
PASADENA: \indent Parahydrogen and synthesis allow dramatically-enhanced nuclear alignment\\
PCR: \indent Polymerase chain reaction\\
PDMS: \indent Polydimethylsiloxane\\
PHIP: \indent Parahydrogen-induced polarization\\
photo-CIDNP: \indent Photo-chemically induced dynamic nuclear polarization\\
PMMA: \indent Poly(methyl methacrylate)\\
rf: \indent Radiofrequency\\
RF-SABRE: \indent Radiofrequency signal amplification by reversible exchange\\
SABRE: \indent Signal amplification by reversible exchange\\
SEOP: \indent Spin-exchange optical pumping\\
SLIC-SABRE: \indent Spin-lock induced crossing signal amplification by reversible exchange\\
TEMPO: \indent (2,2,6,6-tetramethylpiperidin-1-yl)oxyl\\
TEMPOL: \indent 4-hydroxy-2,2,6,6-tetramethylpiperidin-1-oxyl\\
TOF: \indent Time-of-flight\\
ZULF: \indent Zero- to ultralow-field\\

\begin{figure}
	\begin{center}
			\includegraphics[width=8.4cm]{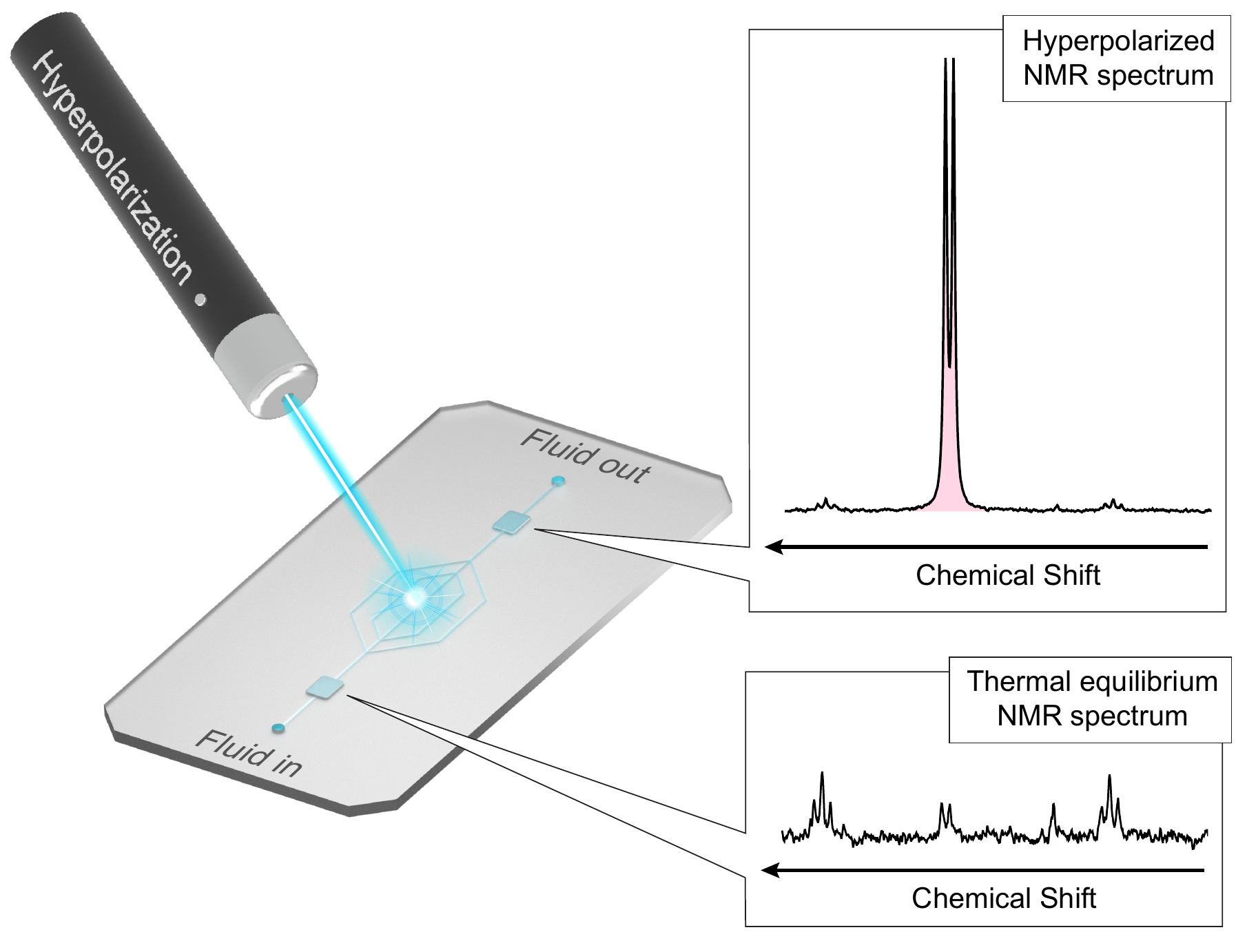}
	\end{center}
	\caption{TOC Figure: Hyperpolarization techniques are used to boost the intensity of selected NMR signals.}
	\label{fig:TOC}
\end{figure}

\end{document}

%% file: symbols.tex
\newcommand{\degC}{${}^\circ\mathrm{C}$}
\newcommand{\para}{\textit{para}}
\newcommand{\Htwo}{\chemical{H\textsubscript{2}}}
\newcommand{\Ntwo}{\chemical{N\textsubscript{2}}}
\newcommand{\carbon}{\chemical{\textsuperscript{13}C}}
\newcommand{\proton}{\chemical{\textsuperscript{1}H}}
\newcommand{\Xe}{\chemical{\textsuperscript{129}Xe}}
\newcommand{\Rb}{\chemical{\textsuperscript{87}Rb}}
\newcommand{\muL}{\ensuremath{\mu\mathrm{l}}}
\newcommand{\muT}{\ensuremath{\mu\mathrm{T}}}
\newcommand{\mum}{\ensuremath{\mu\mathrm{m}}}
\newcommand{\pHtwo}{\textit{p}-\chemical{H_2}}

\newcommand{\muM}{\ensuremath{\mu\mathrm{M}}}